\documentclass[twocolumn,english,aps,prl,superscriptaddress,floats,nobibnotes]{revtex4-1}
\usepackage[latin9]{inputenc}
\usepackage{color}
\usepackage{babel}
\usepackage{bm}
\usepackage{bbm}
\usepackage{amsmath}
\usepackage{amssymb}
\usepackage{graphicx}
\PassOptionsToPackage{version=3}{mhchem}
\usepackage{mhchem}
\usepackage{braket}
\usepackage[unicode=true,
 bookmarks=false,
 breaklinks=false,pdfborder={0 0 1},backref=false,colorlinks=false]
 {hyperref}
\hypersetup{
 colorlinks,linkcolor=blue,citecolor=blue,urlcolor=blue}

\makeatletter

\usepackage{babel}
\PassOptionsToPackage{version=3}{mhchem}

\usepackage{newfloat}
\usepackage{latexsym}
\usepackage{dcolumn}
\usepackage{float}

\usepackage{bm}
\usepackage{babel}

\makeatother

\begin{document}

\title{Longitudinal magnetoconductance and the planar Hall effect in a lattice model of tilted Weyl fermions.}

\author{Azaz Ahmad}
\affiliation{School of Basic Sciences, Indian Institute of Technology Mandi, Mandi 175005, India}

\author{Girish Sharma}
\affiliation{School of Basic Sciences, Indian Institute of Technology Mandi, Mandi 175005, India}

\date{\today}
\begin{abstract}
The experimental verification of chiral anomaly in Weyl semimetals is an active area of investigation in modern condensed matter physics, which typically relies on the  combined signatures of longitudinal magnetoconductance (LMC) along with the planar Hall effect (PHE). It has recently been shown that for weak non-quantizing magnetic fields, a sufficiently strong finite intervalley scattering drives the system to switch the sign of LMC from positive to negative. Here we unravel another  independent source that produces the same effect. Specifically, a smooth lattice cutoff to the linear dispersion, which is ubiquitous in real Weyl materials, introduces nonlinearity in the problem and also drives the system to exhibit negative LMC for non-collinear electric and magnetic fields even in the limit of vanishing intervalley scattering. We examine longitudinal magnetoconductivity and the planar Hall effect semi-analytically for a lattice model of tilted Weyl fermions within the Boltzmann approximation. We independently study the effects of a finite lattice cutoff and tilt parameters and construct phase diagrams in relevant parameter spaces that are relevant for diagnosing chiral anomaly in real Weyl materials. 
\end{abstract}

\maketitle
\section{Introduction}
As dictated by the well-known no-crossing theorem~\cite{neumann1929}, the Bloch bands in a solid typically do no cross each other at any point in the Brillouin zone. Some exceptions to this general rule are Dirac and Weyl materials, where non-trivial topology of the Bloch bands can stabilize the band degenerate point~\cite{volovik2003universe,chiu2016classification,armitage2018weyl,yang2018symmetry,murakami2007phase,murakami2007tuning,burkov2011topological,burkov2011weyl,wan2011topological,xu2011chern,yang2011quantum}.  In a Weyl semimetal (WSM), a band crossing point, also known as a Weyl node, can act as a source or sink of Abelian Berry curvature~\cite{xiao2010berry}. Since the net Berry flux through the Brillouin zone must vanish, the Weyl nodes must occur in multiples of two. The topological nature of the Bloch bands in a WSM gives rise to very interesting physics typically that is absent in conventional condensed matter systems. Some examples include the manifestation of anomalous Hall~\cite{yang2011quantum,burkov2014anomalous} and Nernst~\cite{sharma2016nernst,sharma2017nernst,liang2017anomalous} effects, open Fermi arcs~\cite{wan2011topological}, and the most prominent one being the manifestation of chiral or Adler-Bell-Jackiw anomaly~\cite{adler1969axial,nielsen1981no,nielsen1983adler,bell1969pcac,aji2012adler,zyuzin2012weyl,zyuzin2012weyl,son2012berry,goswami2013axionic,goswami2015optical, fukushima2008chiral}.

Weyl fermions have an associated chirality quantum number that is identical with the integral of the flux of the Berry curvature around a Weyl node. The number of Weyl fermions of a specific chirality remain conserved in the absence of an external gauge or gravitational field coupling. However, in the presence of background gauge fields, such as electric and magnetic fields, the separate number conservation laws for Weyl fermions is violated~\cite{adler1969axial,nielsen1981no,nielsen1983adler}. This is the result of chiral anomaly in Weyl fermions and has its origins rooted in high-energy physics. The verification of chiral anomaly in  Weyl semimetals is one of the most active areas of investigation in condensed matter physics.

Chiral anomaly in WSMs maybe verified by experimental probes such as that measure magnetoconductance~\cite{son2013chiral,kim2014boltzmann,zyuzin2017magnetotransport,he2014quantum,liang2015ultrahigh,zhang2016signatures,li2016chiral,xiong2015evidence,hirschberger2016chiral}, Hall effectt~\cite{nandy2017chiral,kumar2018planar,yang2019w,li2018giant,chen2018planar,li2018giant2,yang2019frustration,pavlosiuk2019negative,singha2018planar}, thermopower~\cite{lundgren2014thermoelectric,sharma2016nernst,sharma2019transverse,das2019berry}, or optical activity~\cite{goswami2015optical}. It was initially concluded that chiral anomaly in WSMs directly correlates with the observation of positive longitudinal magnetoconductance. For example, from elementary field-theory calculations~\cite{fukushima2008chiral}, the chiral chemical potential ($\mu_5$, which is difference between the chemical potential between Weyl nodes of two chiralities) created by the external parallel $\mathbf{E}$ and $\mathbf{B}$ fields in the presence of intervalley scattering is $\mu_5 = 3v_F^3 e^2 \tau_i {E}{B}/4\hbar^2 \mu^2$, where $v_F$, $\tau_i$, and $\mu$ denote the Fermi velocity, scattering time, and the chemical potential, respectively. The corresponding longitudinal current is given by ${j} = e^2 \mu_5 {B}/ 2\pi^2$, which immediately gives us positive longitudinal magnetoconductance. However, a detailed analysis shows that positive longitudinal
magnetoconductance is neither a necessary, nor a sufficient condition to prove the existence of chiral anomaly in WSMs. It has now been well established that both positive or negative magnetoconductance can arise from chiral anomaly in WSMs~\cite{goswami2015axial,lu2015high,chen2016positive,zhang2016linear,shao2019magneto,li2016weyl,ji2018effect,spivak2016magnetotransport,das2019linear,imran2018berry,dantas2018magnetotransport,johansson2019chiral,grushin2016inhomogeneous,cortijo2016linear,sharma2017chiral,knoll2020negative,xiao2020linear,sharma2020sign}. In the presence of strong magnetic field, when Landau quantization is relevant, the sign of magnetoconductance depends on the nature of scattering impurities~\cite{goswami2015axial,lu2015high,chen2016positive,zhang2016linear,shao2019magneto,li2016weyl,ji2018effect}. For weak magnetic fields, it was recently shown that sufficiently strong intervalley scattering can switch the sign of LMC~\cite{knoll2020negative,xiao2020linear}.

In this work we unravel another independent source that produces negative LMC for weak non-collinear electric and magnetic fields even for vanishing intervalley scattering strength. 
Around a Weyl node, the energy dispersion locally behaves as $\epsilon^\chi_\mathbf{k} = \hbar v_F k$, where $v_F$ is the Fermi velocity, while $k$ is the modulus of the wavevector measured from the nodal point. In practice, the linear energy dispersion around a Weyl node is only valid for a small energy window. In a realistic lattice model of Weyl fermions, the bands are no longer linear far apart from the nodal point, and the lattice regularization provides a physical ultraviolet cutoff to the low-energy spectrum. The lattice model of Weyl fermions introduces a source non-linearity in the problem and has important implications in several physical properties. For example, the lattice model of Weyl fermions produces a non-zero Nernst effect~\cite{sharma2016nernst,sharma2017nernst} (as also observed experimentally~\cite{liang2017anomalous}), which is otherwise predicted to vanish in the linear approximation~\cite{lundgren2014thermoelectric}. Here, we semi-analytically examine longitudinal magnetoconductance and the planar Hall effect for a lattice model of Weyl fermions. Earlier works on a lattice model of Weyl semimetals mostly resort to numerical evaluation of various intrinsic quantities such as the Berry curvature and the orbital magnetic moment, as well as transport quantities such as longitudinal conductance or the Hall conductance~\cite{sharma2016nernst,sharma2017nernst,sharma2017chiral,nandy2017chiral,goswami2013axionic}. The lattice model we adopt has exact analytical expressions for the Berry curvature as well as the orbital magnetic moment at all energies. The associated transport quantities are also  evaluated semi-analytically within the Boltzmann formalism. We find that nonlinear lattice effects can produce negative LMC for non-collinear electric and magnetic fields even in the absence of intervalley scattering. Crucially, it is important to account for orbital magnetic moment effects to obtain negative LMC.
We also find that in the presence of finite intervalley scattering, lattice effects drive the system to exhibit negative longitudinal magnetoconductance quickly at a lesser threshold of intervalley scattering as compared to the linearized approximation. 

Further, in realistic materials the Weyl cones not only have a smooth lattice cutoff but are also in generally tilted along a particular direction. We also examine longitudinal magnetoconductance $\sigma_{zz}$ and the planar Hall conductance $\sigma_{xz}$ in the presence of a tilt parameter both parallel and perpendicular to the $z-$direction. When the electric and magnetic fields are aligned parallel to each other, and Wwhen the Weyl cones are tilted along the  direction of the magnetic field, LMC is  quadratic if the cones are oriented in the same direction, and the sign of LMC depends on the strength of intervalley scattering ($\alpha_i$). When the cones are tilted opposite to each other, LMC is found to be linear-in-$B$ with sign depending on the magnitude of the tilt as well as $\alpha_i$. When the cones are tilted perpendicular to the direction of the magnetic field, LMC is found to be quadratic, with the sign again depending on the value of intervalley scattering strength $\alpha_i$. However, more interesting features emerge when LMC is examined for non-collinear electric and magnetic fields, as demonstrated by several phase plots in the $\alpha_i-t_k$ space ($t_k$ being the tilt parameter). We also find that the planar Hall conductance also shows linear-in-$B$ behaviour for tilted Weyl cones oriented opposite to each other, and this linear-in-$B$ behavior is enhanced in the presence of intervalley scattering $\alpha_i$.

\section{Boltzmann formalism for magnetotransport}
We begin with the most general form of a tilted type-I Weyl node of a particular chirality $\chi$, including non-linear effects away from the Weyl node due to lattice regularization. The Hamiltonian expanded around each Weyl point can be expressed as 
\begin{align}
H_\mathbf{k} = \chi E_0 p(a \mathbf{k}\cdot \boldsymbol{\sigma}) + T^\chi_x q(a k_x) + T^\chi_z r(a k_z).
\label{Eq_H1weyl}
\end{align}
In the above expression, $E_0$ is an energy parameter, $T^\chi_x$ and $T^\chi_z$ are tilt parameters along the $x$ and $z$ directions, respectively, $\mathbf{k}$ is the momentum measured relative to the Weyl point, $\boldsymbol{\sigma}$ is the vector of the Pauli matrices. The functions, $p$, $q$, and $r$ are can assume any form as long as $p(0)=q(0)=r(0)=0$, but we choose $p(x) = q(x) = r(x) = \sin(x)$ as prototype of a lattice Weyl node. The corresponding energy dispersion is given by 
\begin{align}
\epsilon^\chi_k=\pm E_0 \sin(ka) +T^{\chi}_z \sin(a k_z) +T^{\chi}_x \sin(a k_x).
\label{Eq_E1kweyl}
\end{align}
Note that for a Weyl node without any tilt the energy bandwidth equals $2E_0$. 

We  study charge transport for weak electric and magnetic fields via the quasiclassical Boltzmann theory and thus the Landau quantization regime will not be relevant for our discussion. A phenomenological Boltzmann equation for the non-equilibrium distribution function $f^\chi_\mathbf{k}$ can be written as~\cite{bruus2004many} 
\begin{align}
\left(\frac{\partial}{\partial t} + \dot{\mathbf{r}}^\chi\cdot \nabla_\mathbf{r}+\dot{\mathbf{k}}^\chi\cdot \nabla_\mathbf{k}\right)f^\chi_\mathbf{k} = \mathcal{I}_{{col}}[f^\chi_\mathbf{k}],
\label{Eq_boltz1}
\end{align}
where the collision term on the right-hand side incorporates the effect of impurity scattering.
In the presence of electric ($\mathbf{E}$) and magnetic ($\mathbf{B}$) fields, the dynamics of the Bloch electrons is modified as~\cite{son2012berry} 
\begin{align}
\dot{\mathbf{r}}^\chi &= \mathcal{D}^\chi \left( \frac{e}{\hbar}(\mathbf{E}\times \boldsymbol{\Omega}^\chi + \frac{e}{\hbar}(\mathbf{v}^\chi\cdot \boldsymbol{\Omega}^\chi) \mathbf{B} + \mathbf{v}_\mathbf{k}^\chi)\right) \nonumber\\
\dot{\mathbf{p}}^\chi &= -e \mathcal{D}^\chi \left( \mathbf{E} + \mathbf{v}_\mathbf{k}^\chi \times \mathbf{B} + \frac{e}{\hbar} (\mathbf{E}\cdot\mathbf{B}) \boldsymbol{\Omega}^\chi \right),
\end{align}
where $\mathbf{v}_\mathbf{k}^\chi$ is the band velocity, $\boldsymbol{\Omega}^\chi = -\chi \mathbf{k} /2k^3$ is the Berry curvature, and $\mathcal{D}^\chi = (1+e\mathbf{B}\cdot\boldsymbol{\Omega}^\chi/\hbar)^{-1}$ is the factor by which the phase space volume is modified to due Berry phase effects. The self-rotation of Bloch wavepacket also gives rise to an orbital magnetic moment (OMM)~\cite{xiao2010berry} that is given by $\mathbf{m}^\chi_\mathbf{k} = -e \chi E_0 \sin(ak) \mathbf{k} /2\hbar k^3$ for the above lattice model (see Appendix A for details). In the presence of magnetic field, the OMM shifts the energy dispersion as $\epsilon^{\chi}_{\mathbf{k}}\rightarrow \epsilon^{\chi}_{\mathbf{k}} - \mathbf{m}^\chi_\mathbf{k}\cdot \mathbf{B}$. Note that the Berry curvature and the orbital magnetic moment are independent of the tilting of the Weyl cones.

The collision integral must take into account scattering between the two Weyl cones (internode, $\chi\Longleftrightarrow\chi'$), as well as scattering withing a Weyl cone (intranode, $\chi\Longleftrightarrow\chi$), and thus $\mathcal{I}_{{col}}[f^\chi_\mathbf{k}]$ can be expressed as 
\begin{align}
\mathcal{I}_{{col}}[f^\chi_\mathbf{k}] = \sum\limits_{\chi'}\sum\limits_{\mathbf{k}'} W^{\chi\chi'}_{\mathbf{k},\mathbf{k}'} (f^{\chi'}_{\mathbf{k}'} - f^\chi_\mathbf{k}),
\end{align}
where the scattering rate $W^{\chi\chi'}_{\mathbf{k},\mathbf{k}'}$ in the first Born approximation is given by~\cite{bruus2004many} 
\begin{align}
W^{\chi\chi'}_{\mathbf{k},\mathbf{k}'} = \frac{2\pi}{\hbar} \frac{n}{\mathcal{V}} |\langle \psi^{\chi'}_{\mathbf{k}'}|U^{\chi\chi'}_{\mathbf{k}\mathbf{k}'}|\psi^\chi_\mathbf{k}\rangle|^2 \delta(\epsilon^{\chi'}_{\mathbf{k}'}-\epsilon_F)
\label{Eq_W_1}
\end{align}
In the above expression $n$ is the impurity concentration, $\mathcal{V}$ is the system volume, $|\psi^\chi_\mathbf{k}\rangle$ is the Weyl spinor wavefunction (obtained by diagonalizing Eq.~\ref{Eq_H1weyl}), $U^{\chi\chi'}_{\mathbf{k}\mathbf{k}'}$ is the scattering potential profile, and $\epsilon_F$ is the Fermi energy. The scattering potential profile $U^{\chi\chi'}_{\mathbf{k}\mathbf{k}'}$ is determined by the nature of impurities (whether charged or uncharged or magnetic). Here we restrict our attention only to non-magnetic point-like scatterers, but particularly distinguish between intervalley and intravalley scattering that can be controlled independently in our formalism. Thus, the scattering matrix is momentum-independent but has a chirality dependence, i.e.,  $U^{\chi\chi'}_{\mathbf{k}\mathbf{k}'} = U^{\chi\chi'}\mathbb{I}$.

The distribution function is assumed to take the form $f^\chi_\mathbf{k} = f_0^\chi + g^\chi_\mathbf{k}$, where $f_0^\chi$ is the equilibrium Fermi-Dirac distribution function and $g^\chi_\mathbf{k}$ indicates the deviation from equilibrium. 
In the steady state, the Boltzmann equation (Eq.~\ref{Eq_boltz1}) takes the following form 
\begin{align}
&\left[\left(\frac{\partial f_0^\chi}{\partial \epsilon^\chi_\mathbf{k}}\right) \mathbf{E}\cdot \left(\mathbf{v}^\chi_\mathbf{k} + \frac{e\mathbf{B}}{\hbar} (\boldsymbol{\Omega}^\chi\cdot \mathbf{v}^\chi_\mathbf{k}) \right)\right]\nonumber\\
 &= -\frac{1}{e \mathcal{D}^\chi}\sum\limits_{\chi'}\sum\limits_{\mathbf{k}'} W^{\chi\chi'}_{\mathbf{k}\mathbf{k}'} (g^\chi_{\mathbf{k}'} - g^\chi_\mathbf{k})
 \label{Eq_boltz2}
\end{align}
The deviation $g^\chi_\mathbf{k}$ is assumed to be small such that its gradient can be neglected and is also assumed to be proportional to the applied electric field 
\begin{align}
g^\chi_\mathbf{k} = e \left(-\frac{\partial f_0^\chi}{\partial \epsilon^\chi_\mathbf{k}}\right) \mathbf{E}\cdot \boldsymbol{\Lambda}^\chi_\mathbf{k}
\end{align}
We will fix the direction of the applied external electric field to be along $+\hat{z}$, i.e., $\mathbf{E} = E\hat{z}$. Therefore only ${\Lambda}^{\chi z}_\mathbf{k}\equiv {\Lambda}^{\chi}_\mathbf{k}$, is relevant. Further, we rotate the magnetic field along the $xz$-plane such that it makes an angle $\gamma$ with respect to the $\hat{x}-$axis, i.e., $\mathbf{B} = B(\cos\gamma,0,\sin\gamma)$. When $\gamma=\pi/2$, the electric and magnetic fields are parallel to each other. When $\gamma\neq \pi/2$, the electric and magnetic fields are non-collinear and this geometry will be useful in analyzing the planar Hall effect, as well as LMC in a non-collinear geometry that has non-trivial implications in a lattice model as well as for tilted Weyl fermions even in the linear approximation.

Keeping terms only up to linear order in the electric field, Eq.~\ref{Eq_boltz2} takes the following form 
\begin{align}
\mathcal{D}^\chi \left[v^{\chi z}_{\mathbf{k}} + \frac{e B}{\hbar} \sin \gamma (\boldsymbol{\Omega}^\chi\cdot \mathbf{v}^\chi_\mathbf{k})\right] = \sum\limits_{\eta}\sum\limits_{\mathbf{k}'} W^{\eta\chi}_{\mathbf{k}\mathbf{k}'} (\Lambda^{\eta}_{\mathbf{k}'} - \Lambda^\chi_\mathbf{k})
\label{Eq_boltz3}
\end{align} 
In order to solve the above equation, we first define valley the scattering rate as follows
\begin{align}
\frac{1}{\tau^\chi_\mathbf{k}} = \mathcal{V} \sum\limits_{\eta} \int{\frac{d^3 \mathbf{k}'}{(2\pi)^3} (\mathcal{D}^\eta_{\mathbf{k}'})^{-1} W^{\eta\chi}_{\mathbf{k}\mathbf{k}'}}
\label{Eq_tau11}
\end{align}
One would assume that when $\gamma=\pi/2$, due to the electric and magnetic field both being parallel to the $\hat{z}$ axis the azimuthal symmetry is retained in the problem. However, due to the tilting of the Weyl cones the azimuthal symmetry is destroyed even for parallel electric and magnetic fields, and therefore the above integration (and all other subsequent integrations) must be performed both over $\theta$ and $\phi$ when either (i) the Weyl cones are tilted and/or (ii) $\gamma\neq\pi/2$. Note that finite lattice effects by themselves do not break azimuthal symmetry.  The radial integration is simplified due to the delta-function in Eq.~\ref{Eq_W_1}.

Substituting the scattering rate from Eq.~\ref{Eq_W_1} in the above equation, we have 
\begin{widetext}
\begin{align}
\frac{1}{\tau^\chi_\mathbf{k}} = \frac{\mathcal{V}N}{8\pi^2 \hbar} \sum\limits_{\eta} |U^{\chi\eta}|^2 \iiint{(k')^2 \sin \theta' \mathcal{G}^{\chi\eta}(\theta,\phi,\theta',\phi') \delta(\epsilon^{\eta}_{\mathbf{k}'}-\epsilon_F)(\mathcal{D}^\eta_{\mathbf{k}'})^{-1}dk'd\theta'd\phi'},
\label{Eq_tau1}
\end{align}
\end{widetext}
where $N$ now indicates the total number of impurities, and $ \mathcal{G}^{\chi\eta}(\theta,\phi,\theta',\phi') = (1+\chi\eta(\cos\theta \cos\theta' + \sin\theta\sin\theta' \cos(\phi-\phi')))$ is the Weyl chirality factor defined by the overlap of the wavefunctions. Since quasiclassical Boltzmann theory is valid away from the nodal point such that $\mu^2\gg \hbar v_F^2 e B$, therefore without any loss of generality we will assume that the chemical potential lies in the conduction band. 

Including orbital magnetic moment effects, the energy dispersion $\epsilon^\chi_\mathbf{k}$ is in general a function of several parameters including the chirality index, i.e.,  $\epsilon^\chi_\mathbf{k}= \epsilon^\chi_\mathbf{k}(E_0,k,a,\chi,B,\theta,\gamma)$. This equation has to be inverted tin order to find a constant energy contour $k^\chi = k^\chi(E_0,\epsilon^\chi_\mathbf{k},a,B,\theta,\gamma)$. For the case of lattice Weyl fermions, a closed-form analytical solution is not feasible and we will resolve to a numerical solution for $k^\chi$. For tilted Weyl fermions in the linearized spectrum approximation, it is possible to invert the equation as will be shown shortly. 

The three-dimensional integral in Eq.~\ref{Eq_tau1} is then reduced to just integration in $\phi'$ and $\theta'$. The scattering time ${\tau^\chi_\mathbf{k}}$ depends on the chemical potential ($\mu$), and is a function of the angular variables $\theta$ and $\phi$. 

\begin{align}
\frac{1}{\tau^\chi_\mu(\theta,\phi)} = \mathcal{V} \sum\limits_{\eta} \iint{\frac{\beta^{\chi\eta}(k')^3}{|\mathbf{v}^\eta_{\mathbf{k}'}\cdot \mathbf{k}'^\eta|}\sin\theta'\mathcal{G}^{\chi\eta}(\mathcal{D}^\eta_{\mathbf{k}'})^{-1} d\theta'd\phi'},
\label{Eq_tau2}
\end{align}
where the prefactor $\beta^{\chi\eta} = N|U^{\chi\eta}|^2 / 4\pi^2 \hbar^2$. The Boltzmann equation (Eq~\ref{Eq_boltz3}) assumes the form  
\begin{align}
&h^\chi_\mu(\theta,\phi) + \frac{\Lambda^\chi_\mu(\theta,\phi)}{\tau^\chi_\mu(\theta,\phi)} =\nonumber\\ &\mathcal{V}\sum_\eta \iint {\frac{\beta^{\chi\eta}(k')^3}{|\mathbf{v}^\eta_{\mathbf{k}'}\cdot \mathbf{k}'^\eta|} \sin\theta'\mathcal{G}^{\chi\eta}(\mathcal{D}^\eta_{\mathbf{k}'})^{-1}\Lambda^\eta_{\mu}(\theta',\phi') d\theta'd\phi'}
\label{Eq_boltz4}
\end{align}
We make the following ansatz for $\Lambda^\chi_\mu(\theta,\phi)$
\begin{align}
\Lambda^\chi_\mu(\theta,\phi) &= (\lambda^\chi - h^\chi_\mu(\theta,\phi) + a^\chi \cos\theta +\nonumber\\
&b^\chi \sin\theta\cos\phi + c^\chi \sin\theta\sin\phi)\tau^\chi_\mu(\theta,\phi),
\label{Eq_Lambda_1}
\end{align}
where we solve for the eight unknowns ($\lambda^{\pm 1}, a^{\pm 1}, b^{\pm 1}, c^{\pm 1}$). The L.H.S in Eq.~\ref{Eq_boltz4} simplifies to $\lambda^\chi + a^\chi \cos\theta + b^\chi \sin\theta\cos\phi + c^\chi \sin\theta\sin\phi$. The R.H.S of Eq.~\ref{Eq_boltz4} simplifies to
\begin{align}
\mathcal{V}\sum_\eta \beta^{\chi\eta} \iint &f^{\eta} (\theta',\phi') \mathcal{G}^{\chi\eta} (\lambda^\eta - h^\eta_\mu(\theta',\phi') + a^\eta \cos\theta' +\nonumber\\
	&b^\eta \sin\theta'\cos\phi' + c^\eta \sin\theta'\sin\phi')d\theta'd\phi',
	\label{Eq_boltz5_rhs}
\end{align}
where the function
\begin{align}
f^{\eta} (\theta',\phi') = \frac{(k')^3}{|\mathbf{v}^\eta_{\mathbf{k}'}\cdot \mathbf{k}'^\eta|} \sin\theta' (\mathcal{D}^\eta_{\mathbf{k}'})^{-1} \tau^\chi_\mu(\theta',\phi')
\label{Eq_f_eta}
\end{align}
The above equations, when written down explicitly take the form of seven simultaneous equations to be solved for eight variables (see Appendix B for details). The last constraint comes from the particle number conservation 
\begin{align}
\sum\limits_{\chi}\sum\limits_{\mathbf{k}} g^\chi_\mathbf{k} = 0
\label{Eq_sumgk}
\end{align}
Thus Eq.~\ref{Eq_Lambda_1}, Eq.~\ref{Eq_boltz5_rhs}, Eq.~\ref{Eq_f_eta} and Eq.~\ref{Eq_sumgk} can be solved together with Eq~\ref{Eq_tau2}, simultaneously for the eight unknowns ($\lambda^{\pm 1}, a^{\pm 1}, b^{\pm 1}, c^{\pm 1}$). Due to the complicated nature of the problem, the associated two dimensional integrals w.r.t \{$\theta'$, $\phi'$\}, and the solution of the simultaneous equations are all performed numerically. 
Before we proceed further, we will divide our results into two broad classes. The first class considers the effects of introducing a natural lattice cutoff for Weyl fermions without considering tilting of the Weyl cones. In the second class, we consider effects due to  tilting the Weyl cones in the linearized spectrum approximation, that is without considering effects due to a finite lattice cutoff. Although our formalism can handle the generic case of tilted lattice Weyl fermion, the reason for this division is because effects due to lattice and due to tilting of the Weyl cones can in fact be considered independent of each other, and linearized approximation speeds up the numerical computation. The combined effect from the two gives the net result.

\begin{figure*}
    \centering
    \includegraphics[width=0.49\columnwidth]{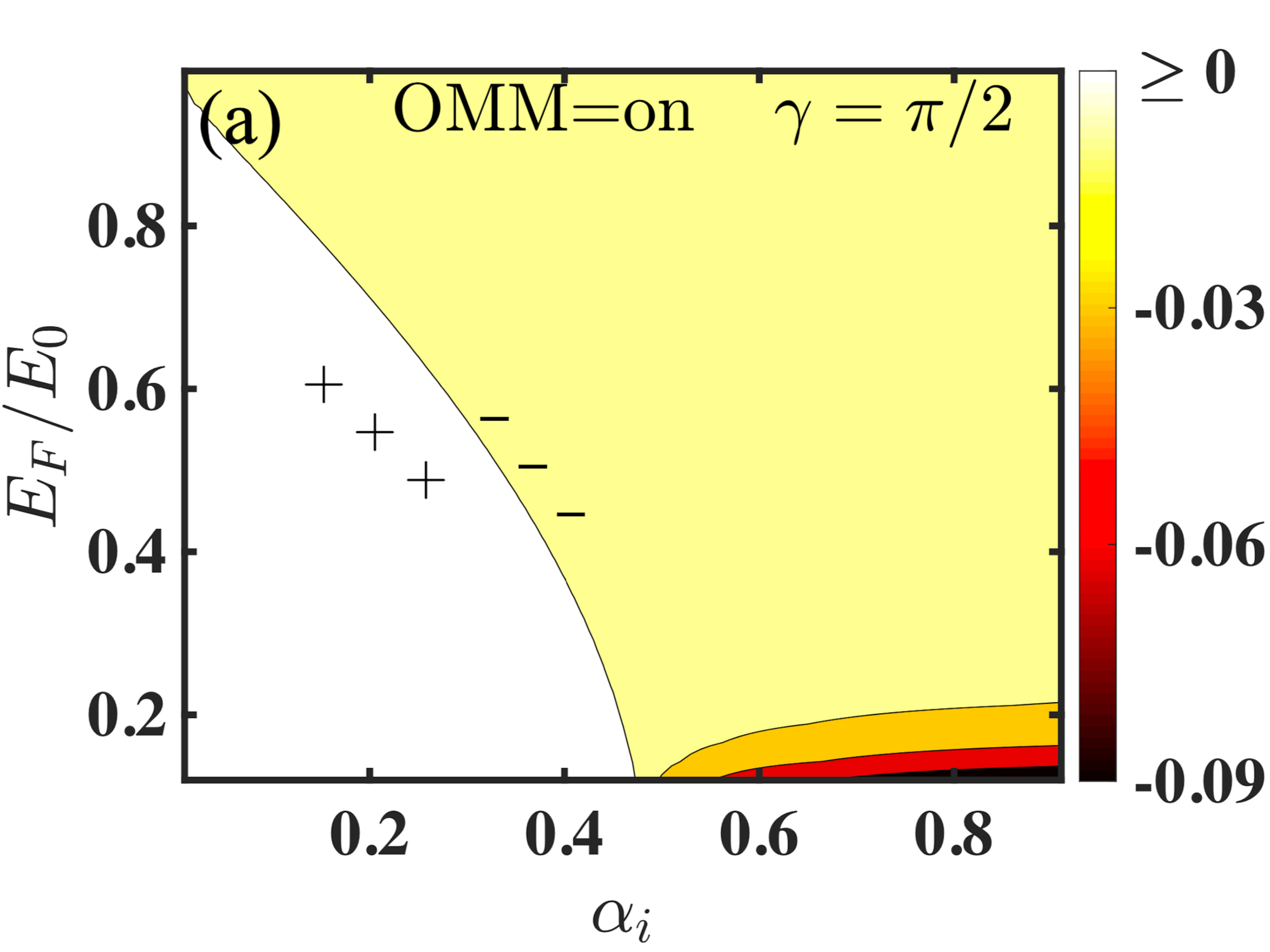}
    \includegraphics[width=0.49\columnwidth]{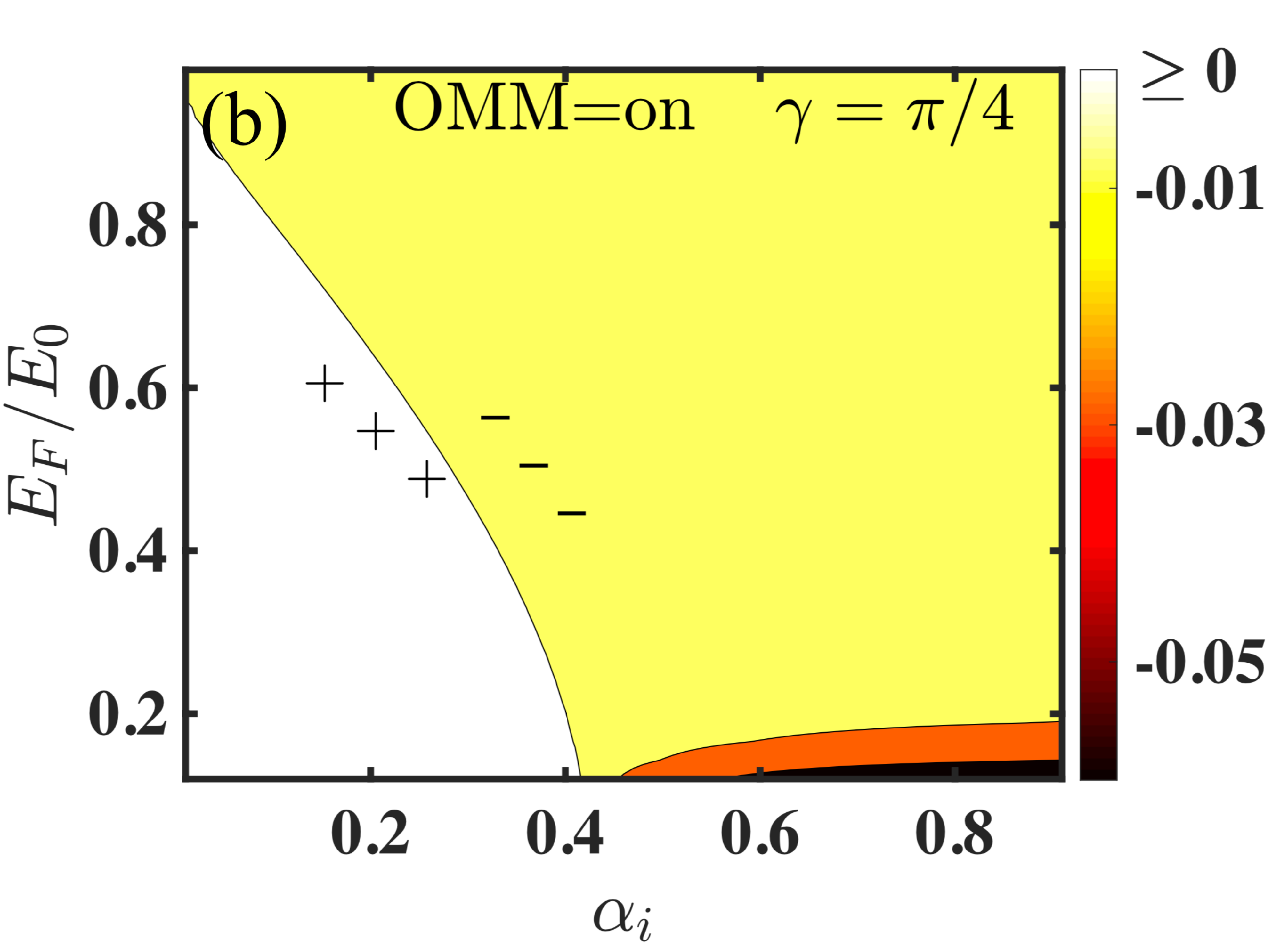}
    \includegraphics[width=0.49\columnwidth]{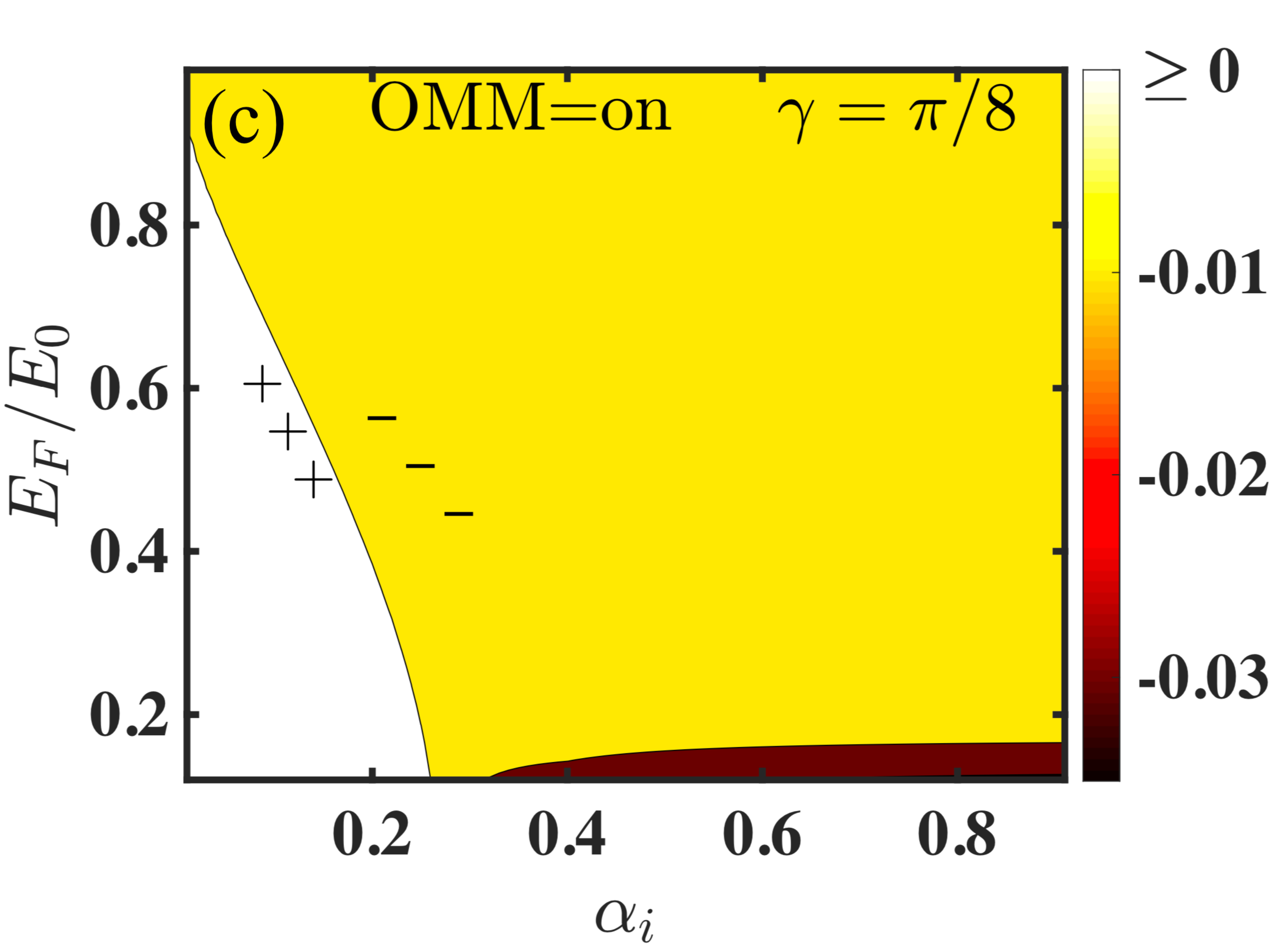}
    \includegraphics[width=0.49\columnwidth]{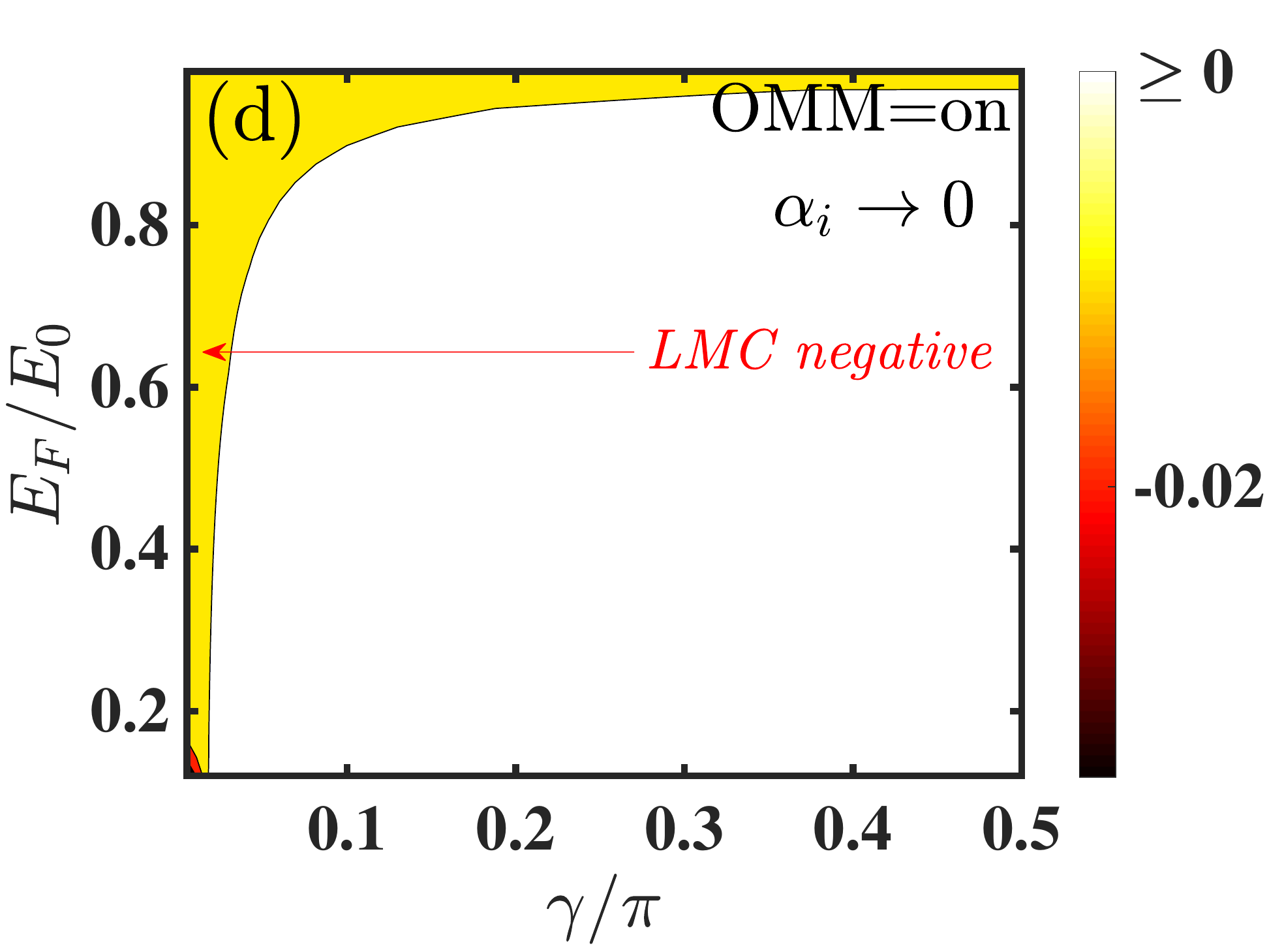}
    \caption{(a)-(c) Phase plot of the quadratic coefficient of the longitudinal magnetoconductance for a lattice model of untilted Weyl fermions as a function of Fermi energy and intervalley scattering strength $\alpha_i$ for various different angles of the magnetic field. We explicitly map the zero-LMC contour in the $E_F-\alpha_i$ space where the change in sign of LMC occurs. At higher Fermi energies the switching of LMC sign from positive to negative happens at a lower threshold of $\alpha_i=\alpha_i^c$ due to nonlinear lattice effects. Secondly, orienting the magnetic field direction away from the electric field also lowers the threshold value of $\alpha_i^c$. (d) Quadratic LMC coefficient in the limit of vanishing intervalley scattering strength $\alpha_i$ as a function of the Fermi energy and angle of the magnetic field.}
    \label{fig:lattice1}
\end{figure*}

\subsection{Non-collinear $\mathbf{E}$ and $\mathbf{B}$ fields without tilting of the Weyl cones for lattice Weyl fermions}
Including orbital magnetic moment effects, the energy dispersion assumes the form of the following transcendental equation
\begin{align}
2\hbar k^2\epsilon^\chi_k&= 2\hbar k^2 E_0 \sin(ka) +\nonumber\\ 
&e\chi E_0 \sin(ak) B(\cos(\theta)\sin\gamma + \sin\theta\cos\phi\cos\gamma).
\label{Eq_weyl_latt2}
\end{align}
The above equation has no closed-form solution for the momentum $k^\chi$, and therefore the constant Fermi energy contour in $k-$space is evaluated numerically. The semi-classical band velocities evaluated in spherical polar coordinates are
\begin{align}
v_k^\chi &= \frac{E_0 a \cos(ak)}{\hbar} - \frac{u_2^\chi \cos(ak) \beta_{\theta\phi}}{\hbar ak^2} + \frac{2 u_2^\chi \sin(ak) \beta_{\theta\phi}}{\hbar a^2k^3}\nonumber\\
v_\theta^\chi &= \frac{u_2^\chi \sin(ak) (-d\beta_{\theta\phi}/d\theta)}{\hbar a^2 k^3}; \hspace{10mm}
v^\chi_\phi = \frac{u_2^\chi (-d\beta_{\theta\phi}/d\phi)}{\hbar a^2 k^3 \sin\theta}\nonumber\\
u_2^\chi &= -e \chi E_0 B a^2/2\hbar,
\end{align}
where $\beta_{\theta\phi} = (\sin\theta\cos\phi\cos\gamma+\cos\theta\sin\gamma)$.

\subsection{Non-collinear $\mathbf{E}$ and $\mathbf{B}$ fields with tilting of the Weyl cones in the linear approximation}
Since tilting and lattice cutoff effects are physically independent of each other, we treat these effects separately. Linearizing the Hamiltonian in Eq.~\ref{Eq_H1weyl} around the nodal point, we obtain 
\begin{align}
    H_\mathbf{k} = \chi \hbar v_F \mathbf{k}\cdot\boldsymbol{\sigma} + t^\chi_x k_x + t^\chi_z k_z,
\end{align}
where we define $v_F = aE_0/\hbar$, $t^\chi_i = T^\chi_i a$.
The expression for the constant energy contour becomes 
\begin{align}
{ k^\chi=\frac{\epsilon^{\chi}_{\mathbf{k}}+{\sqrt{(\epsilon_\mathbf{k}^{\chi})^2-l^\chi\chi \xi e v_F B\beta_{\theta\phi}}}}{l^\chi}},
\end{align}
where $l\chi=2\hbar v_F + 2 t^{\chi}_z \cos{
\theta}+2t^{\chi}_x \sin{\theta}\cos{\phi}$, while the semiclassical velocities take the following form 
\begin{align}
v^{\chi}_x&=v_F\frac{k_x}{k}+\frac{t^{\chi}_x}{\hbar}+\frac{v_2^\chi}{k^2}\left(\cos\gamma\left({1}-\frac{2k^2_x}{k^2}\right)-\frac{2\sin\gamma k_x k_z}{k^2}\right),\nonumber\\
v^{\chi}_y&=v_F\frac{k_y}{k}+\frac{v_2^\chi}{k^2}\left(\cos{\gamma}\left(\frac{-2k_x k_y}{k^2}\right)+\sin{\gamma}\left(\frac{-2k_y k_z}{k^2}\right)\right),\nonumber\\
v^{\chi}_z&=v_F\frac{k_z}{k}+\frac{t^{\chi}_z}{\hbar}+\frac{v_2^\chi}{k^2}\left(\frac{-2\cos{\gamma}k_x k_z}{k^2}+\sin{\gamma}\left(1-\frac{2k^2_z}{k^2}\right)\right),\nonumber\\
v_2^\chi&=\frac{\chi e v_F B}{2 \hbar}.
\end{align}

\section{Results}
\subsection{LMC for lattice Weyl semimetal}
We first discuss the results for the lattice model of a Weyl semimetal without considering the effects of tilting of the Weyl cones. Since the effects of tilting of the Weyl cones are independent of lattice effects, tilting of the Weyl cones will be considered subsequently. After obtaining the constant energy contour $k^\chi$ numerically using Eq.~\ref{Eq_weyl_latt2}, we solve for the non-equilibrium distribution function $g_\mathbf{k}^\chi$ using the procedure described in the previous section. This is done specifically for each value of $\gamma,\alpha_i, B$, and $E_F$. The obtained LMC is found to be quadratic in magnetic field, and thus we expand the LMC as $\sigma_{zz}(\alpha_i,B,\gamma,E_F)=\sigma_{zz0}(\alpha_i,\gamma,E_F) + \sigma_{zz2}(\alpha_i,\gamma,E_F) B^2$. The linear-in-B term $\sigma_{zz1}$ (which is zero here) will become crucial for our analysis when we introduce tilting of Weyl fermions, as discussed in the next subsection. The longitudinal magnetoconductance switches sign from positive to negative at a critical value of $\alpha_i^c(\gamma,E_F)$, i.e., the coefficient $\sigma_{zz2}$ becomes negative when $\alpha_i>\alpha_i^c(\gamma,E_F)$ as shown in Fig.~\ref{fig:lattice1}. At a fixed relative orientation of the magnetic field (as determined by the angle $\gamma$), the threshold of $\alpha_i^c$ decreases as the Fermi energy is increased. Note that in the linear approximation one expects a straight line contour separating positive and negative LMC areas with a constant $\alpha_i^c$ as a function of $E_F$, however, non-linear lattice effects lower the critical value of $\alpha_i^c$ highlighting the fact that lattice effects can drive the system to exhibit negative LMC . The explicit zero-LMC contour is plotted in Fig.~\ref{fig:lattice1}  separates positive and negative LMC region. 

An interesting feature emerges when the magnetic field is oriented further away from the electric field, i.e., the angle $\gamma$ is shifted away from $\pi/2$. The overall region of positive LMC is shrunk further in this case. Note from Fig.~\ref{fig:lattice1} (a)-(c) that even when $\alpha_i=0$, i.e., in the absence of any intervalley scattering, there is an upper energy cutoff  beyond which LMC becomes negative. This feature is specifically highlighted in Fig.~\ref{fig:lattice1}(d) where we plot the quadratic coefficient $\sigma_{zz2}$ as a function of the Fermi energy and angle $\gamma$ in the limit of vanishing intervalley scattering strength $\alpha_i$. This specifically points out the fact that lattice effects in Weyl fermions can independently produce negative LMC even in the absence of a finite intervalley scattering, a previously unknown result. For parallel electric and magnetic fields ($\gamma=\pi/2$) the LMC is primarily positive even for higher Fermi energies (when lattice effects become important) and becomes negative only at very high Fermi energies near the band edge. When the magnetic field is oriented away from the electric field ($\gamma\rightarrow 0$), small nonlinear lattice effects even $E_F$ is small can produce negative LMC. 

The planar Hall effect on the other hand does not display any sign change due to nonlinear lattice effects and displays the standard $\sin (2\gamma)$ trend as a function of the angle $\gamma$. Thus we do not explicitly plot this behavior. PHE will be discussed in detail for tilted Weyl fermions subsequently. 

\begin{figure*}
    \centering
    \includegraphics[width=0.49\columnwidth]{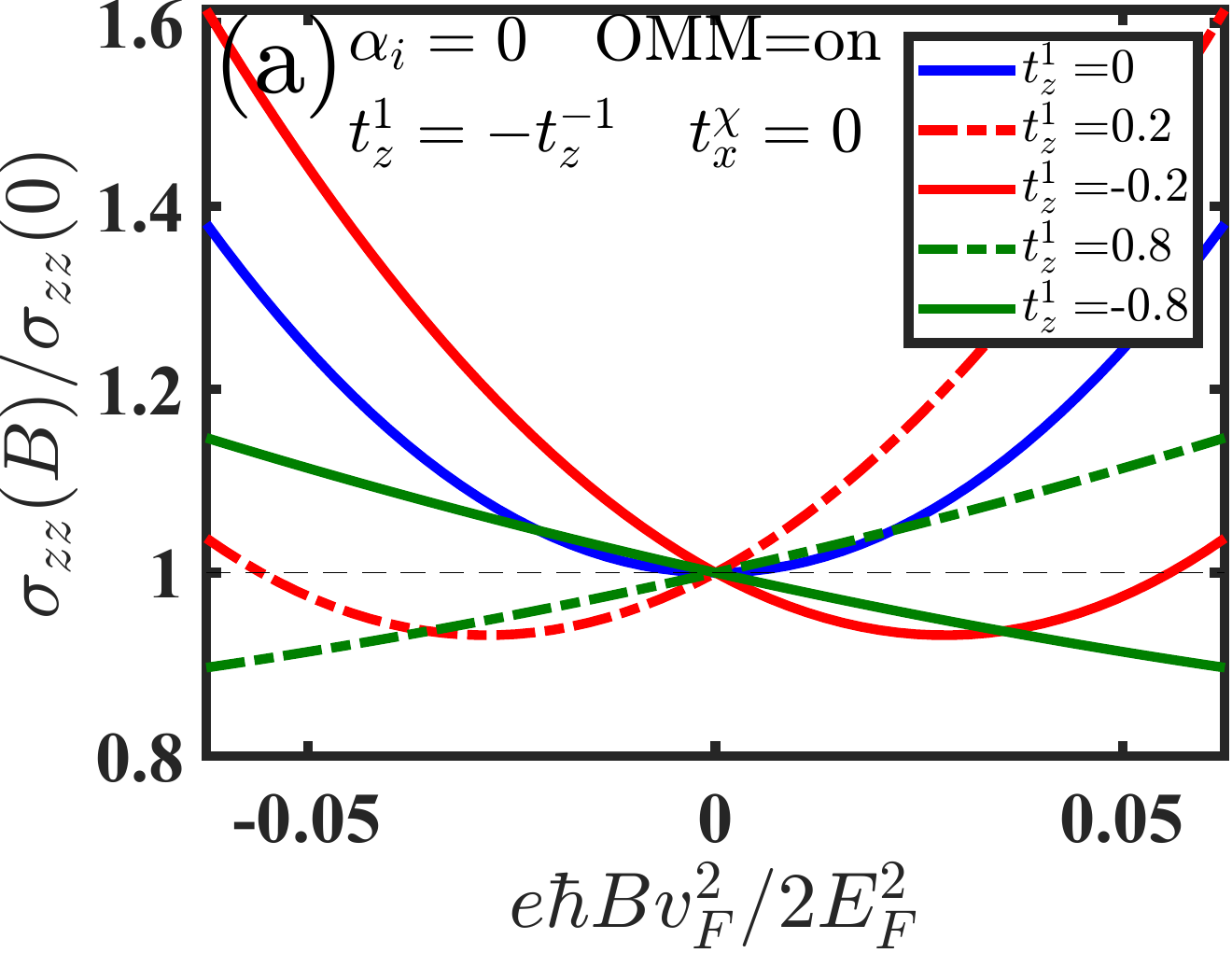}
    \includegraphics[width=0.49\columnwidth]{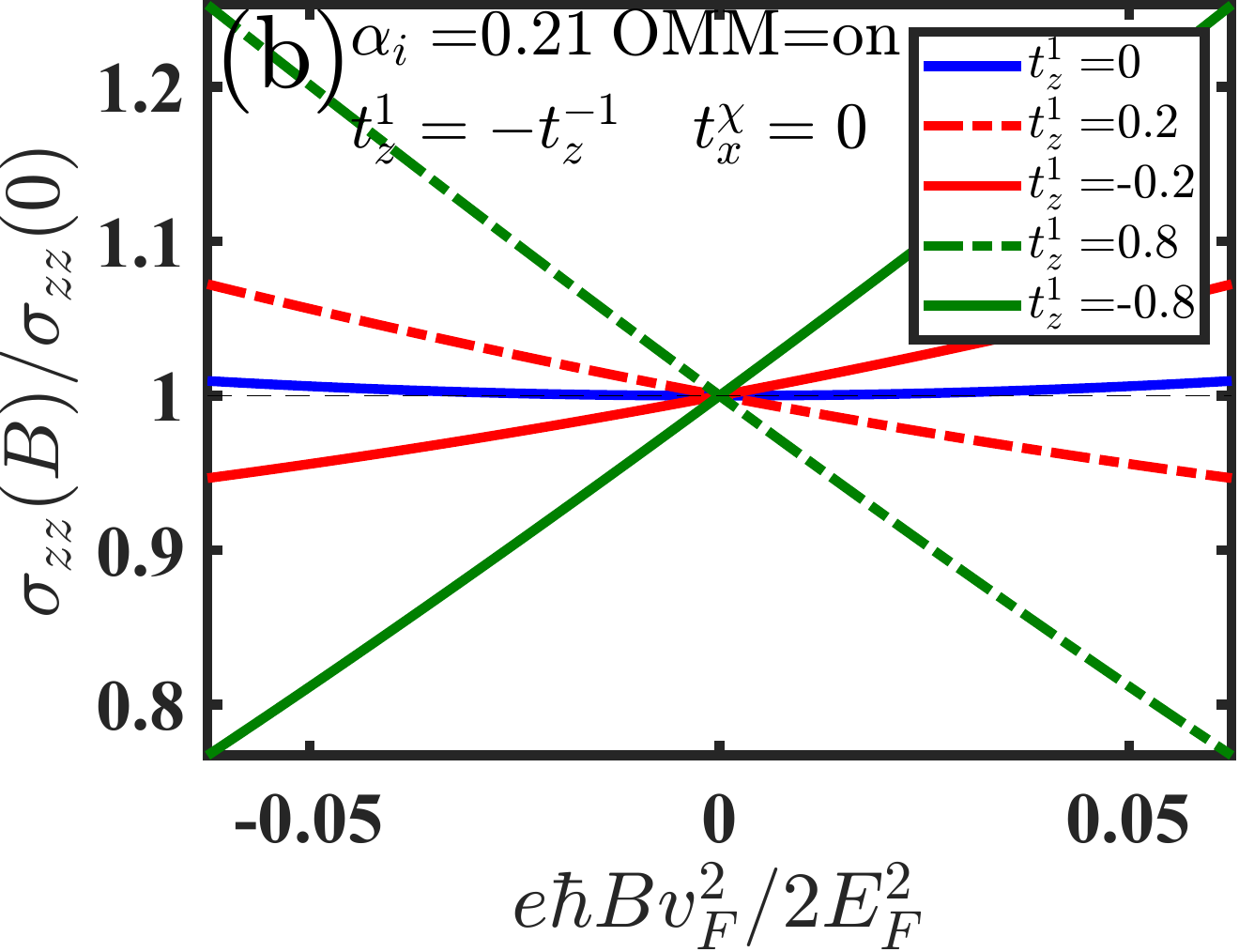}
    \includegraphics[width=0.49\columnwidth]{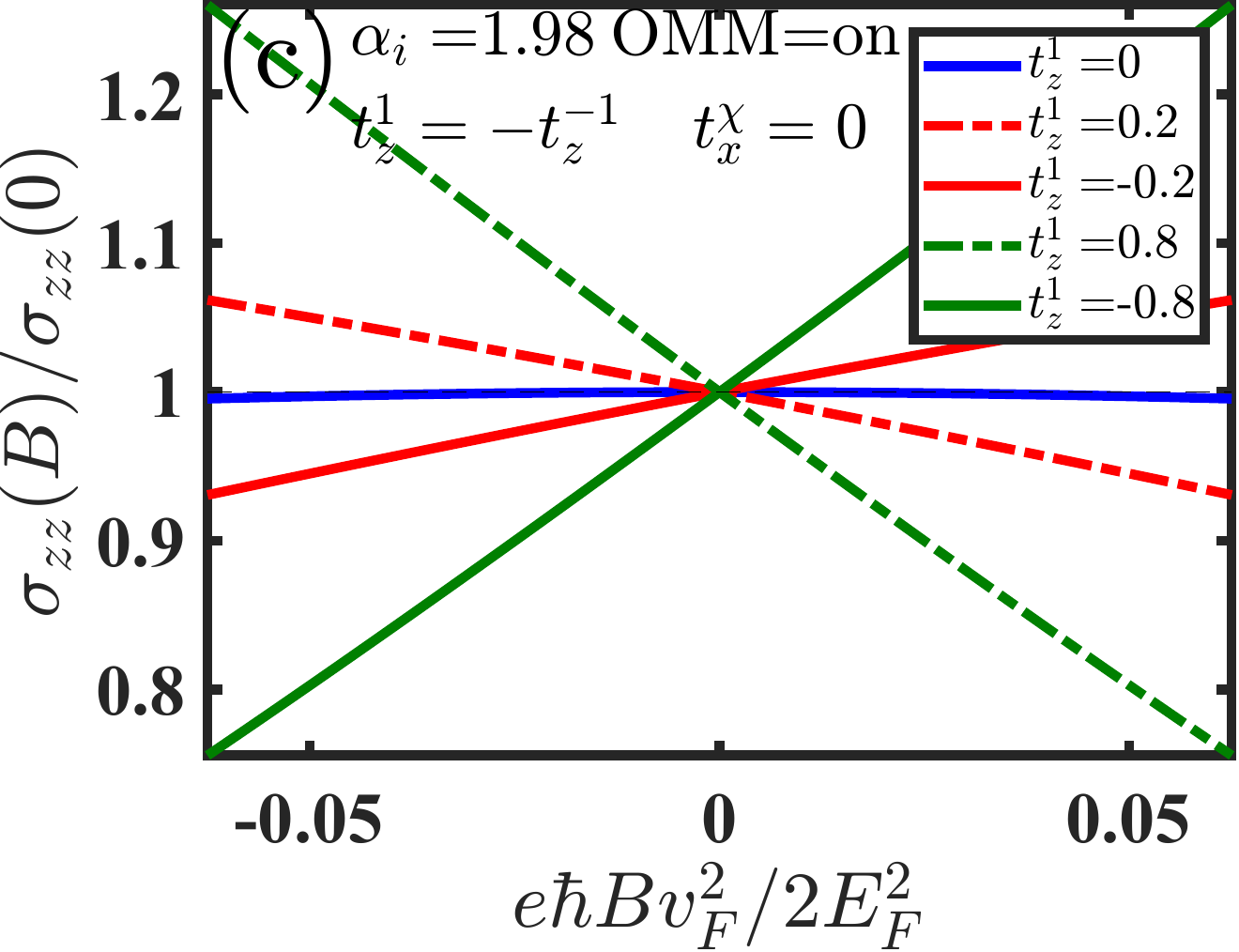}
    \includegraphics[width=0.53\columnwidth]{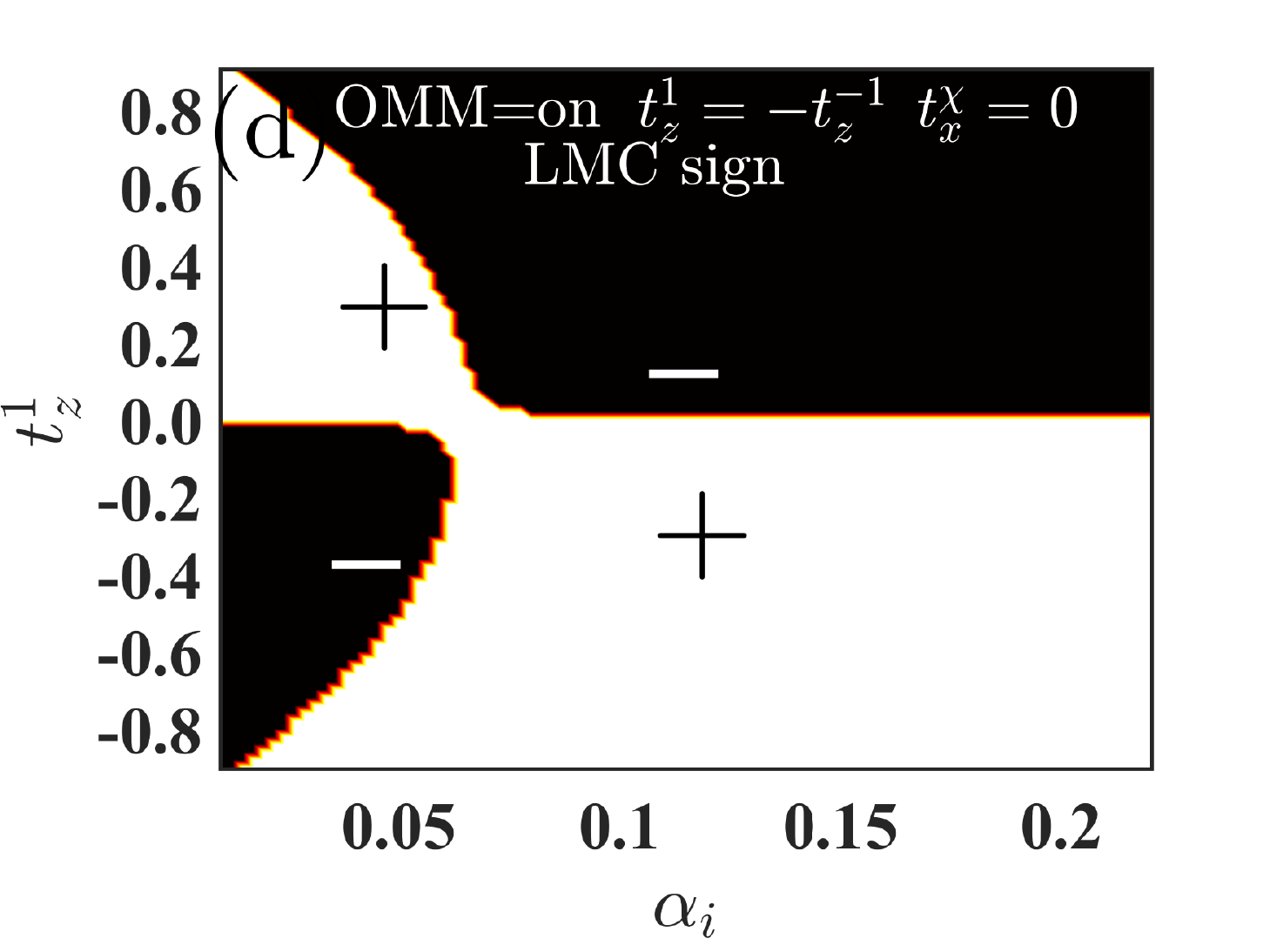}
    \includegraphics[width=0.5\columnwidth]{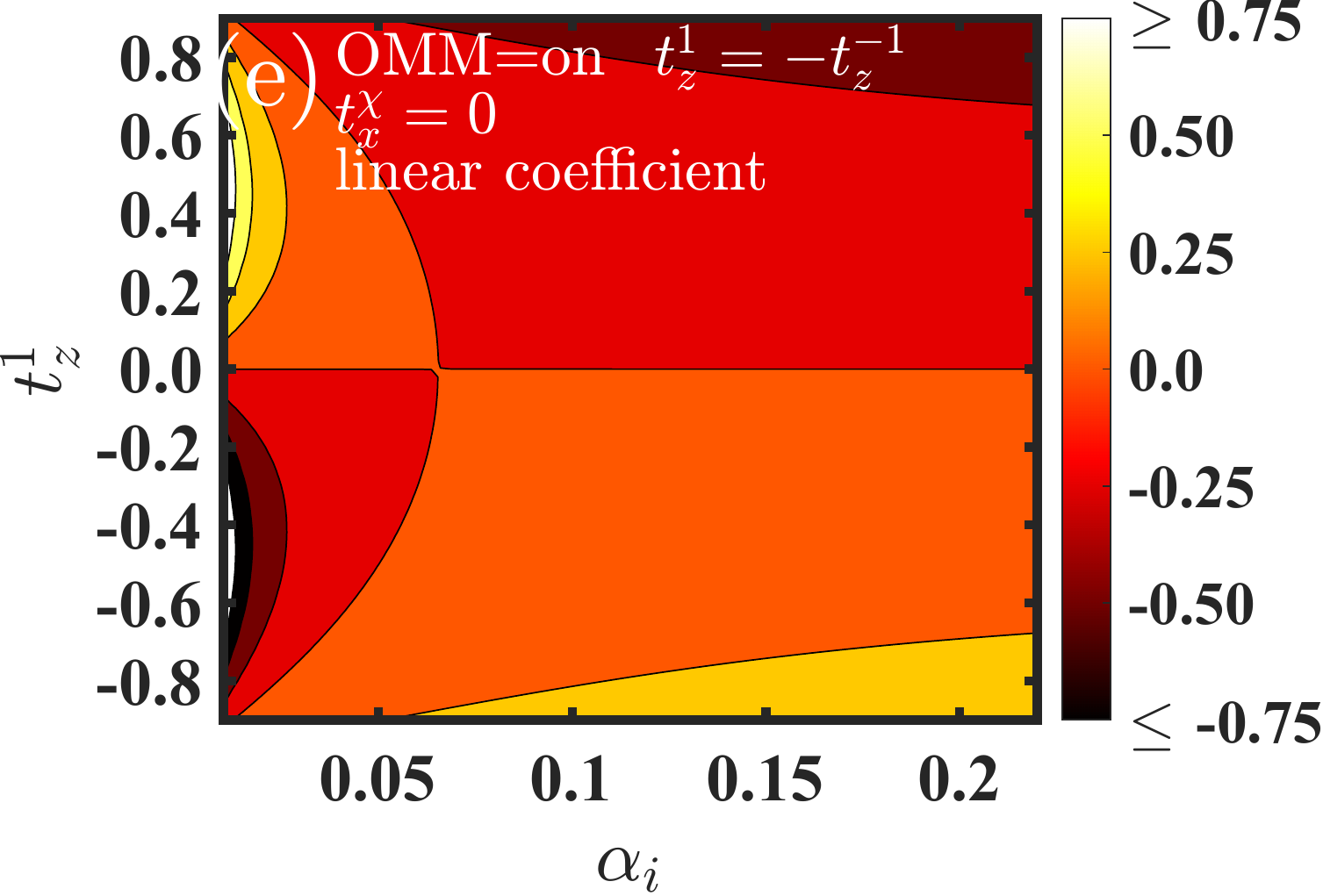}
    \includegraphics[width=0.5\columnwidth]{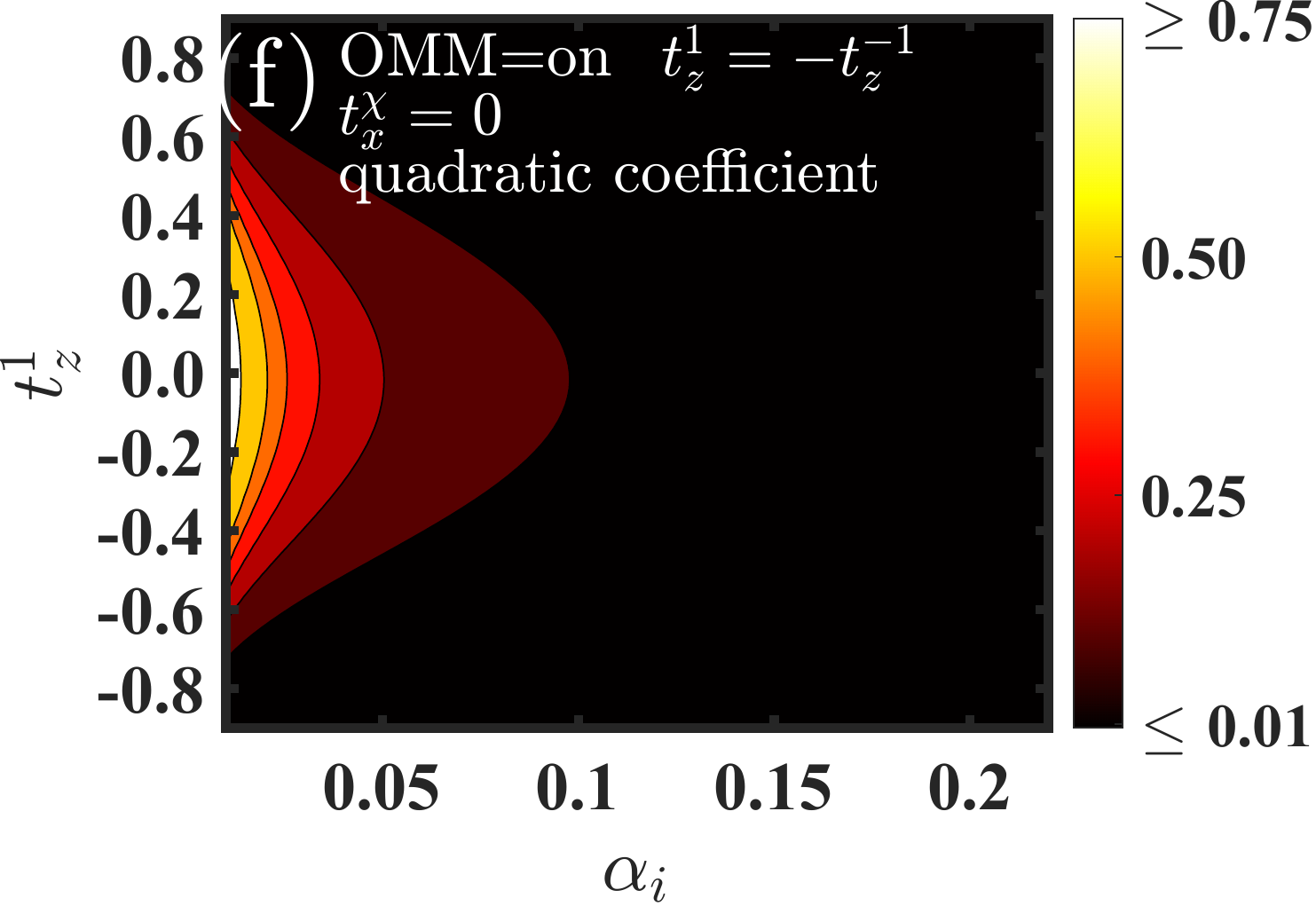}
    \includegraphics[width=0.5\columnwidth]{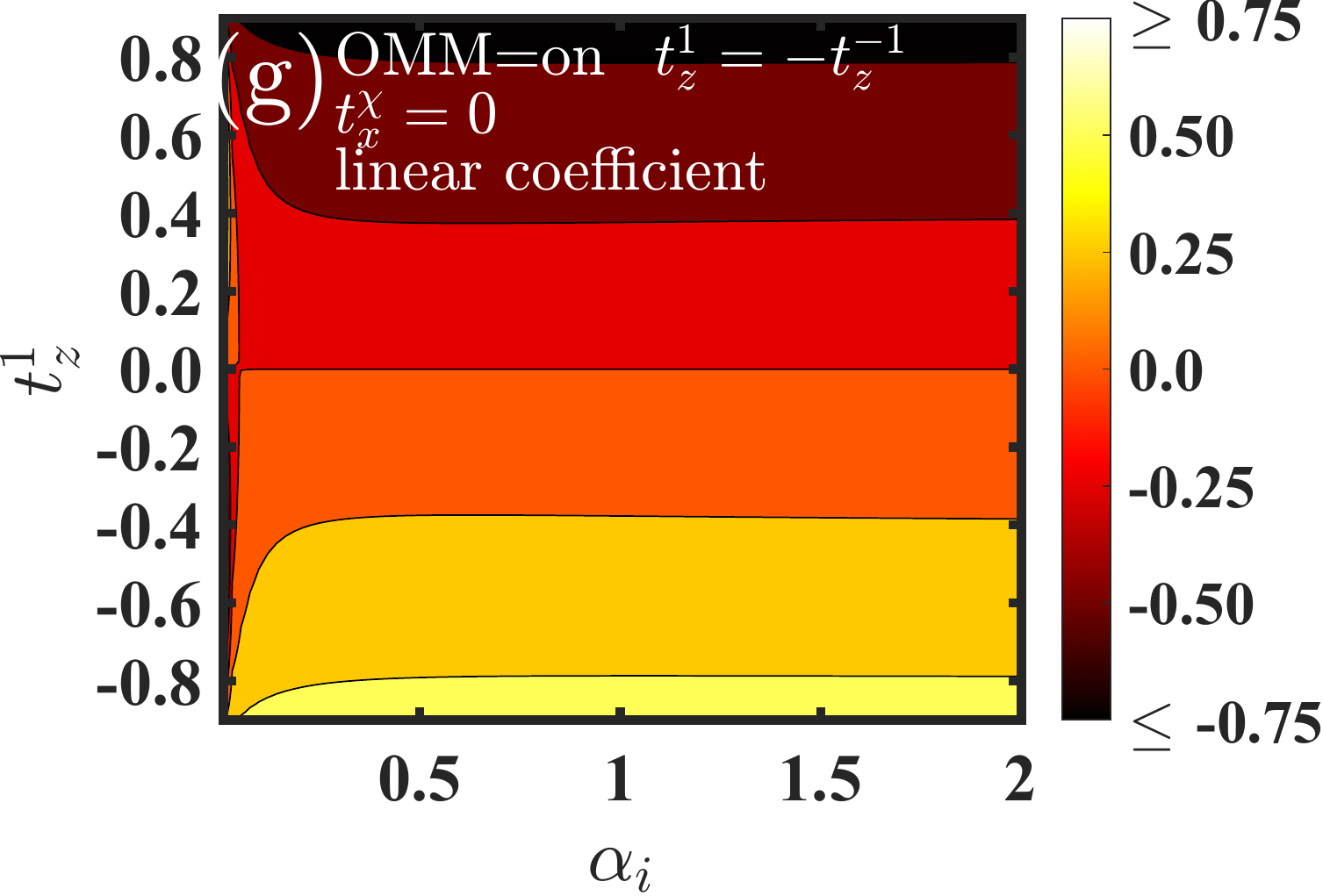}
    \includegraphics[width=0.5\columnwidth]{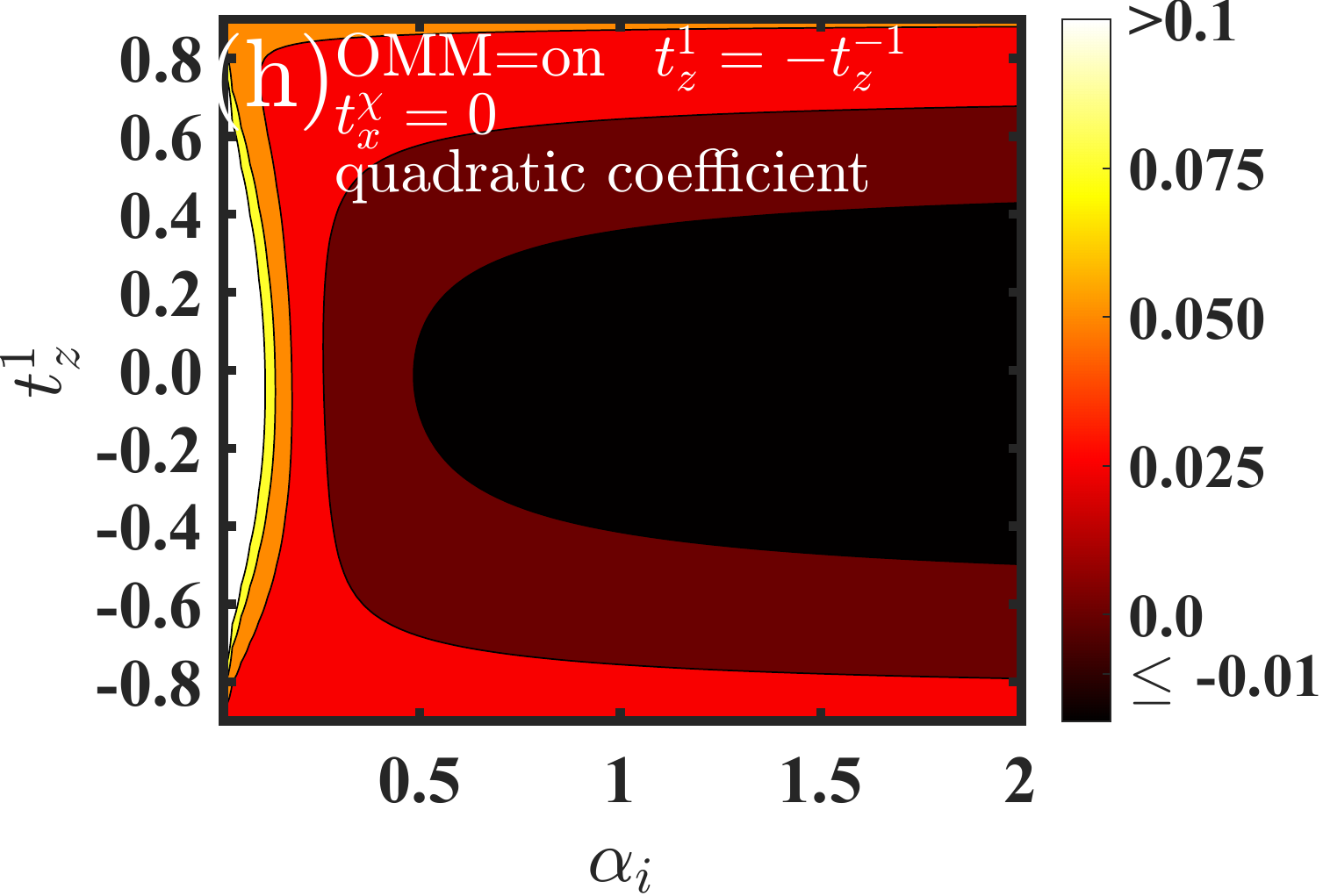}
    \caption{Longitudinal magnetoconductance $\sigma_{zz}(B)$ in the case when the Weyl cones are titled in the direction of the magnetic field ($\hat{z}$) axis, but are oriented opposite to each other ($t_z^1=-t_z^{-1}$).  (a) LMC as a function of magnetic field for various tilt parameters in the absence of intervalley scattering ($\alpha_i=0$). For a finite small tilt $t_z^1$ the LMC is asymmetric about zero magnetic field, but still appears to be quadratic. When the tilt is large, LMC is predominantly linear-in-$B$. (b) and (c) LMC in the presence of a finite intervalley scattering $\alpha_i$. LMC switches sign with the inclusion of $\alpha_i$. (d) The sign of the LMC is plotted as a function of $\alpha_i$ and $t_z^1$ in the limit of $B\rightarrow 0^+$ clearly indicating areas of positive and negative LMC.  (e)-(f) Linear ($\sigma_{zz1}$) and quadratic ($\sigma_{zz2}$) coefficient of the LMC. Below $\alpha_i\sim 0.05$, the coefficients are similar in magnitude and LMC has an overall quadratic trend. for large enough $\alpha_i$ the linear coefficient dominates over the quadratic coefficient leading to an overall linear-in-$B$ LMC as well as a change in sign of LMC. (g)-(h) LMC in the limit of large $\alpha_i$. Note that the phase plots are for positive magnetic field. When $B$ reverses sign, the LMC switches sign as well as it is dependent on the orientation of the tilt with respect to the magnetic field. }
    \label{fig:szz_tiltz_opp_ommon}
\end{figure*}

\begin{figure*}
    \centering
    \includegraphics[width=0.49\columnwidth]{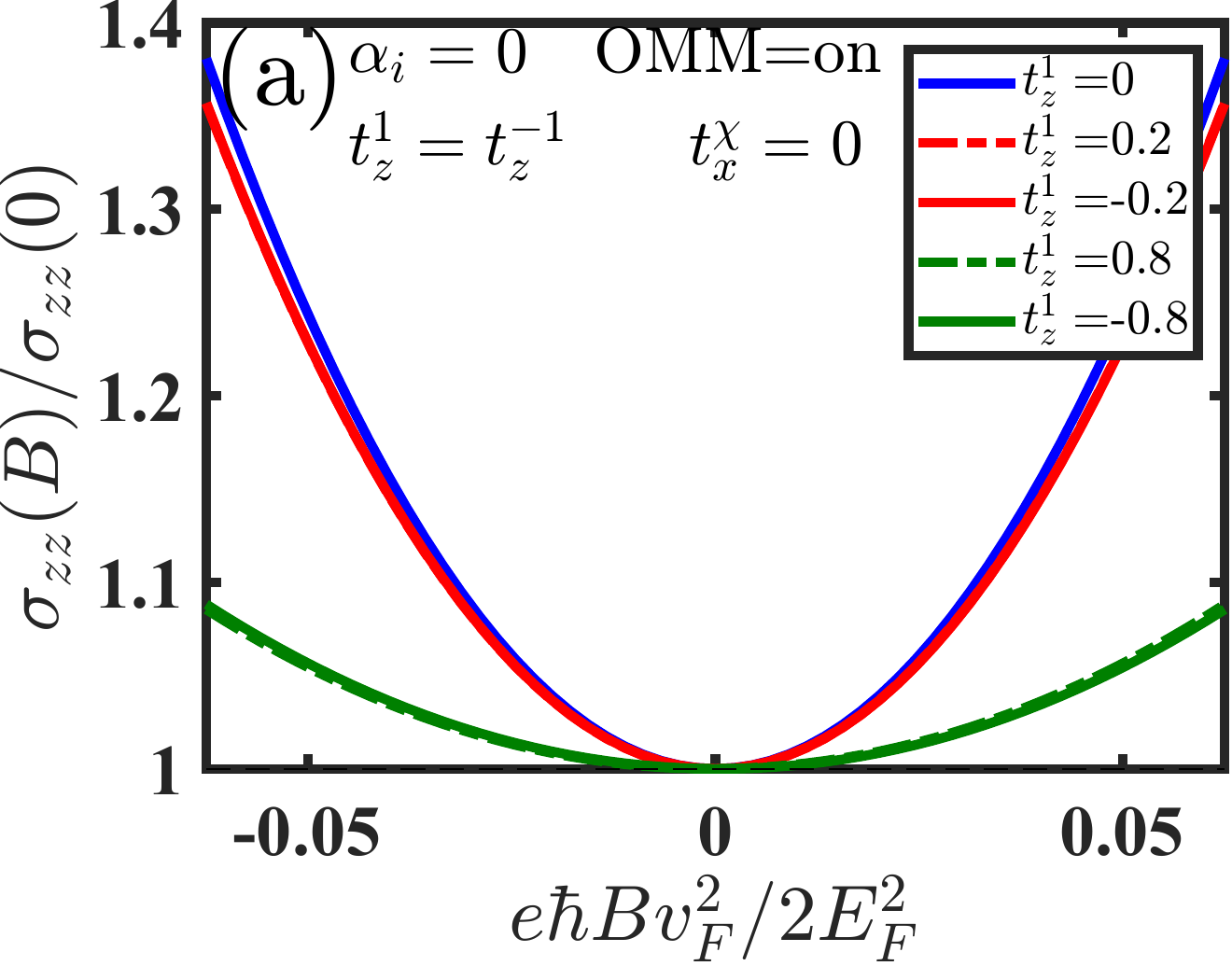}
    \includegraphics[width=0.49\columnwidth]{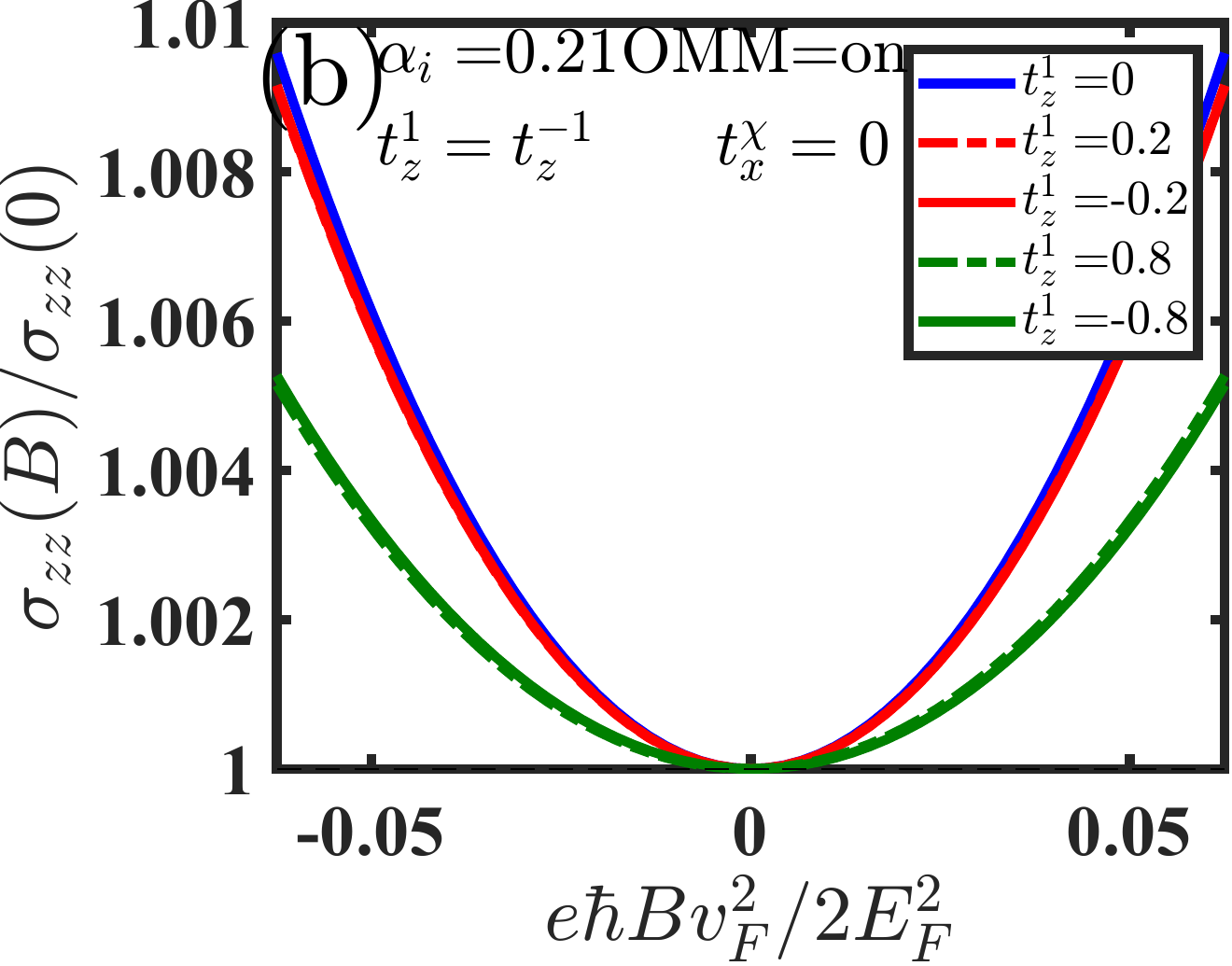}
    \includegraphics[width=0.49\columnwidth]{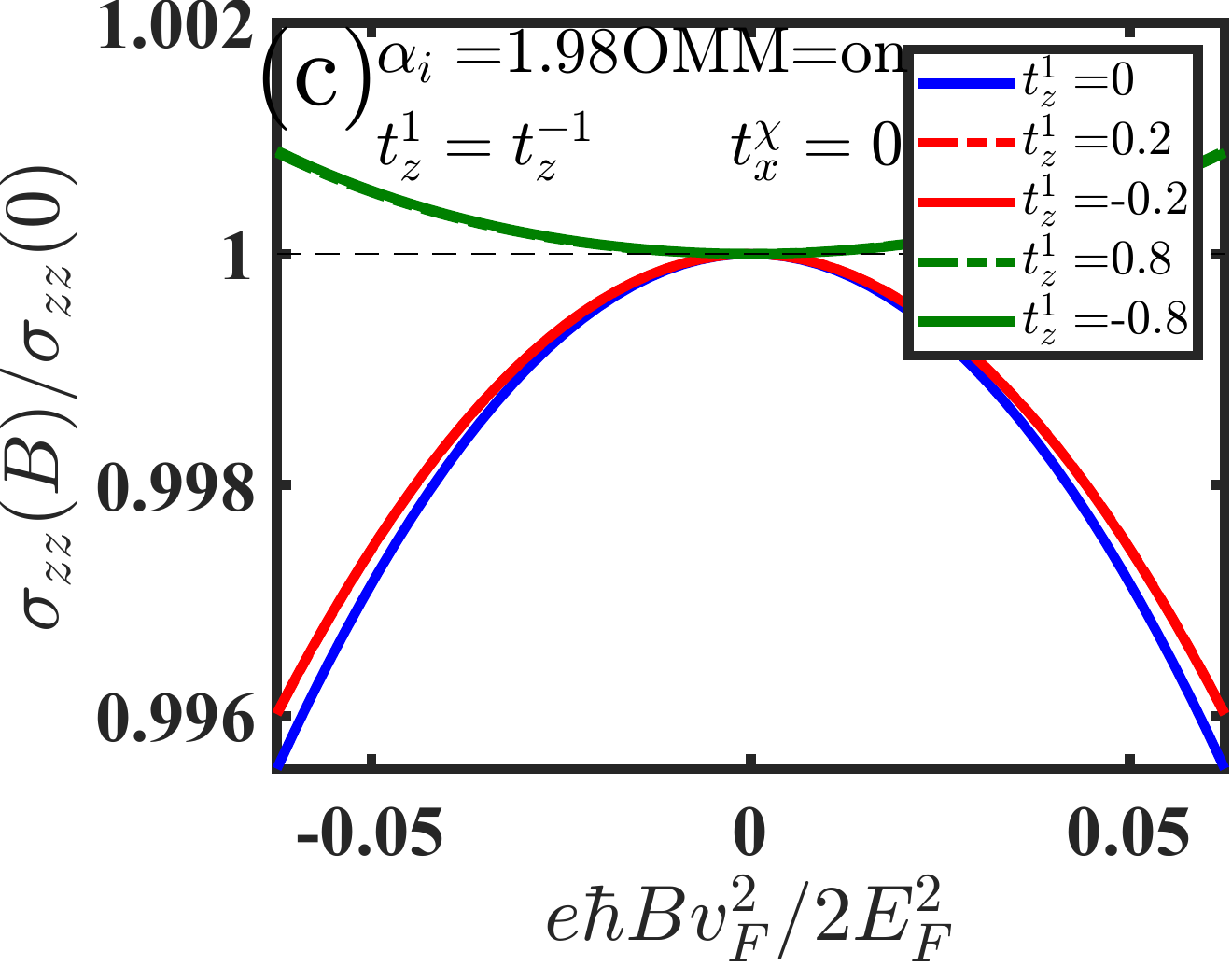}
    \includegraphics[width=0.53\columnwidth]{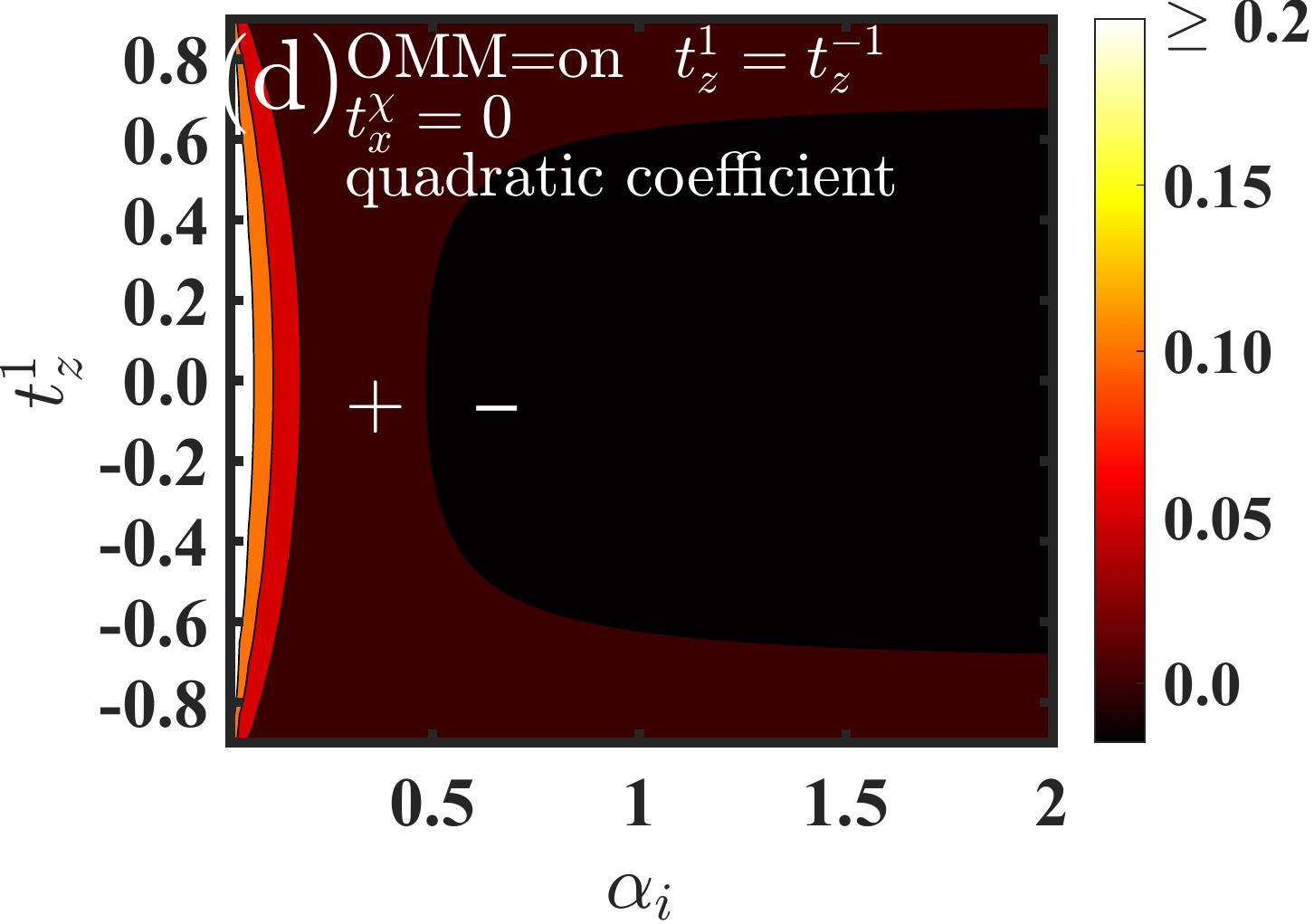}
    \caption{Longitudinal magnetoconductance $\sigma_{zz}(B)$ in the case when the Weyl cones are titled in the direction of the magnetic field ($\hat{z}$) axis, and are oriented in the same direction  to each other ($t_z^1=t_z^{-1}$).  (a) LMC as a function of magnetic field for various tilt parameters in the absence of intervalley scattering ($\alpha_i=0$). (b) and (c) LMC in the presence of a finite intervalley scattering $\alpha_i$. LMC switches sign with the inclusion of $\alpha_i$ whenever $\alpha_i>\alpha_i^c(t_z)$. (d) The quadratic coefficient of LMC is plotted as a function of $\alpha_i$ and $t_z^1$. The sign of the coefficient also corresponds to the sign of LMC. The contour separating positive and negative LMC regions is also clearly shown.}
    \label{fig:szz_tiltz_same_ommon}
\end{figure*}

\subsection{LMC in tilted Weyl semimetal when $\gamma=\pi/2$ }
First we discuss the case of tilted Weyl fermions when the electric and magnetic fields are held parallel to each other, i.e., $\gamma=\pi/2$. In this case the PHE contribution is expected to vanish and hence only LMC is discussed. 
\subsubsection{The case when $t_z^1=-t_z^{-1}\neq 0$ and $t_x^\chi=0$}
Fig.~\ref{fig:szz_tiltz_opp_ommon} presents the results of LMC $\sigma_{zz}(B)$ as a function of magnetic field when the two Weyl cones are oppositely titled with respect to each other, but in the direction of the magnetic field, i.e. $t_z^1 = -t_z^{-1}$, and $t_x^\chi=0$. In the absence of any intervalley scattering ($\alpha_i=0$) and tilt ($t_z^\chi=0$), the LMC is always positive, quadratic in $B$, and symmetric about $B=0$ as expected.  Now keeping $\alpha_i=0$, a finite tilt introduces a linear-in-$B$ term in the LMC and a corresponding asymmetry around $B=0$, i.e. the value LMC depends on the direction of magnetic field, or more generally it is dependent on the orientation of the magnetic field with respect to the direction of the tilt. The $B-$linear term survives because the tilts of the Weyl cones are opposite to each other. 
For higher tilt values ($\sim~\geq 0.4$) the linear-in-$B$ term dominates over the quadratic term and the LMC is observed to be linear in the relevant range of the magnetic field. Now, when the signs of the tilts are interchanged at each valley, i.e., $t_z^\chi\rightarrow t_z^{-\chi}$, the behavior with respect to $B=0$ reverses as well, i.e., $\sigma_{zz}(t_z^1, B) = \sigma_{zz}(-t_z^1, -B)$ and $\sigma_{zz}(t_z^1, B) = -\sigma_{zz}(-t_z^1, B)$. This implies that the sign of LMC can be positive or negative depending on  the orientation of the magnetic field with respect to the tilt in the Weyl cone. 

In the presence of finite intervalley scattering $\alpha_i$, there is a qualitative change in LMC: beyond a critical value $\alpha^c_i(t_z^1)$, LMC switches sign. For example, note that $\sigma_{zz}(t_z^1=0.2, B>0)$ to be positive when $\alpha_i=0$, but $\sigma_{zz}(t_z^1=0.2, B>0)$ becomes negative when $\alpha_i=0.2$. In order to better understand this behavior, we expand the longitudinal magnetoconductance as $\sigma_{zz}(t_z^1, \alpha_i, B) = \sigma_{zz0}(t_z^1,\alpha_i) + \sigma_{zz1}(t_z^1,\alpha_i) B + \sigma_{zz2}(t_z^1,\alpha_i) B^2$, where each coefficient $\sigma_{zzj}$ corresponds to the $j^{th}$ order in the magnetic field. The evaluated LMC as a function of the magnetic field is then fit according to the above equation to obtain the coefficients $\sigma_{zzj}(t_z^1,\alpha_i)$. The linear and quadratic coefficients are plotted in Fig.~\ref{fig:szz_tiltz_opp_ommon}(e-h).
When $\alpha_i$ is small ($\sim \leq 0.05$) the linear ($\sigma_{zz1}$) and quadratic ($\sigma_{zz2}$) coefficients are similar in their magnitude, and therefore the behavior with respect to the magnetic field has both linear and quadratic trend. When $\alpha_i$ crosses threshold value $\alpha^c_i(t_z^1)$ the linear coefficient dominates and LMC switches sign as a function the magnetic field. Note that there is a special case of $t_z^1=0$, where the linear coefficient is always zero and the LMC switches sign when $\alpha_i=0.5$~\cite{sharma2020sign}. However, for even small values of $t_z^1$, the linear coefficient dominates over the quadratic coefficient and the sign reversal in LMC occurs below $\alpha_i=0.5$. The sign of the LMC is also plotted in Fig.~\ref{fig:szz_tiltz_opp_ommon}(d) as a function of $\alpha_i$ and $t_z^1$ in the limit of $B\rightarrow 0^+$. Note that for small values of $B$ and tilt, LMC can show both positive and negative behavior depending on the fact of $B\rightarrow 0$ or if $B$ is away from zero (see Fig.~\ref{fig:szz_tiltz_opp_ommon}(a) when $t_z^1=0.2$). Finally, we also comment on the effect of non-collinear $\mathbf{E}$ and $\mathbf{B}$ fields. Qualitatively, we find no difference from Fig.~\ref{fig:szz_tiltz_opp_ommon} even when the angle $\gamma\neq \pi/2$. However, when $\gamma\neq \pi/2$, along with LMC we also have a finite planar Hall conductivity, which is discussed later.

\begin{figure*}
    \centering
    \includegraphics[width=0.49\columnwidth]{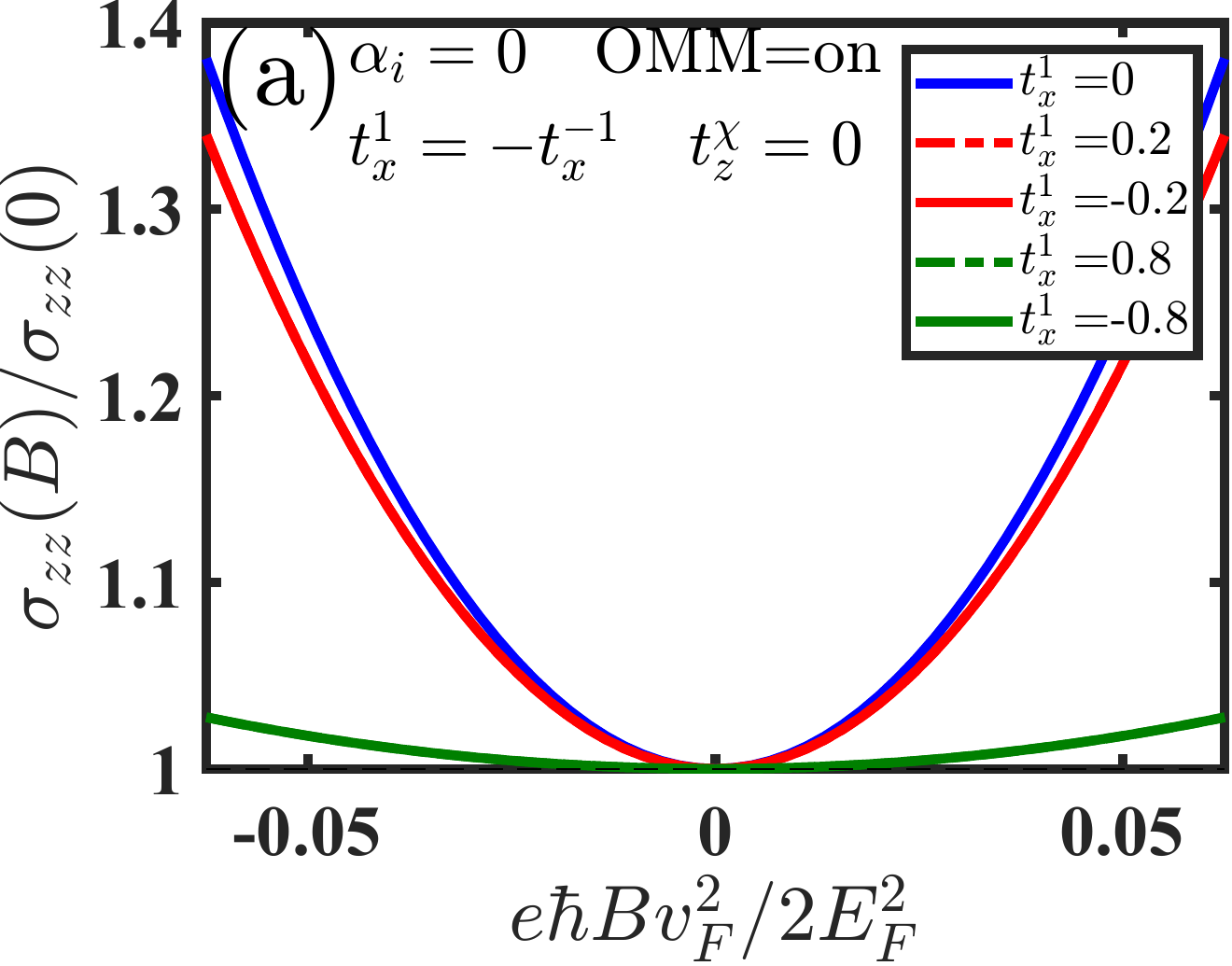}
    \includegraphics[width=0.49\columnwidth]{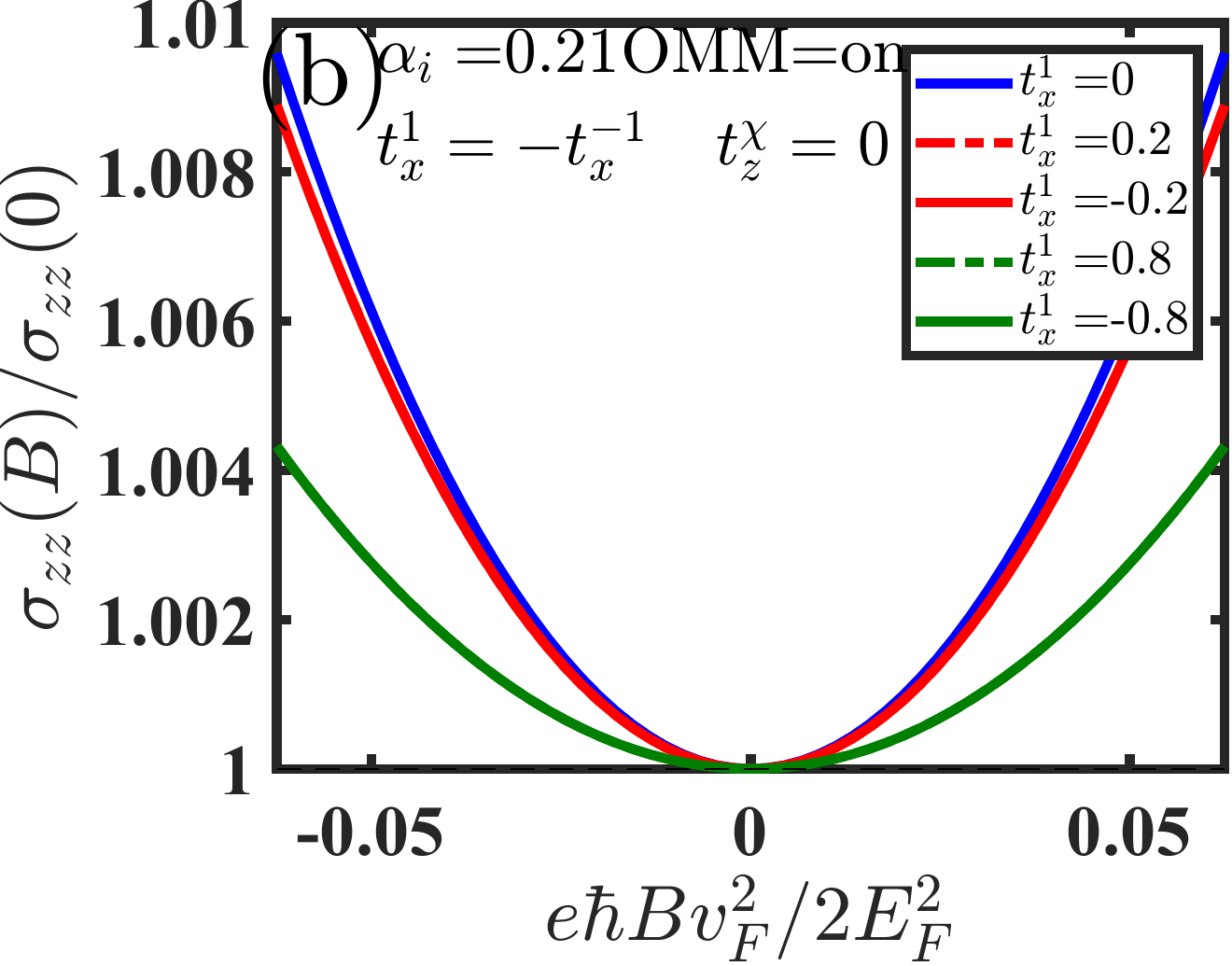}
    \includegraphics[width=0.49\columnwidth]{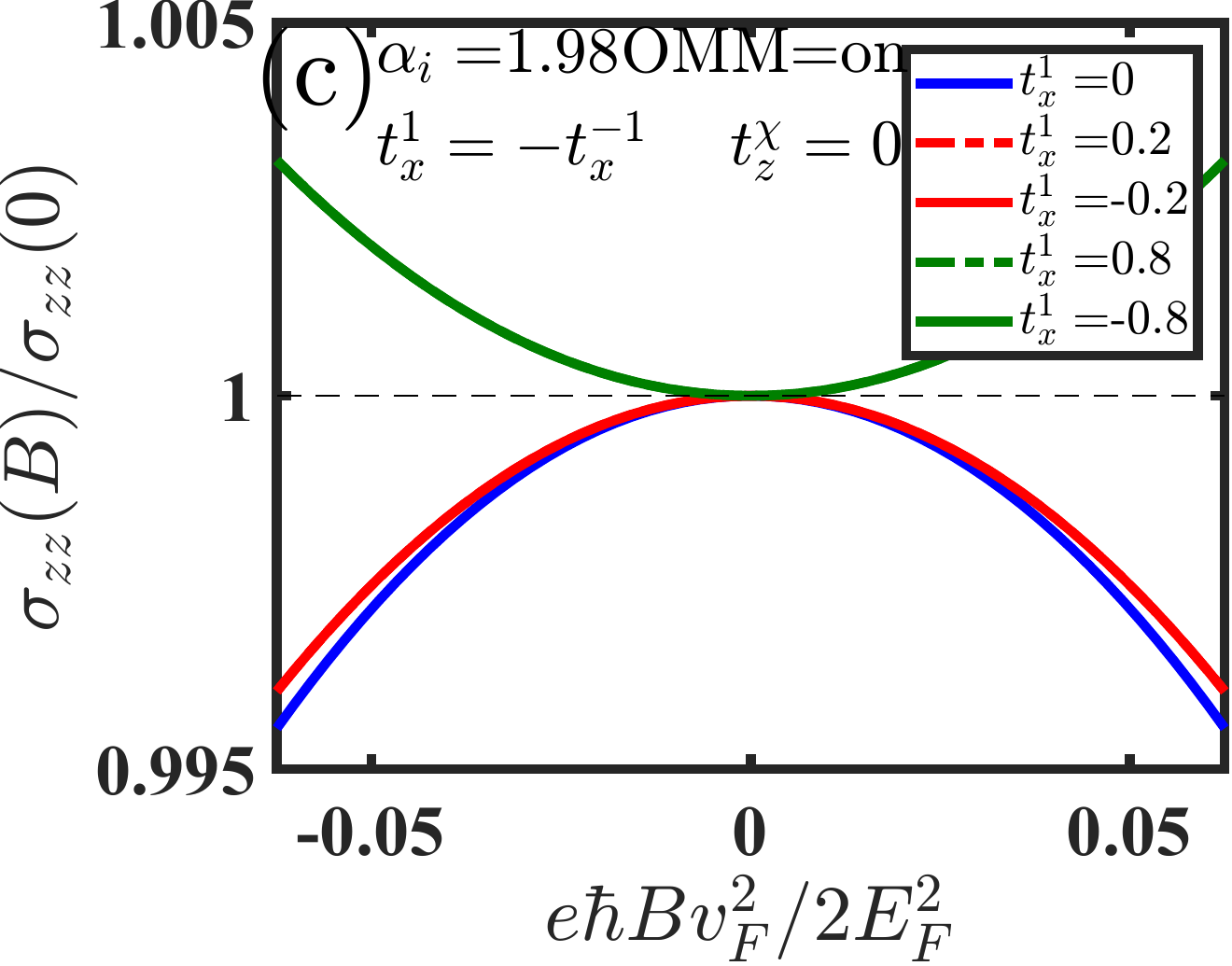}
    \includegraphics[width=0.53\columnwidth]{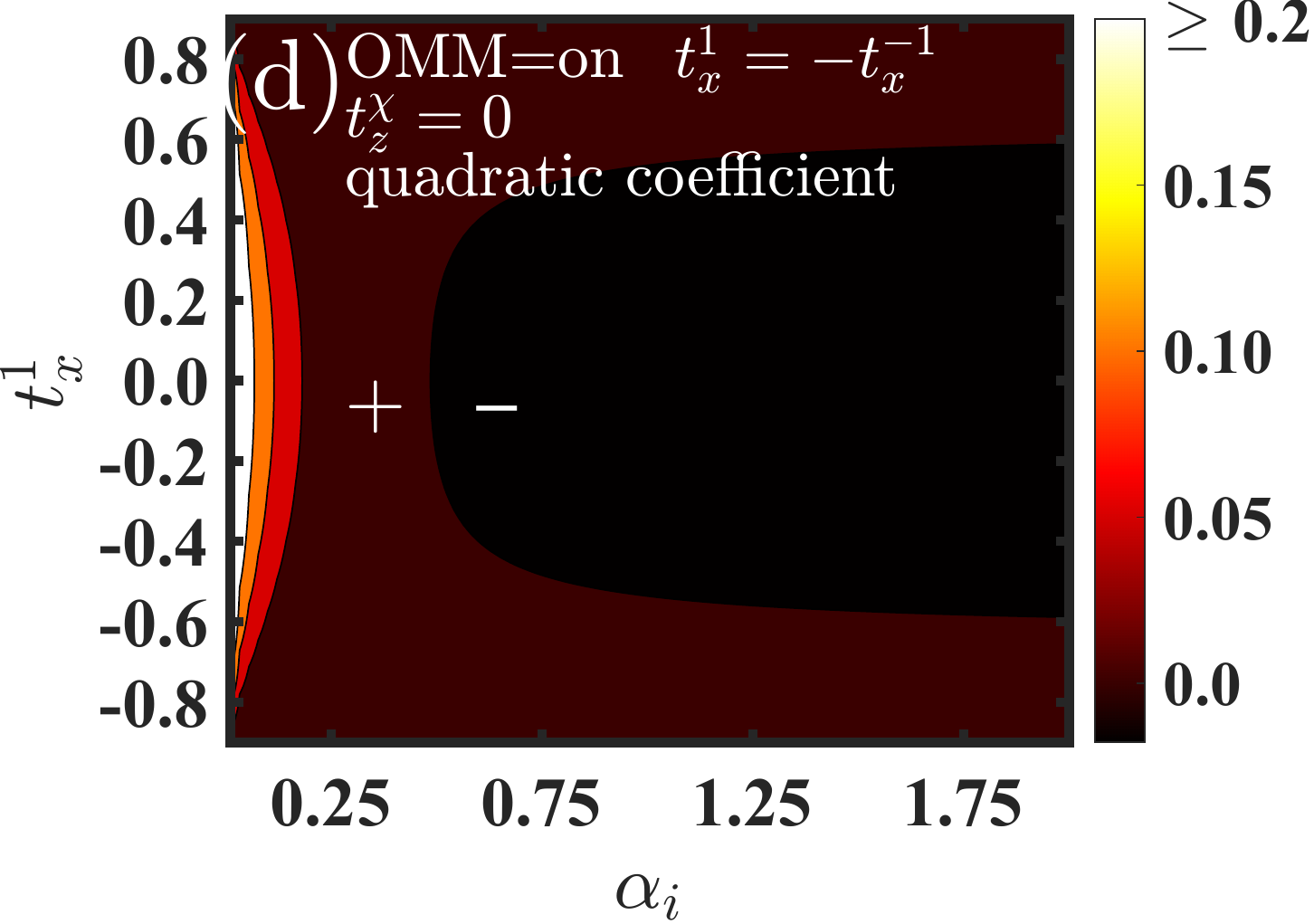}
    \caption{Longitudinal magnetoconductance $\sigma_{zz}(B)$ in the case when the Weyl cones are titled perpendicular to the direction of the magnetic field (i.e. along $\hat{x}$) axis, and are oppositely oriented  to each other ($t_x^1=-t_x^{-1}$).  (a) LMC as a function of magnetic field for various tilt parameters in the absence of intervalley scattering ($\alpha_i=0$). (b) and (c) LMC in the presence of a finite intervalley scattering $\alpha_i$. LMC switches sign with the inclusion of $\alpha_i$ whenever $\alpha_i>\alpha_i^c(t_x)$. (d) The quadratic coefficient of LMC is plotted as a function of $\alpha_i$ and $t_x^1$. The sign of the coefficient also corresponds to the sign of LMC. The contour separating positive and negative LMC regions is also clearly shown.}
    \label{fig:szz_tiltx_opp_ommon}
\end{figure*}


\subsubsection{The case when $t_z^1=t_z^{-1}\neq 0$ and $t_x^\chi=0$}
Fig.~\ref{fig:szz_tiltz_same_ommon} presents the results of LMC as a function of magnetic field when the two Weyl cones are titled in the same direction with respect to each other in the direction of the magnetic field, i.e. $t_z^1 = t_z^{-1}$, and $t_x^\chi=0$. Note that LMC is always quadratic in $B$, because the $B-$linear coefficients cancel out (as they appear with a chirality sign that is opposite for the two Weyl cones). 
When the intervalley scattering $\alpha_i$ is small, LMC is always positive. For large $\alpha_i$, we note that when the tilt ($t_z^1$) magnitude is small, LMC changes sign from positive to negative, but remains positive when the tilt parameter is large enough. The critical value of intervalley scattering ($\alpha_i^c$) where the change in sign occurs is dependent on the tilt parameter, i.e., $\alpha_i^c = \alpha_i^c(t_z^1)$. 
In Fig.~\ref{fig:szz_tiltz_same_ommon}(d) we present the phase plot of the quadratic coefficient $\sigma_{zz2}$. The sign of the quadratic coefficient corresponds to the sign of LMC in this case as the linear-in-$B$ term is absent. We also map out the contour in $\alpha_i-t_z^1$ space where the change in sign of LMC occurs. When $\alpha_i\gtrsim 0.5$ and $|t_z^1|\lesssim 0.6$, LMC is observed to be negative,  but remains positive and has a weak dependence on $\alpha_i$ when $|t_z^1|\gtrsim 0.6$. The LMC is determined by the interplay of $\alpha_i$ and $t_z^1$ and the tilt parameter opposes the change in sign of LMC due to intervalley scattering and its contribution dominates when $|t_z^1|\gtrsim 0.6$. 
This behavior of the quadratic coefficient is similar to that observed in Fig.~\ref{fig:szz_tiltz_opp_ommon} (h), but the presence of a linear-in-$B$ coefficient in the previous case changes the qualitative behavior of LMC. 

\begin{figure*}
    \centering
    \includegraphics[width=0.49\columnwidth]{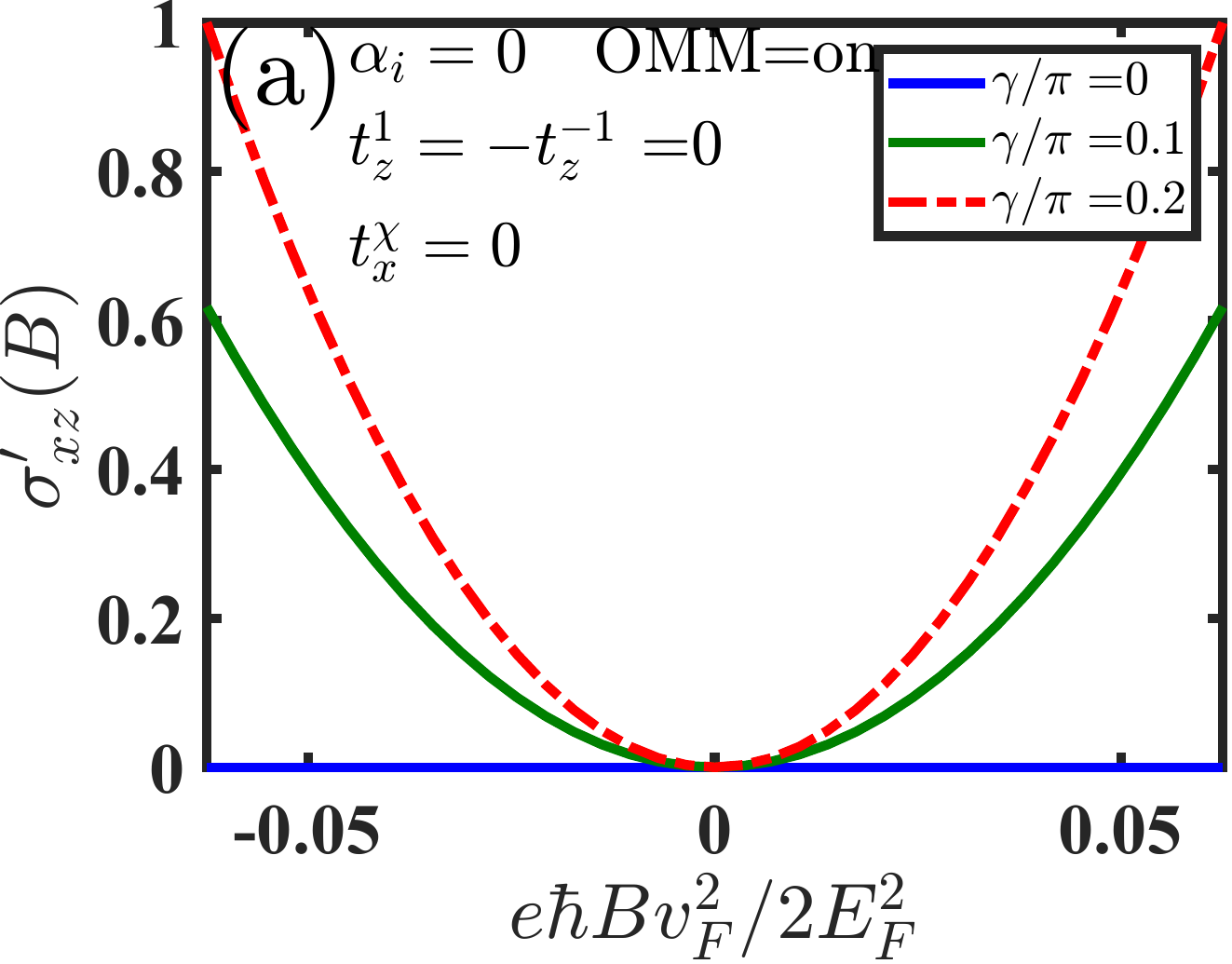}
     \includegraphics[width=0.49\columnwidth]{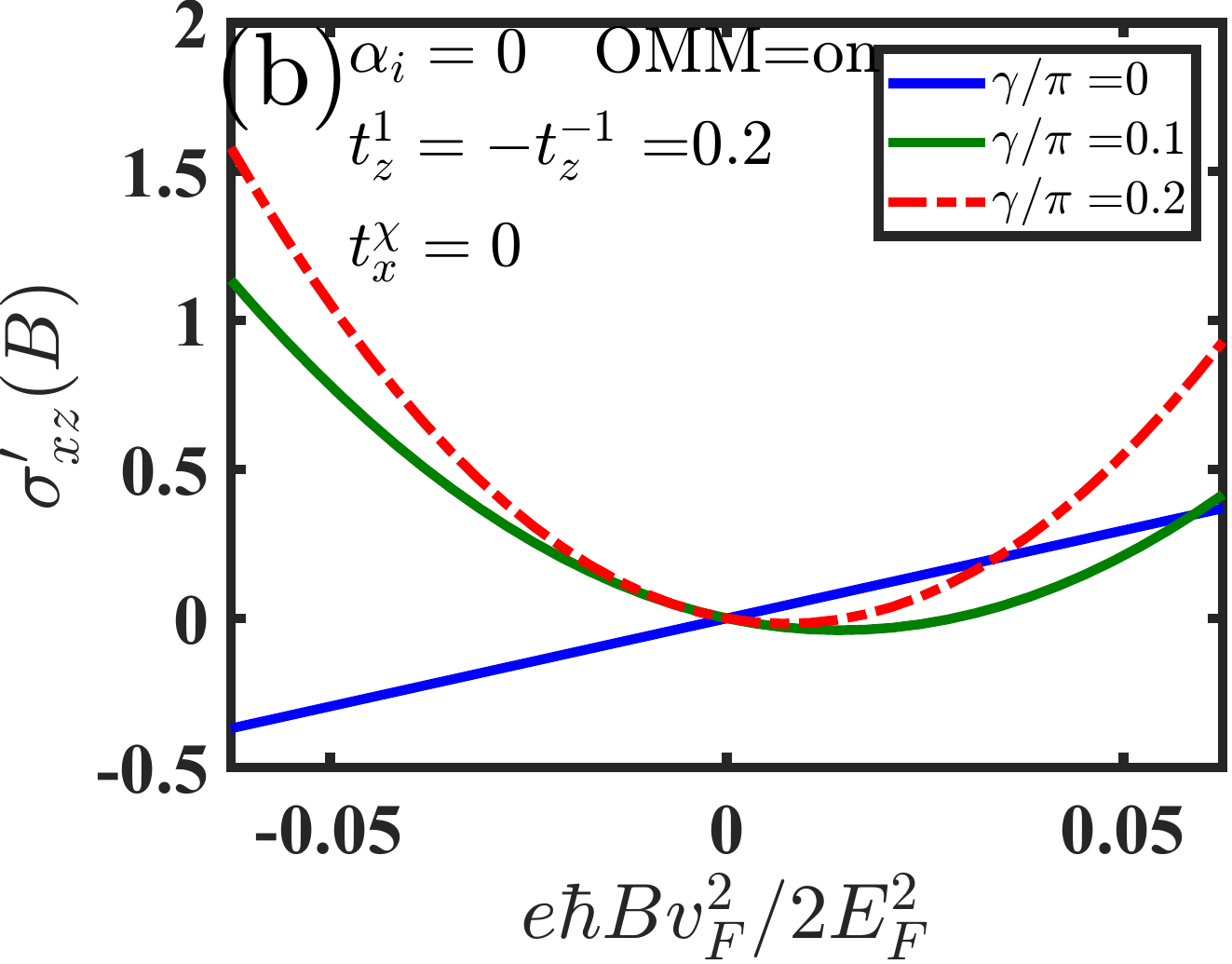}
    \includegraphics[width=0.49\columnwidth]{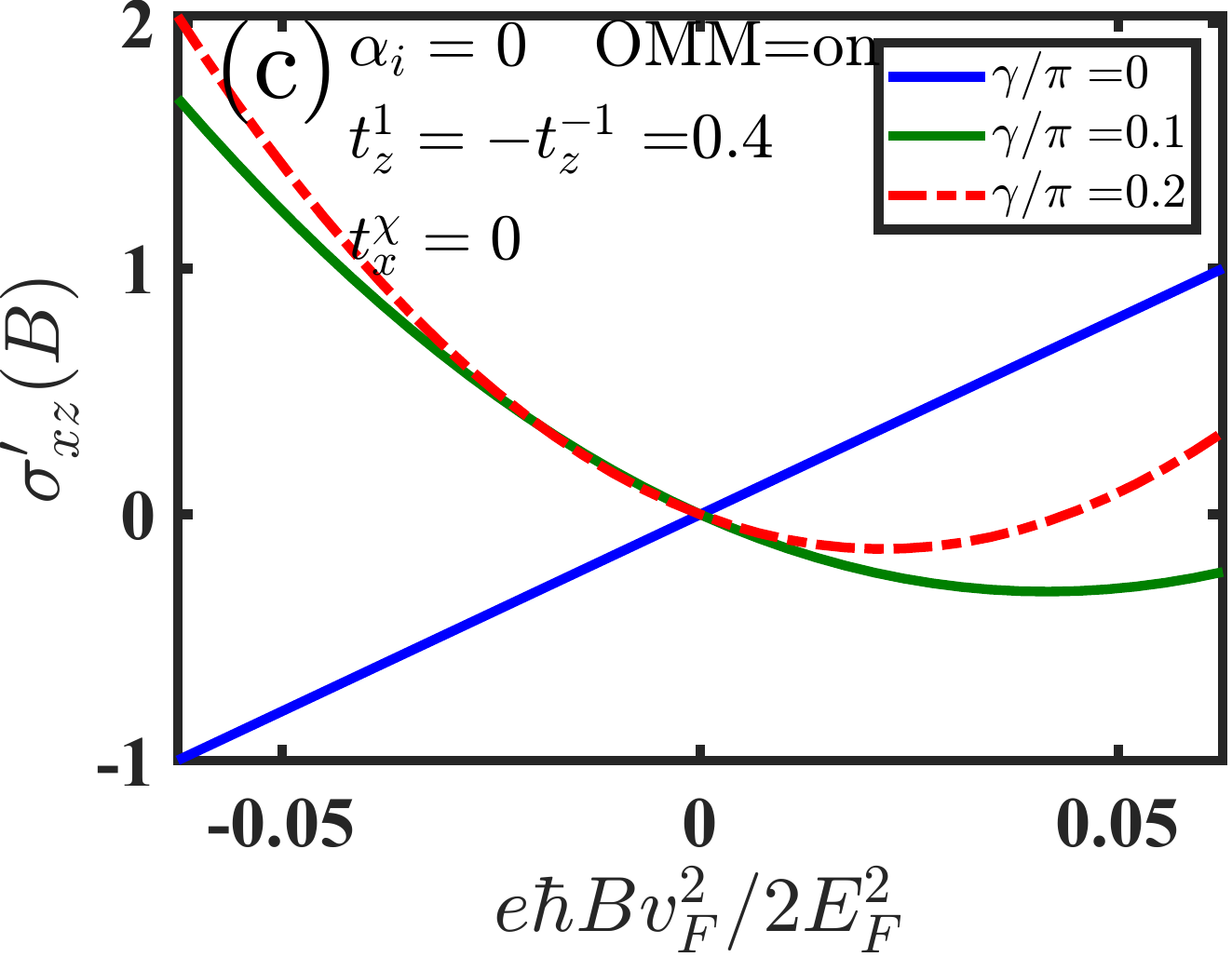}
    \includegraphics[width=0.49\columnwidth]{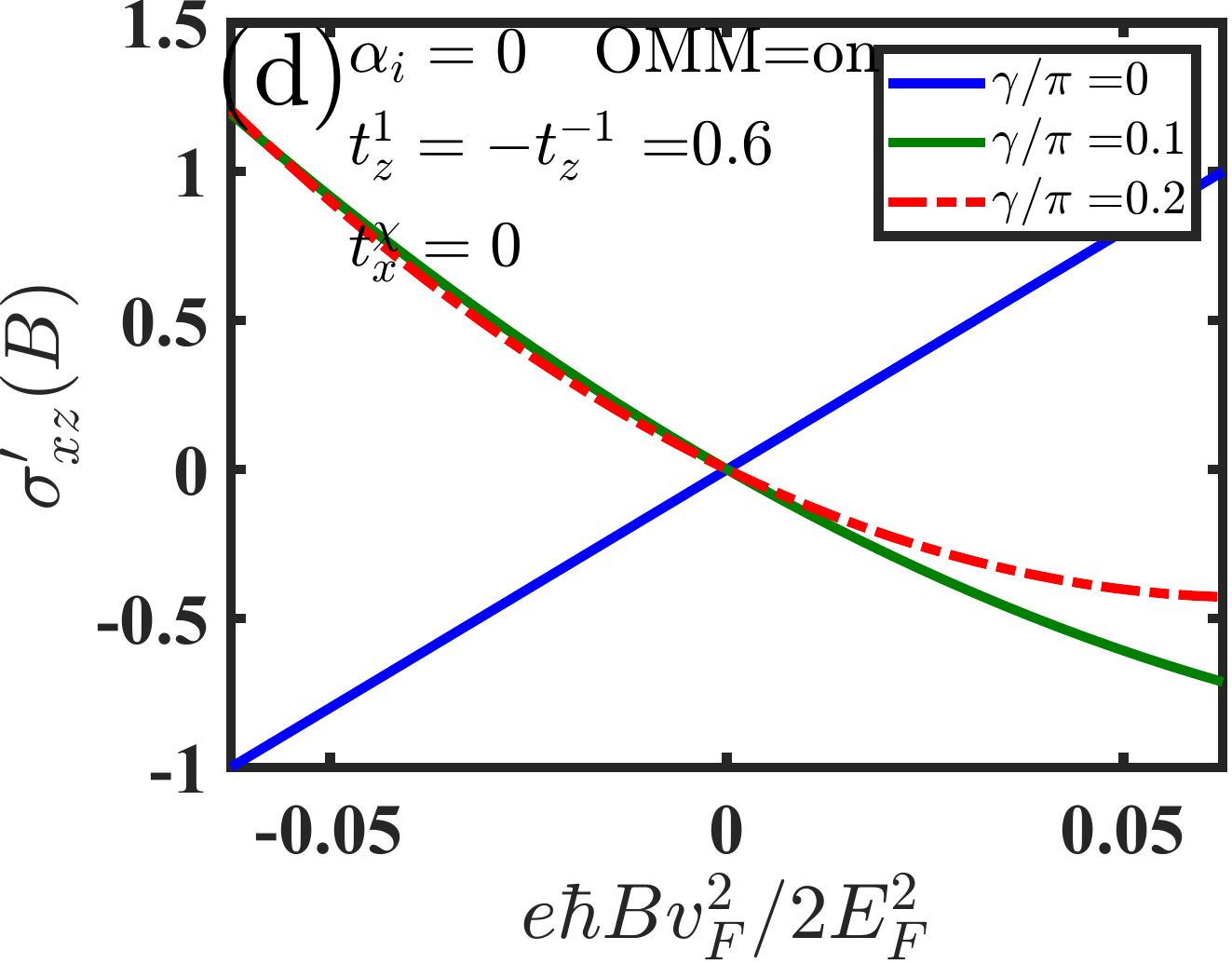}
    \includegraphics[width=0.49\columnwidth]{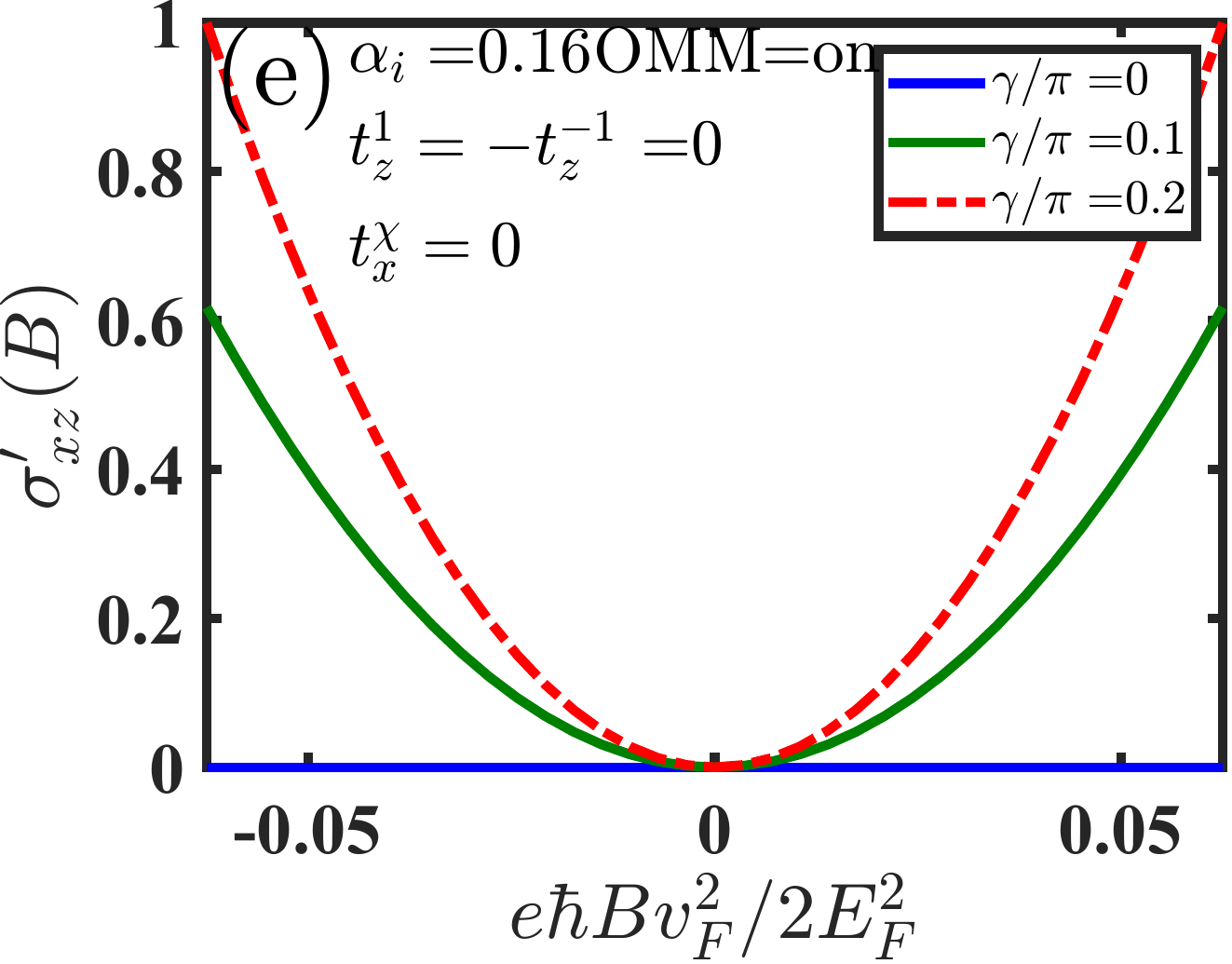}
    \includegraphics[width=0.49\columnwidth]{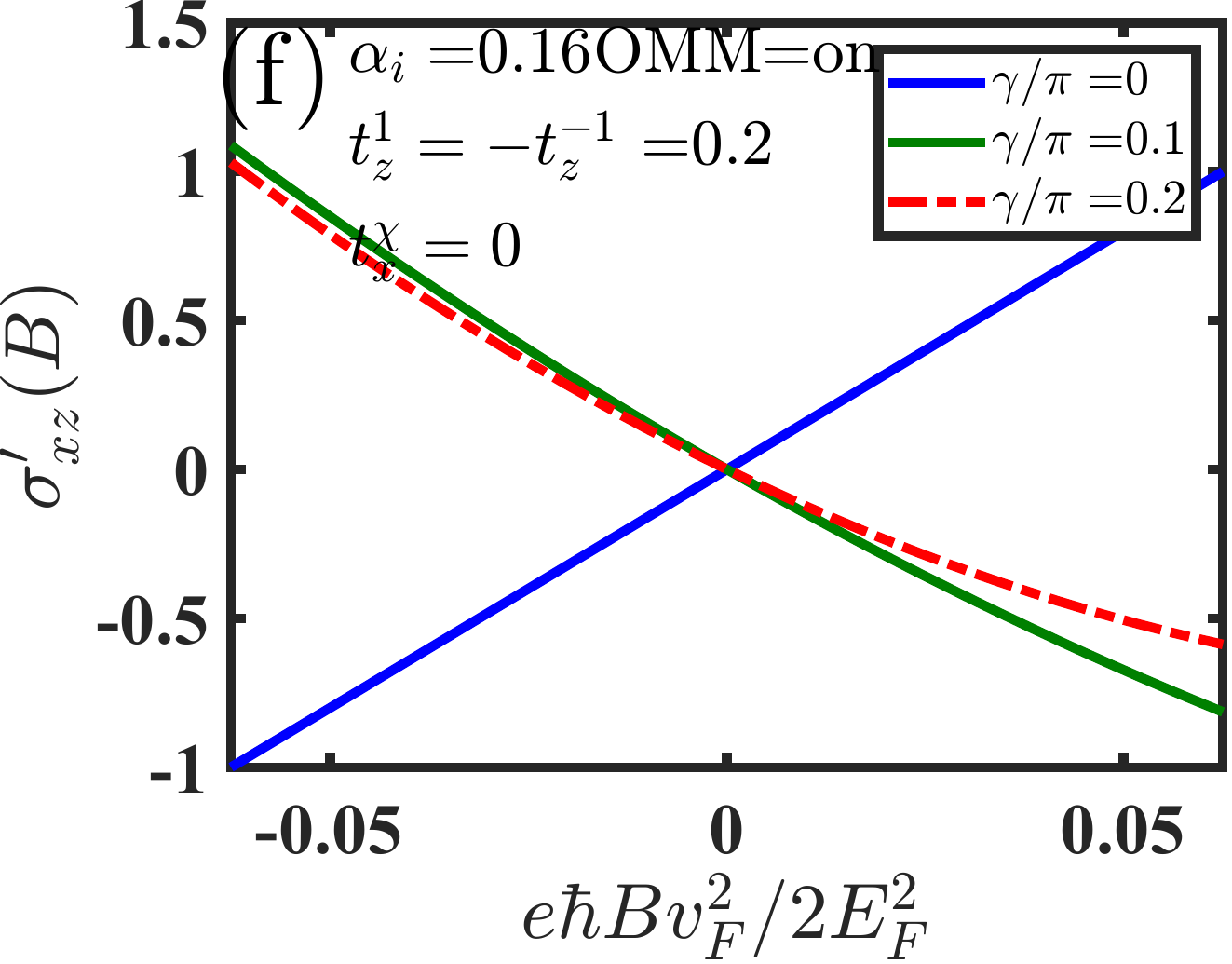}
    \includegraphics[width=0.49\columnwidth]{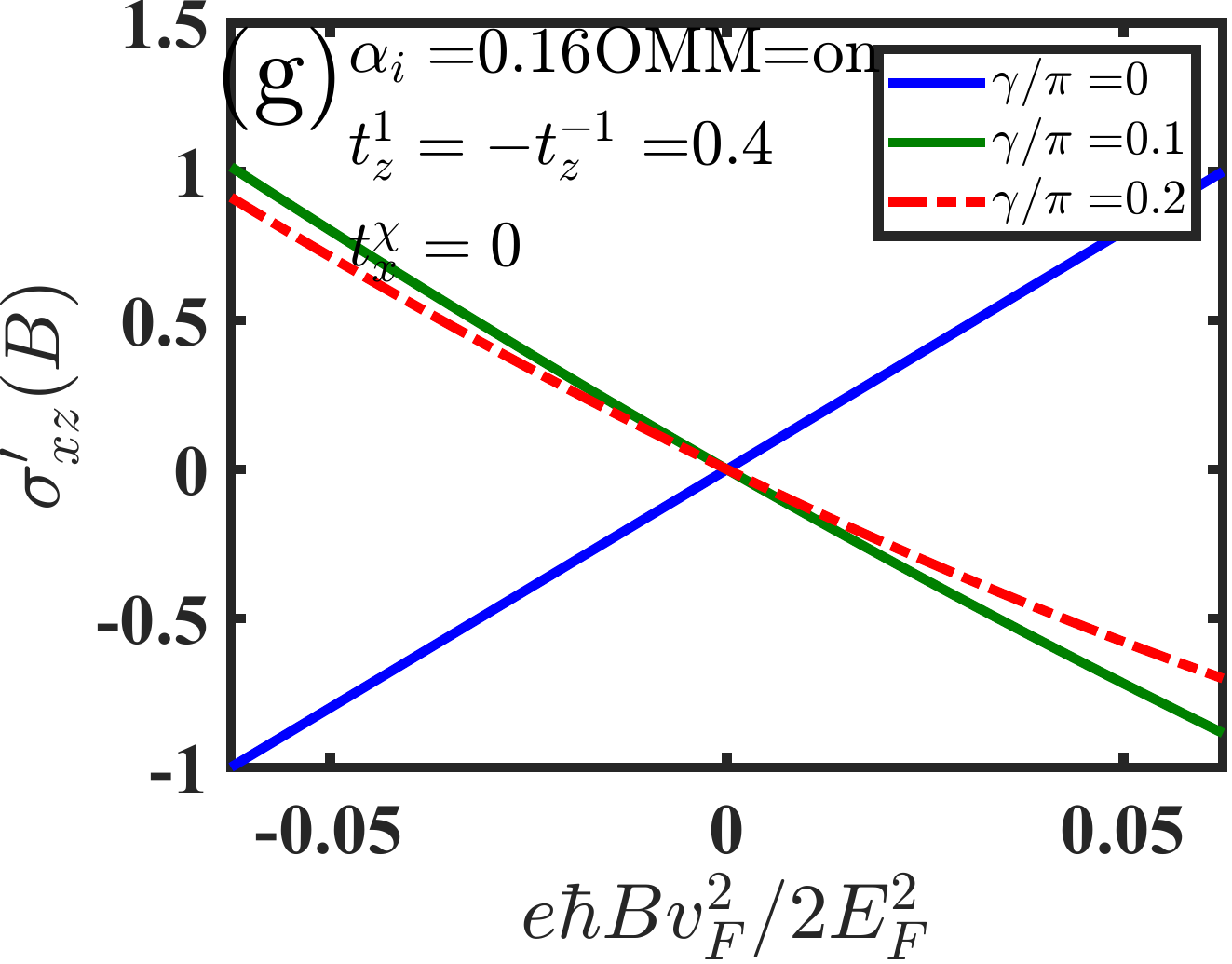}
    \includegraphics[width=0.49\columnwidth]{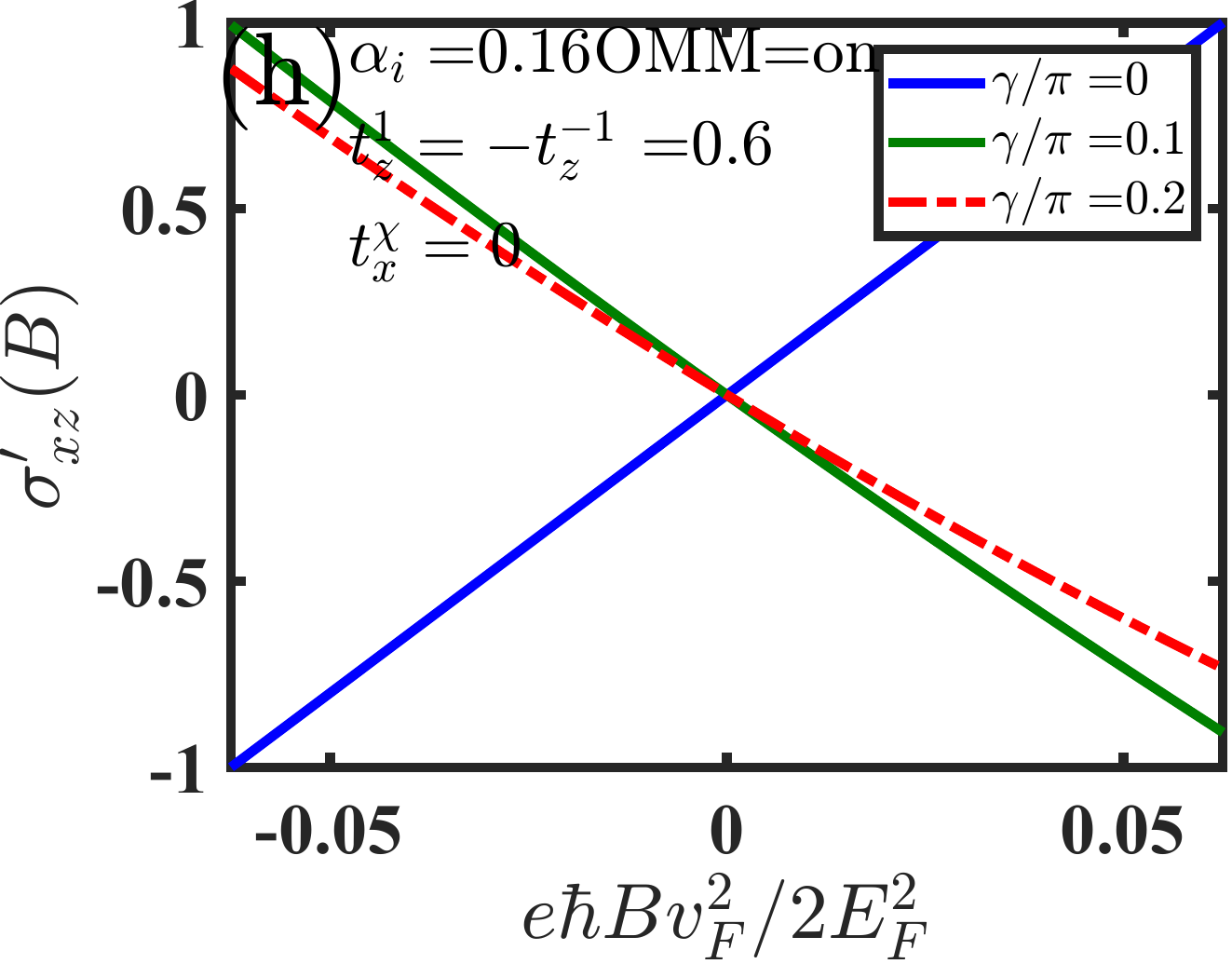}
    \caption{Normalized planar Hall conductivity $\sigma_{xz}'$ (prime indicating that the value is normalized with respect to the value at 0.5T) as a function of the magnetic field for different values of the tilt parameter $t_z^\chi$ (oppositely tilted Weyl cones) and at angles $\gamma$. In (a)-(d), the intervalley strength is zero. A finite tilt is observed to add a $B$-linear component that shifts the minima of $\sigma_{xz}'$ away from $B=0$. For a higher tilt value, the behavior is linear for all relevant range of magnetic field. In (e)-(h), we apply a finite intervalley scattering strength $\alpha_i$. This enhances the $B$-linear contribution, however only in the presence of a finite tilt.}
    \label{fig:sxz_vs_B_tiltz_opp_omm_on}
\end{figure*}
\begin{figure*}
    \centering
    \includegraphics[width=0.49\columnwidth]{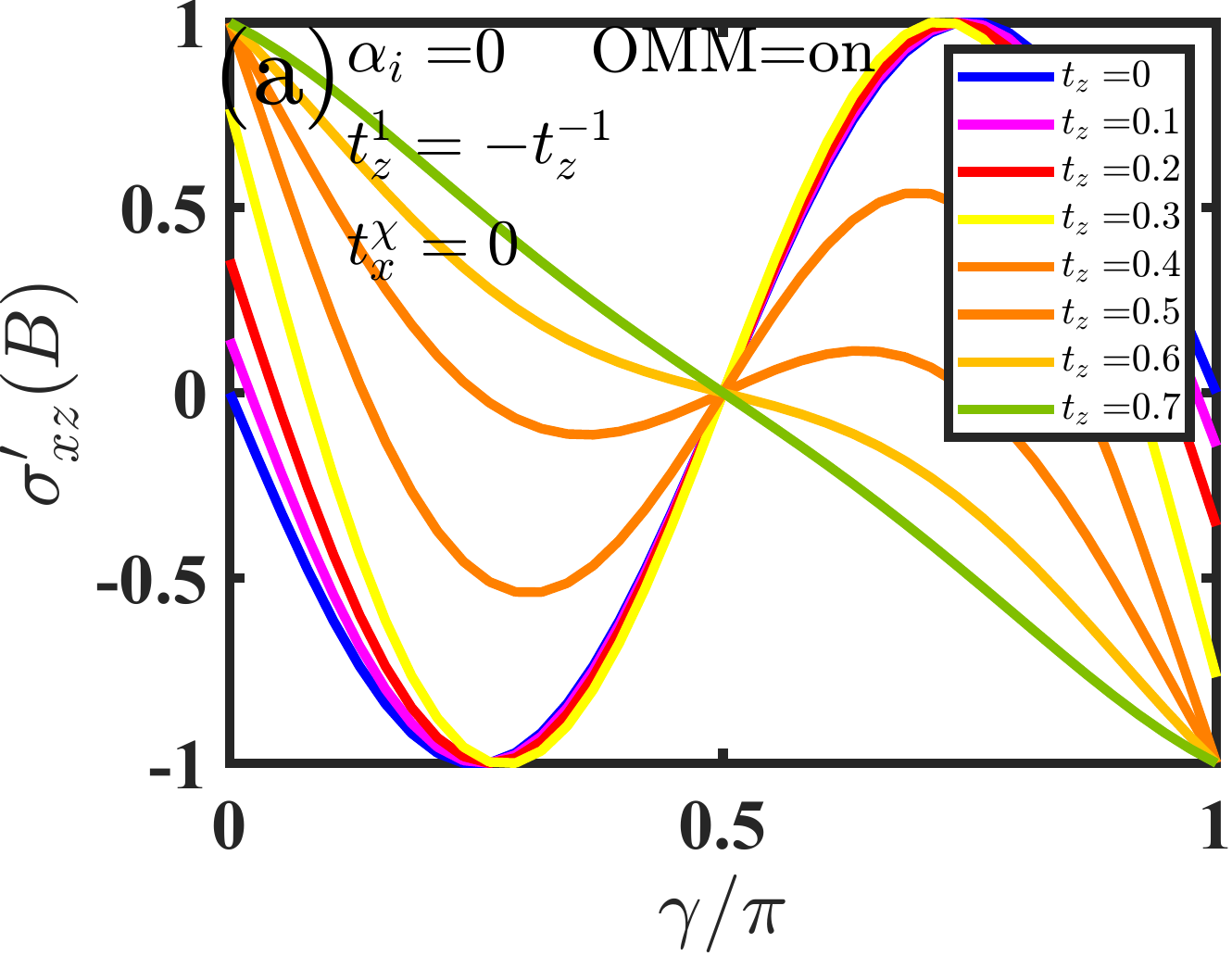}
    \includegraphics[width=0.49\columnwidth]{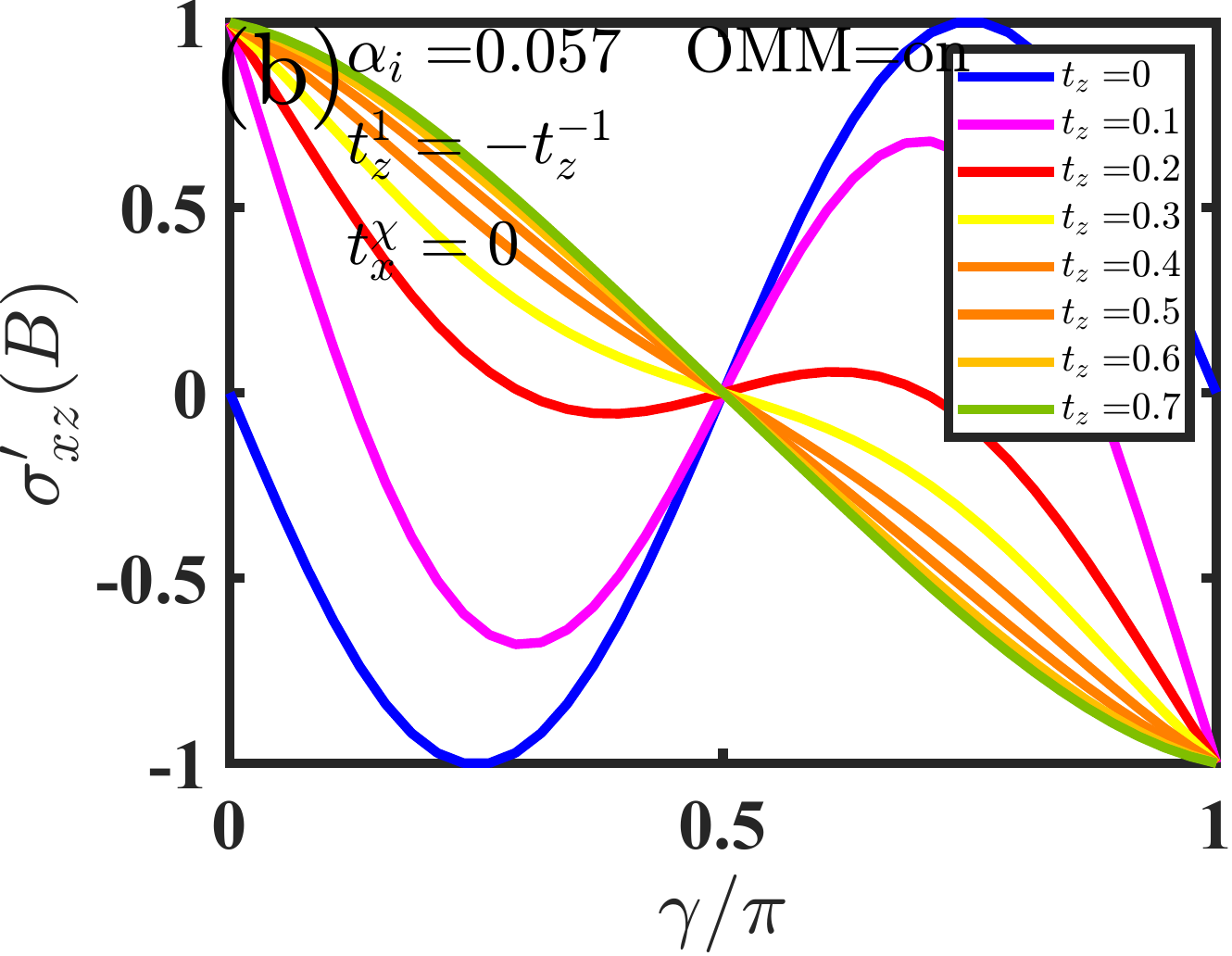}
    \includegraphics[width=0.49\columnwidth]{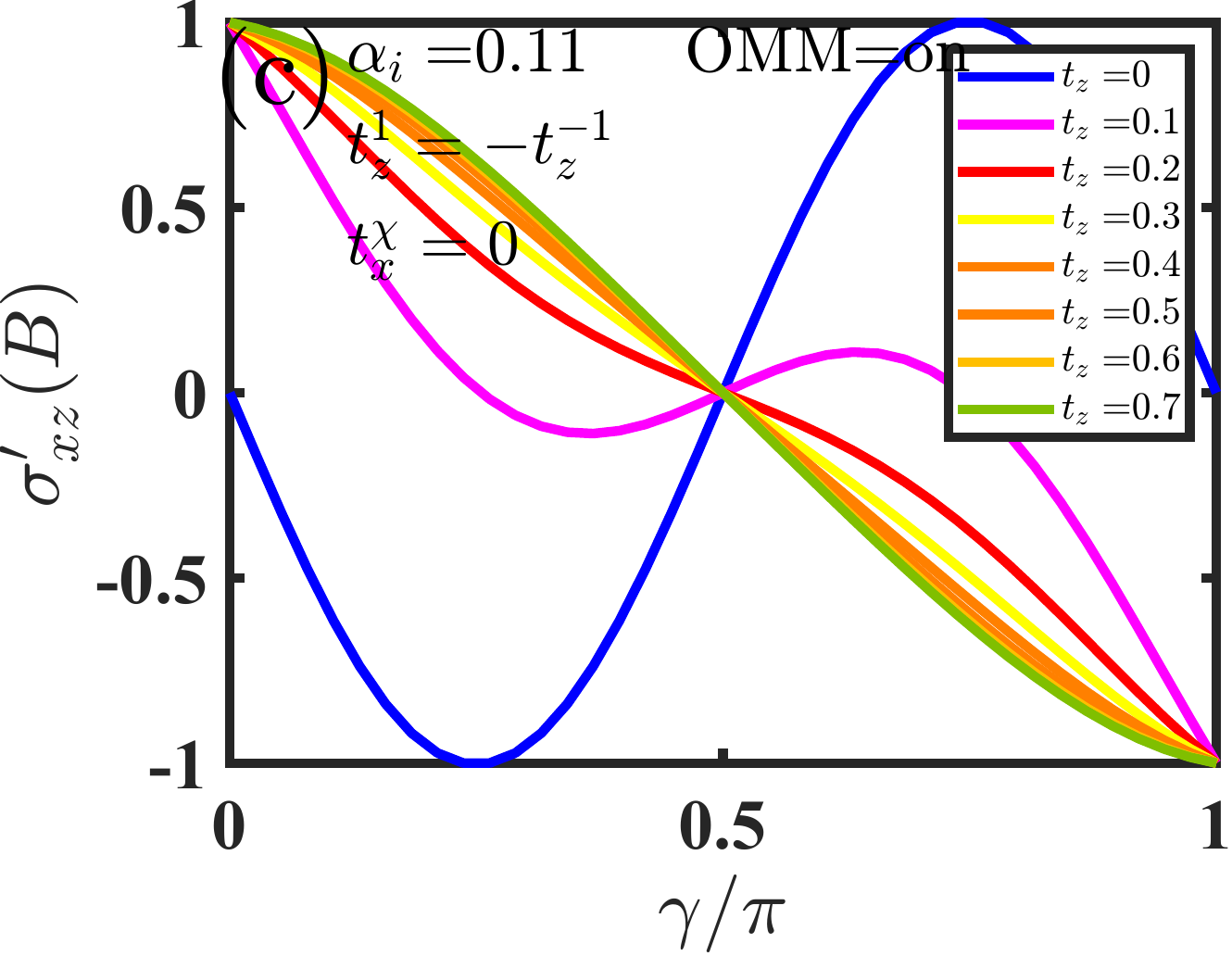}
    \includegraphics[width=0.49\columnwidth]{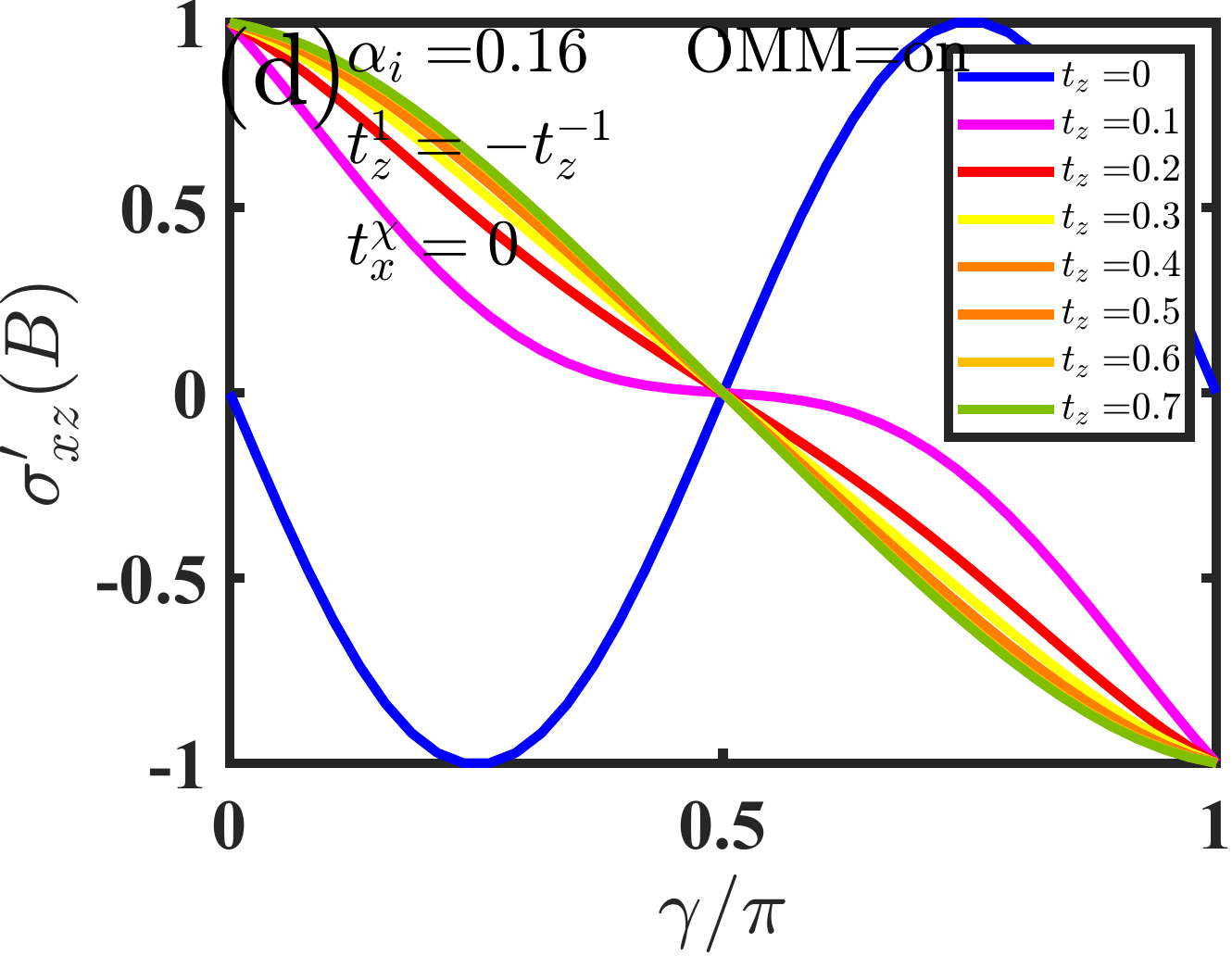}
    \caption{Normalized planar Hall conductivity ($\sigma_{xz}'$) as a function of the angle $\gamma$ for several values of tilt parameter $t_z$ for oppositely tilted Weyl cones. In the absence of tilt the behavior follows the trend $\sin(2\gamma)$, while in the presence of tilt, a $\cos\gamma$ component is added. Beyond a critical $t_z^c$, the $\cos\gamma$ term dominates and $\sigma_{xz}'(\pi/2 + \epsilon)$ changes from positive to negative, where $\epsilon$ is a small positive angle.
    A finite intervalley scattering further enhances the $\cos\gamma$ trend (however only in the presence of a finite tilt). It's effect is to lower the critical tilt $t_z^c$ where the sign change occurs.}
    \label{Fig_sxz_vs_gamma_tiltz_opp_omm_on}
\end{figure*}

\begin{figure*}
    \centering
    \includegraphics[width=0.49\columnwidth]{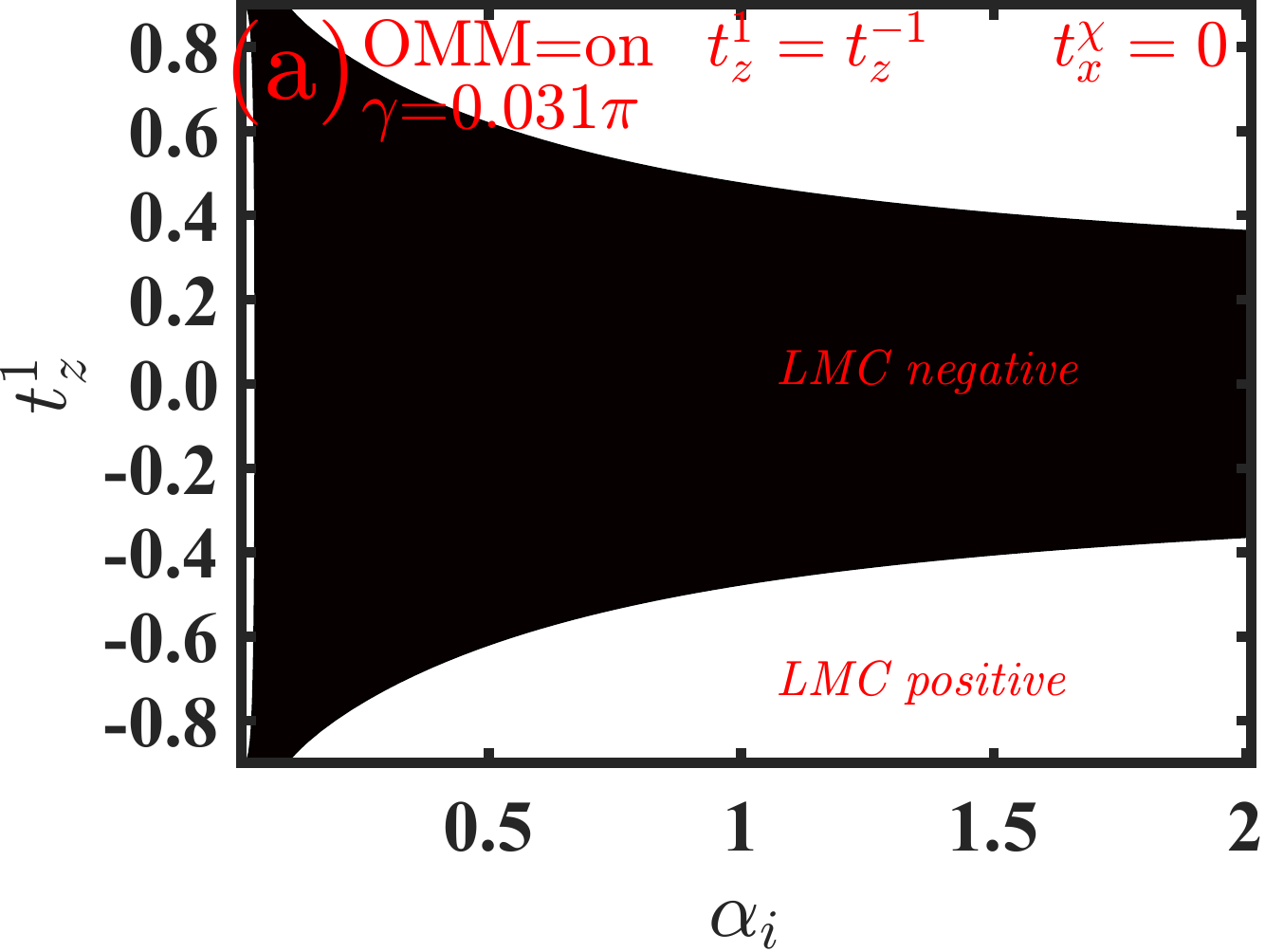}
    \includegraphics[width=0.49\columnwidth]{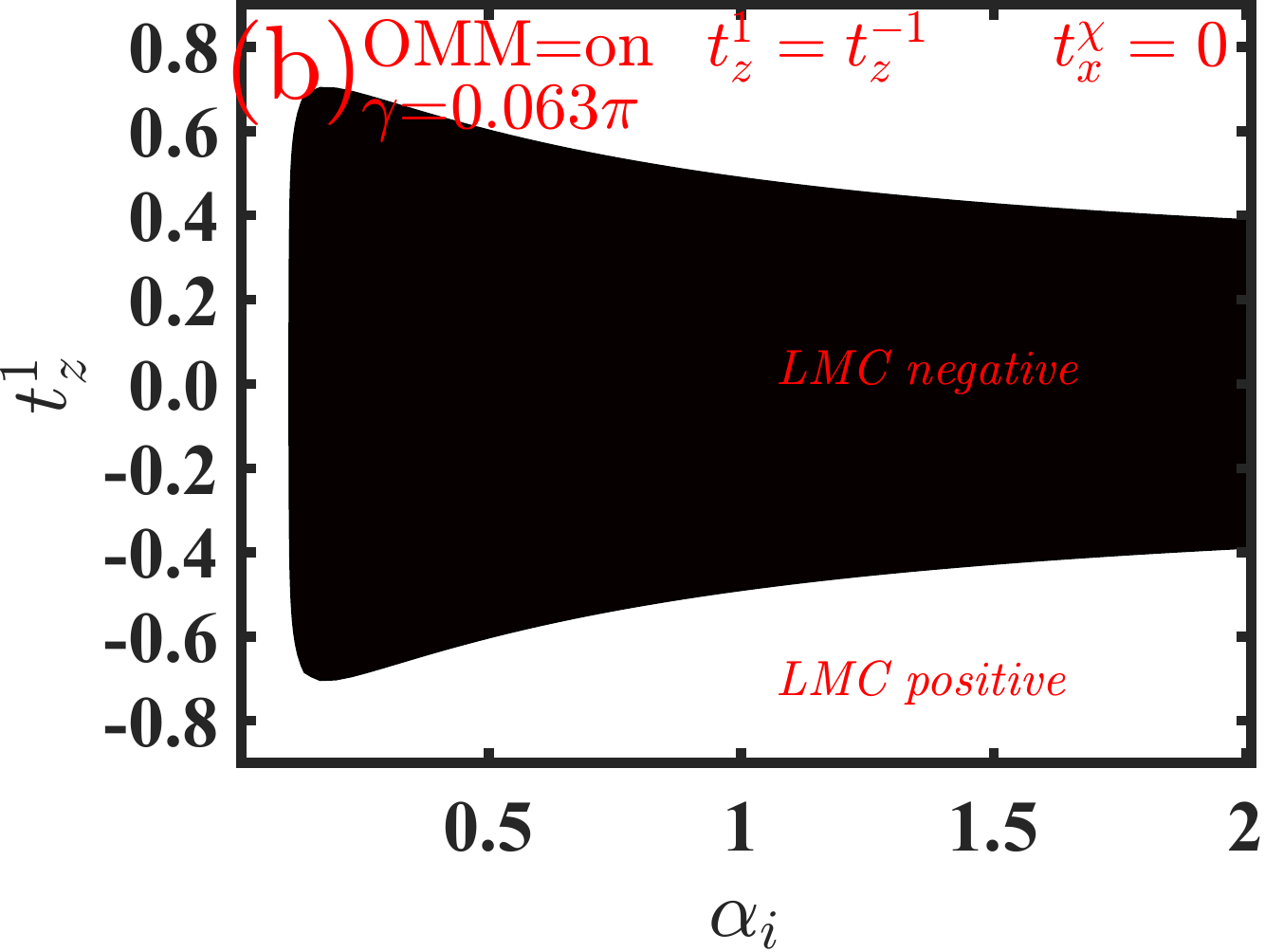}
    \includegraphics[width=0.49\columnwidth]{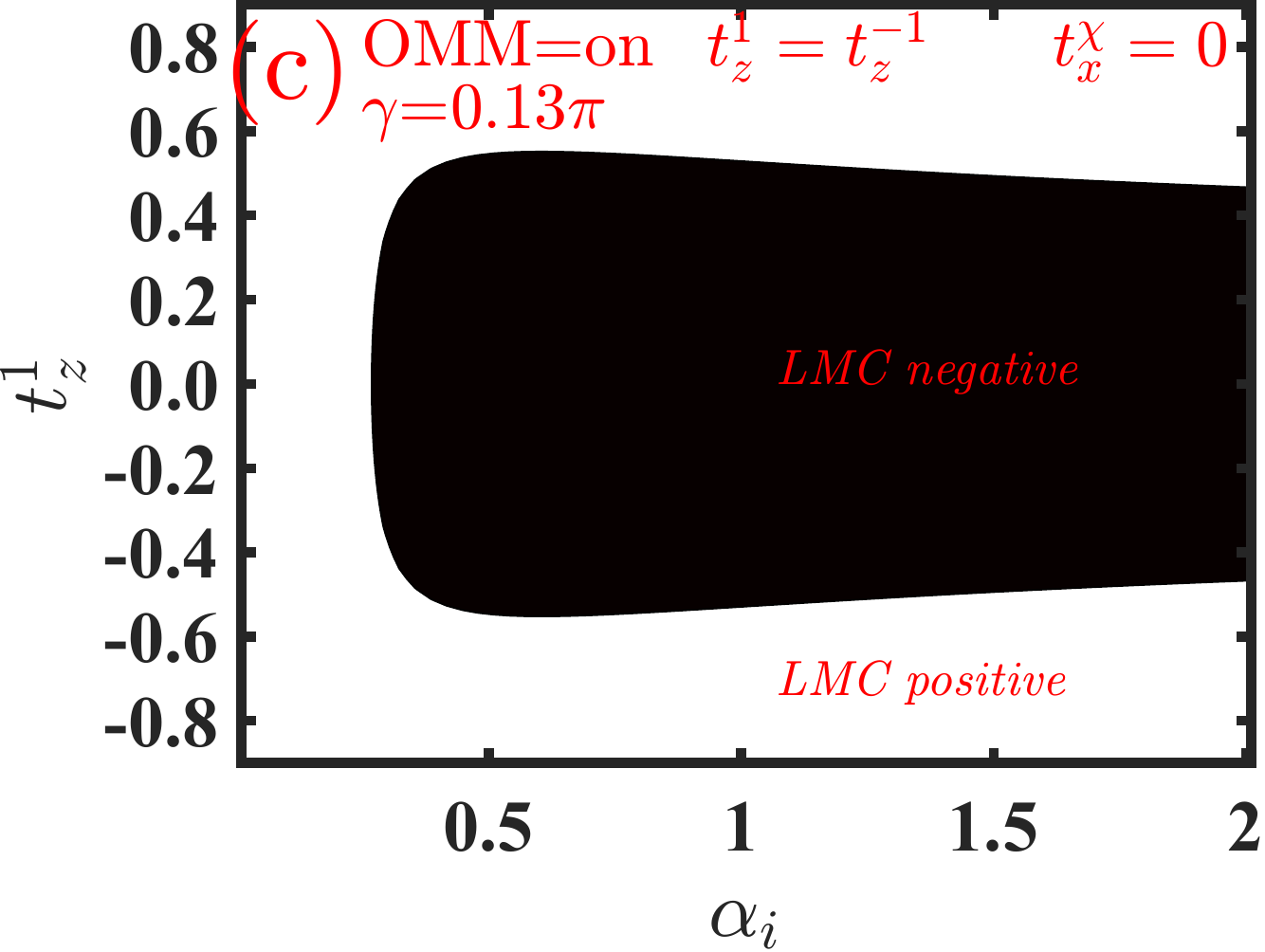}
    \includegraphics[width=0.49\columnwidth]{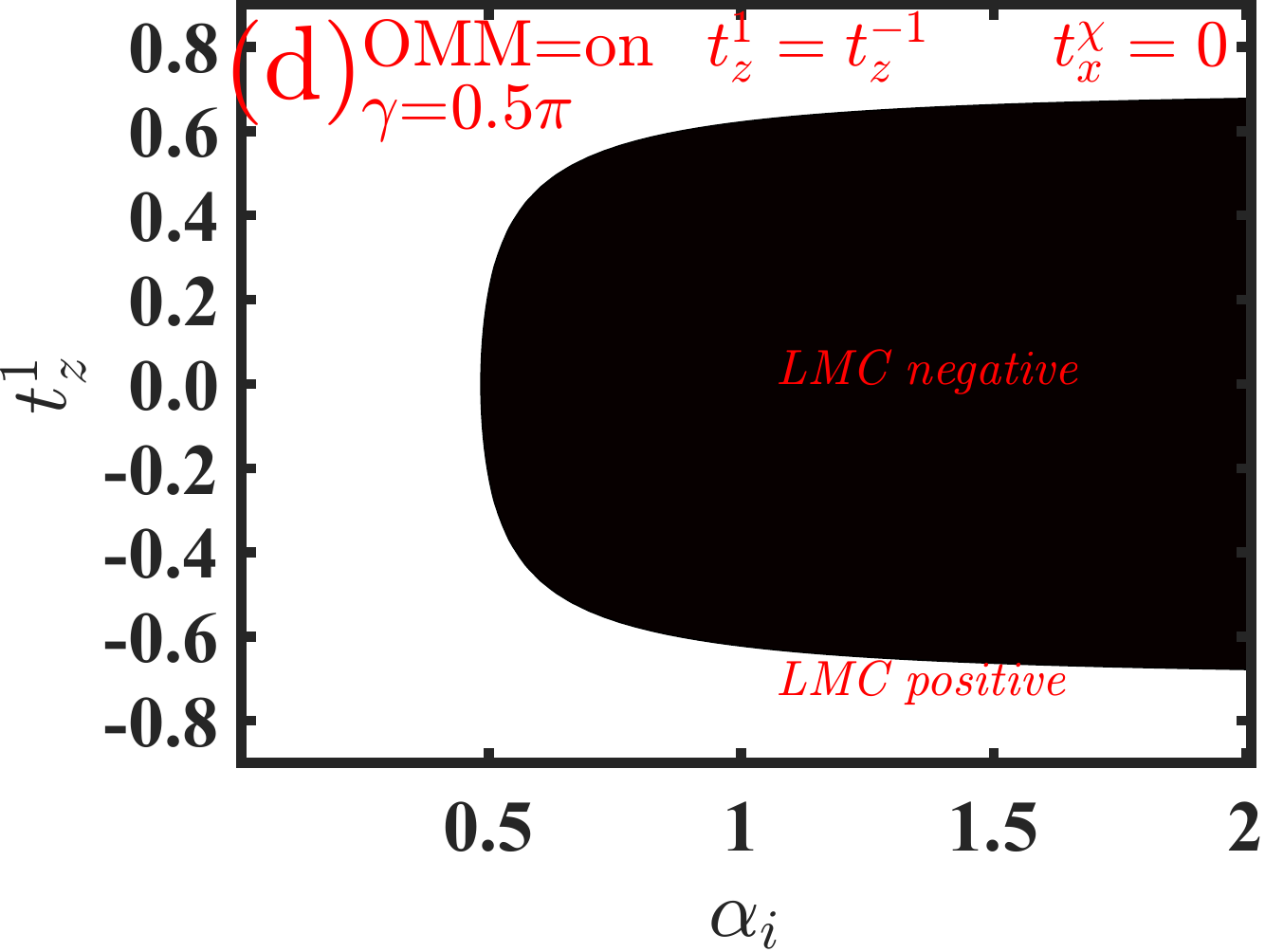}
    \caption{The sign of LMC $\sigma_{zz}$ when the Weyl cones are tilted in the same direction ($t_z^1=t_z^{-1}\neq 0$ and $t_x^\chi=0$) for different angles of the magnetic field. As $\gamma\rightarrow\pi/2$ (parallel $\mathbf{E}$ and $\mathbf{B}$ fields), we recover the result presented in Fig.~\ref{fig:szz_tiltz_same_ommon} and the shape of zero LMC contour (line separating the black and white regions) is like a rotated $U$. When $\gamma$ is directed away from $\pi/2$ the shape of the zero LMC contour looks like a curved trapezoid.}
    \label{Fig_lmc_sign_t1z_ai_tiltz_same_omm_on_gm}
\end{figure*}

\begin{figure*}
    \centering
    \includegraphics[width=0.49\columnwidth]{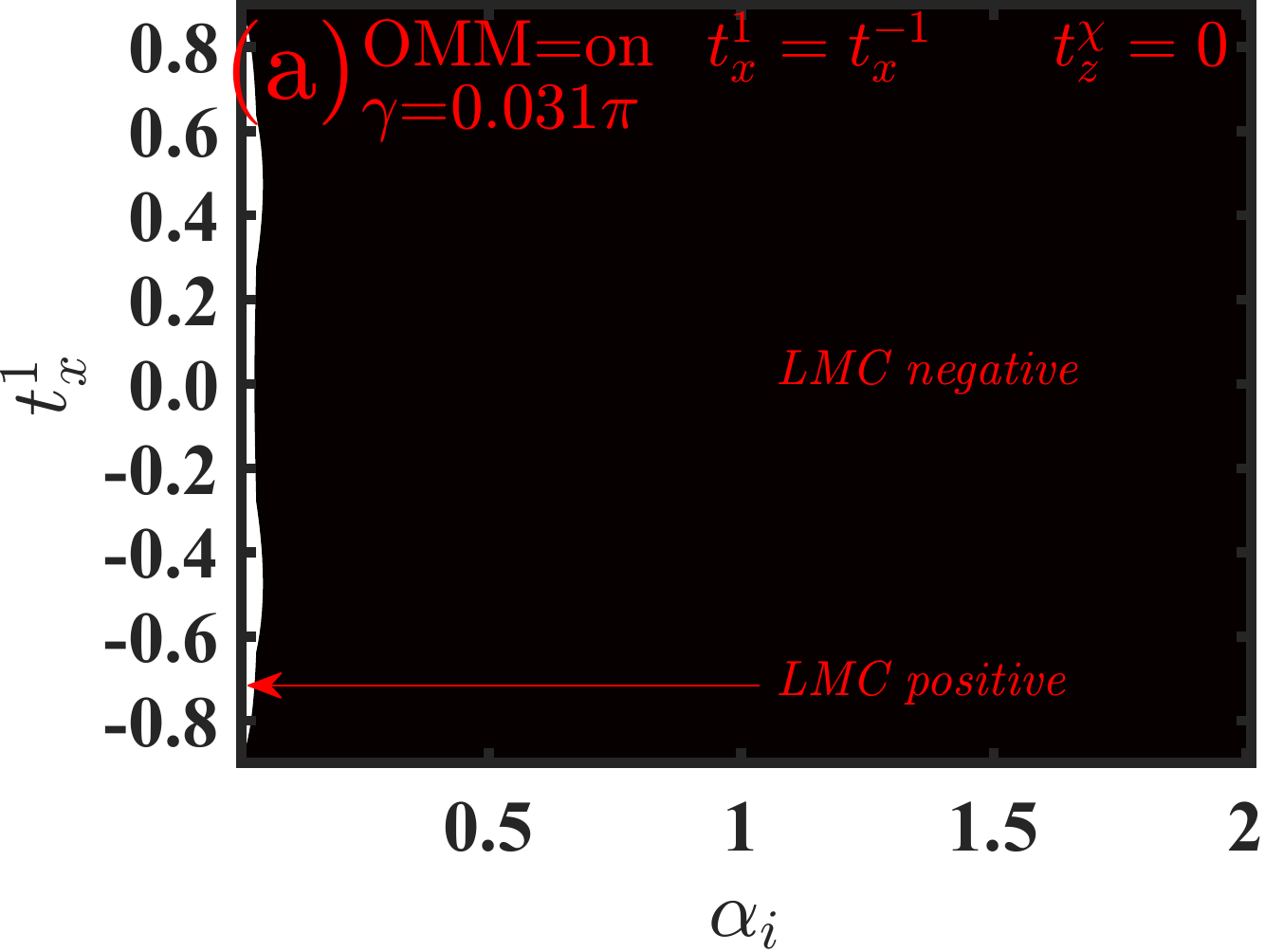}
    \includegraphics[width=0.49\columnwidth]{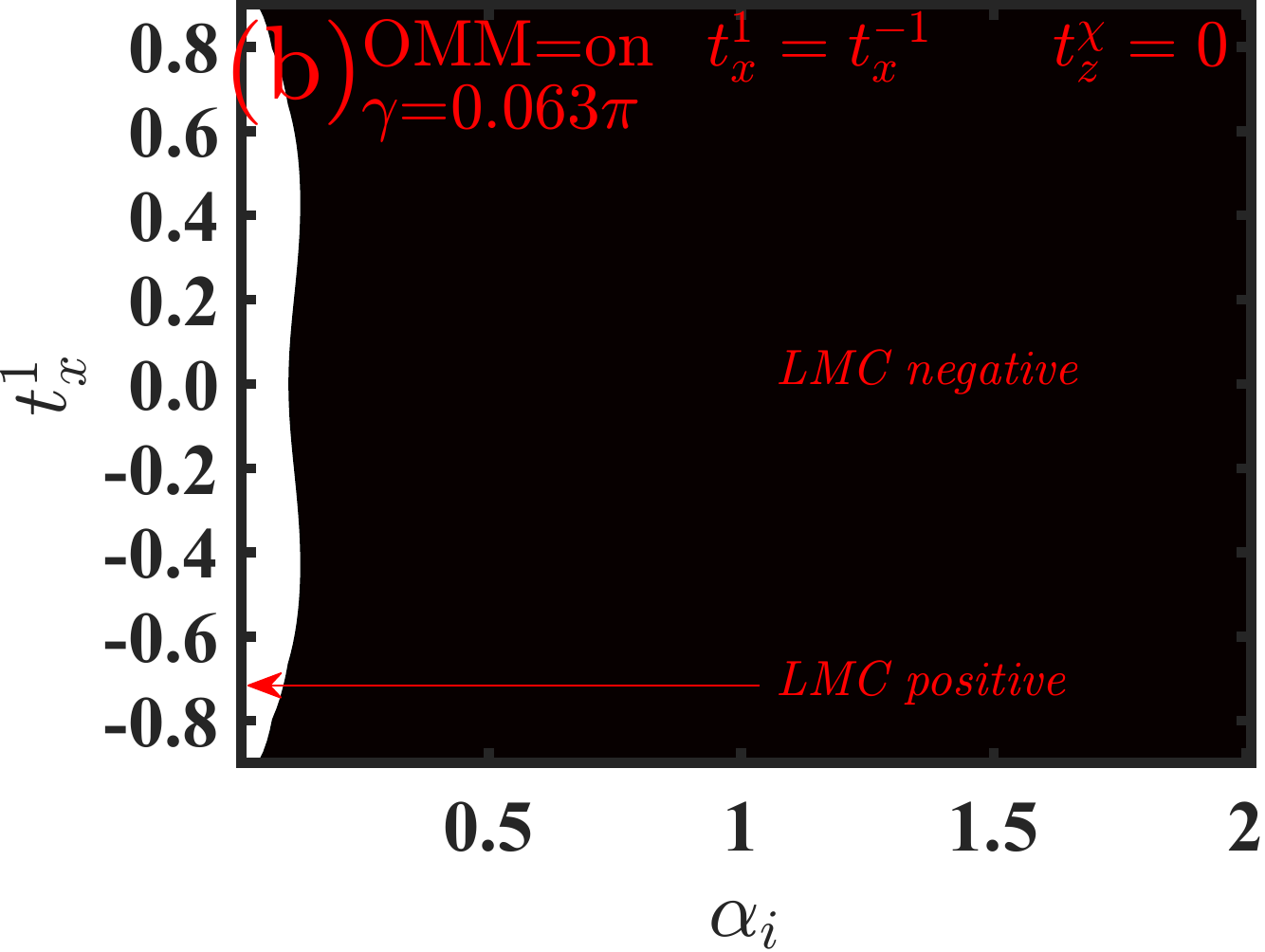}
    \includegraphics[width=0.49\columnwidth]{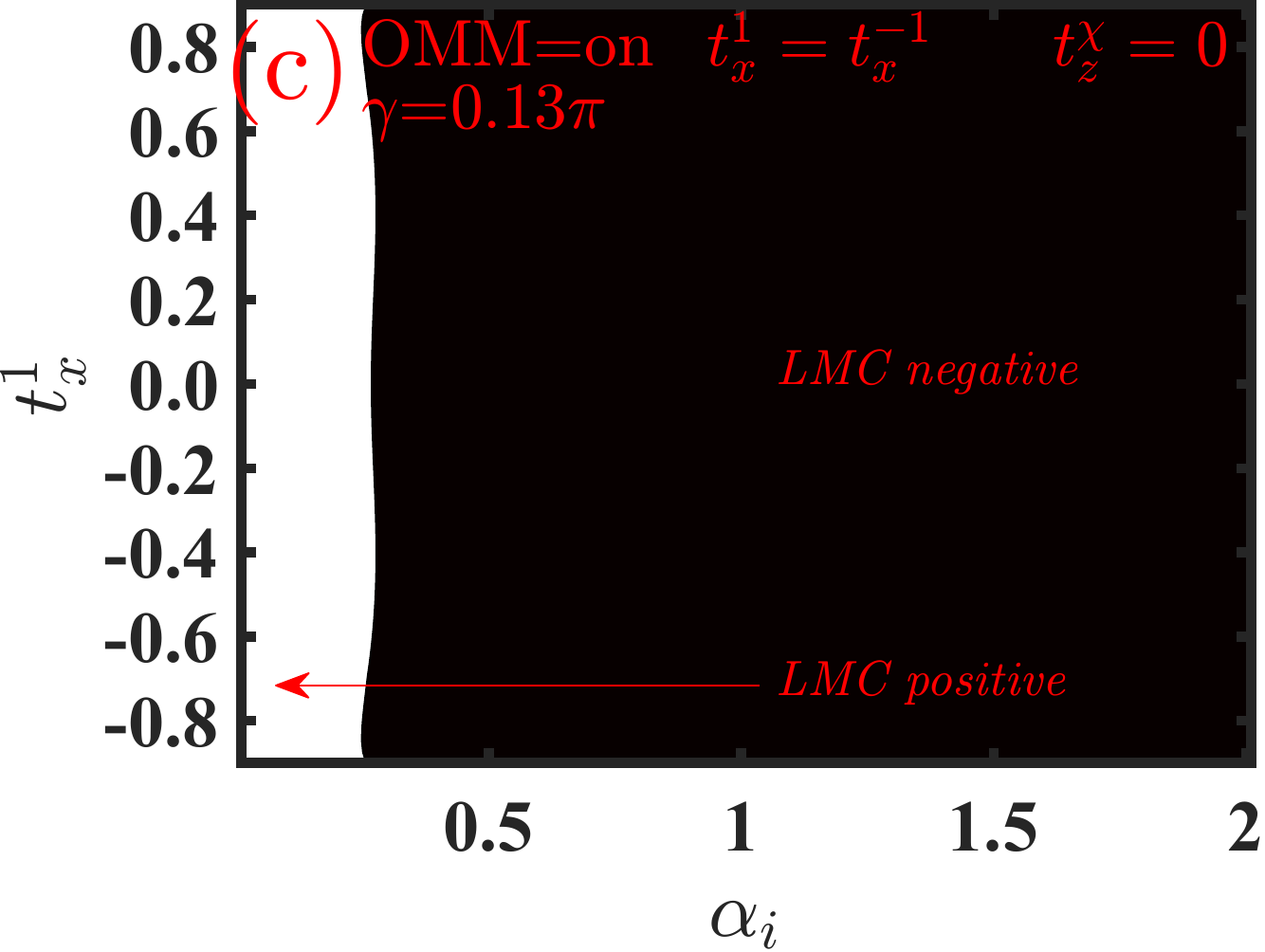}
    \includegraphics[width=0.49\columnwidth]{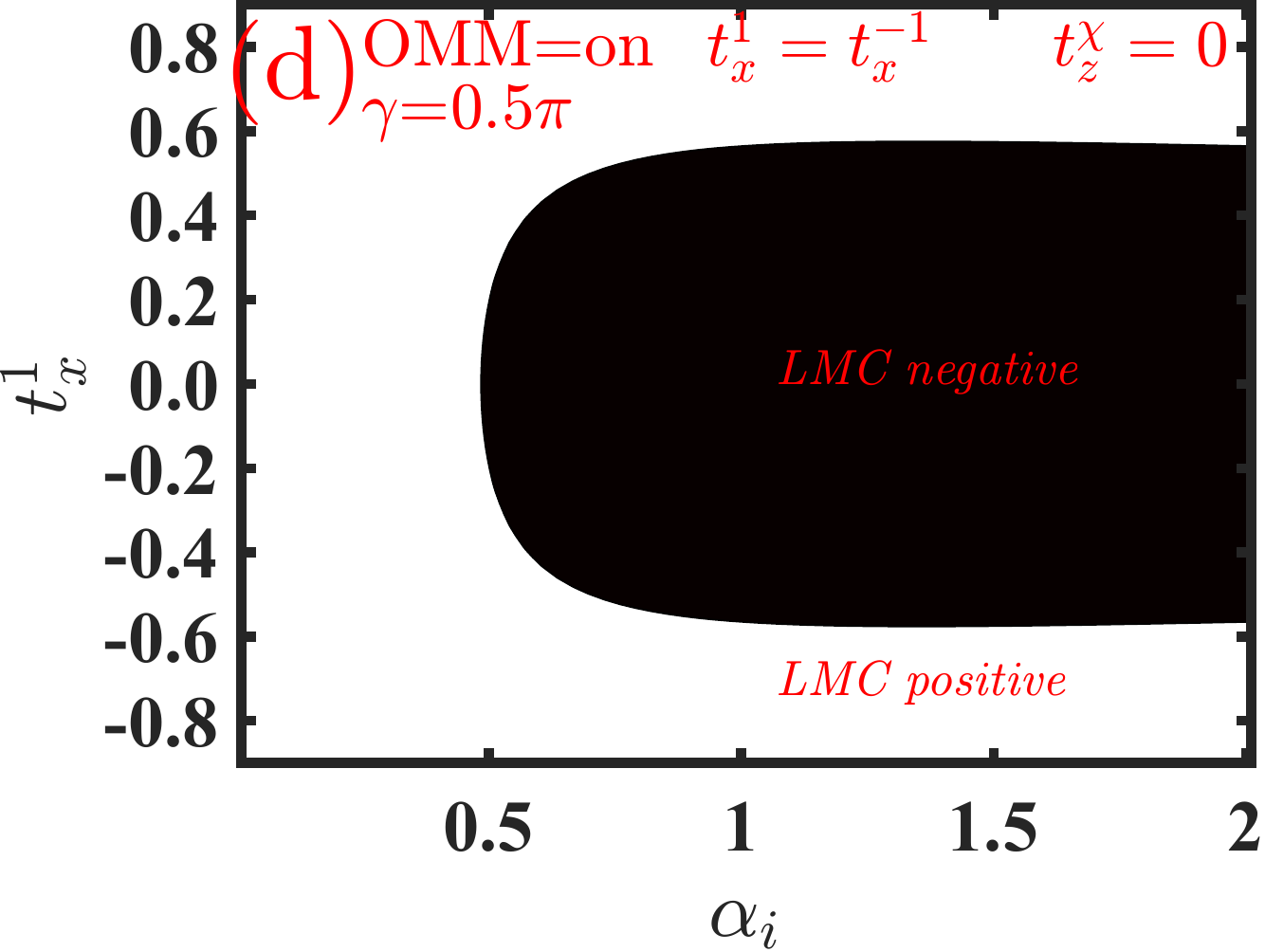}    
    \caption{The sign of LMC $\sigma_{zz}$ when the Weyl cones are tilted in the same direction ($t_x^1=t_x^{-1}\neq 0$ and $t_z^\chi=0$) for different angles of the magnetic field. As $\gamma\rightarrow\pi/2$ (parallel $\mathbf{E}$ and $\mathbf{B}$ fields), we recover the result presented in Fig.~\ref{fig:szz_tiltx_opp_ommon} and the shape of zero LMC contour (line separating the black and white regions) is like a rotated $U$. When $\gamma$ is directed away from $\pi/2$ the region of negative LMC is seen to expand out.}
    \label{Fig+lmc_sign_t1x_ai_tiltx_same_omm_on_gm}
\end{figure*}

\begin{figure*}
    \centering
    \includegraphics[width=0.49\columnwidth]{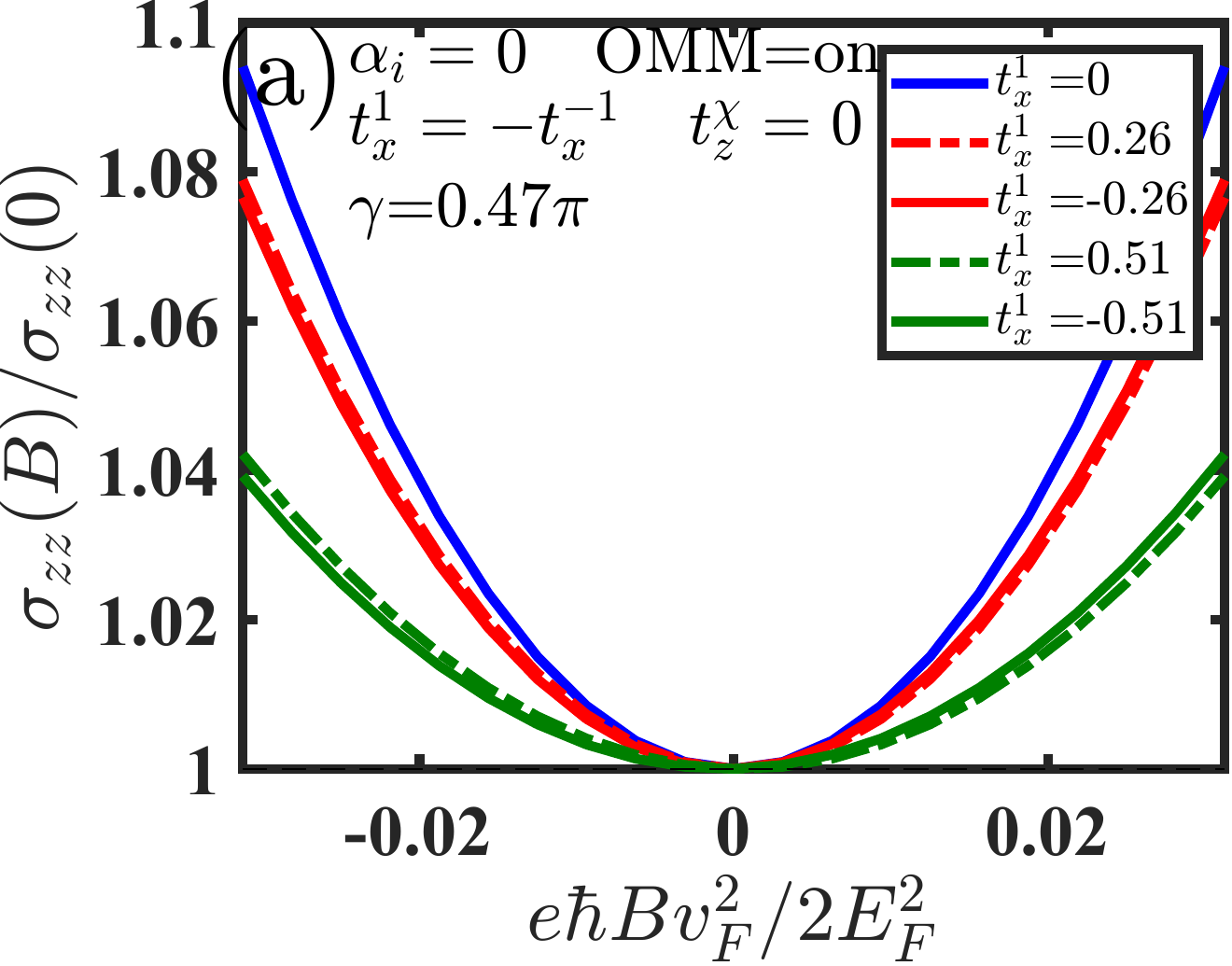}
    \includegraphics[width=0.49\columnwidth]{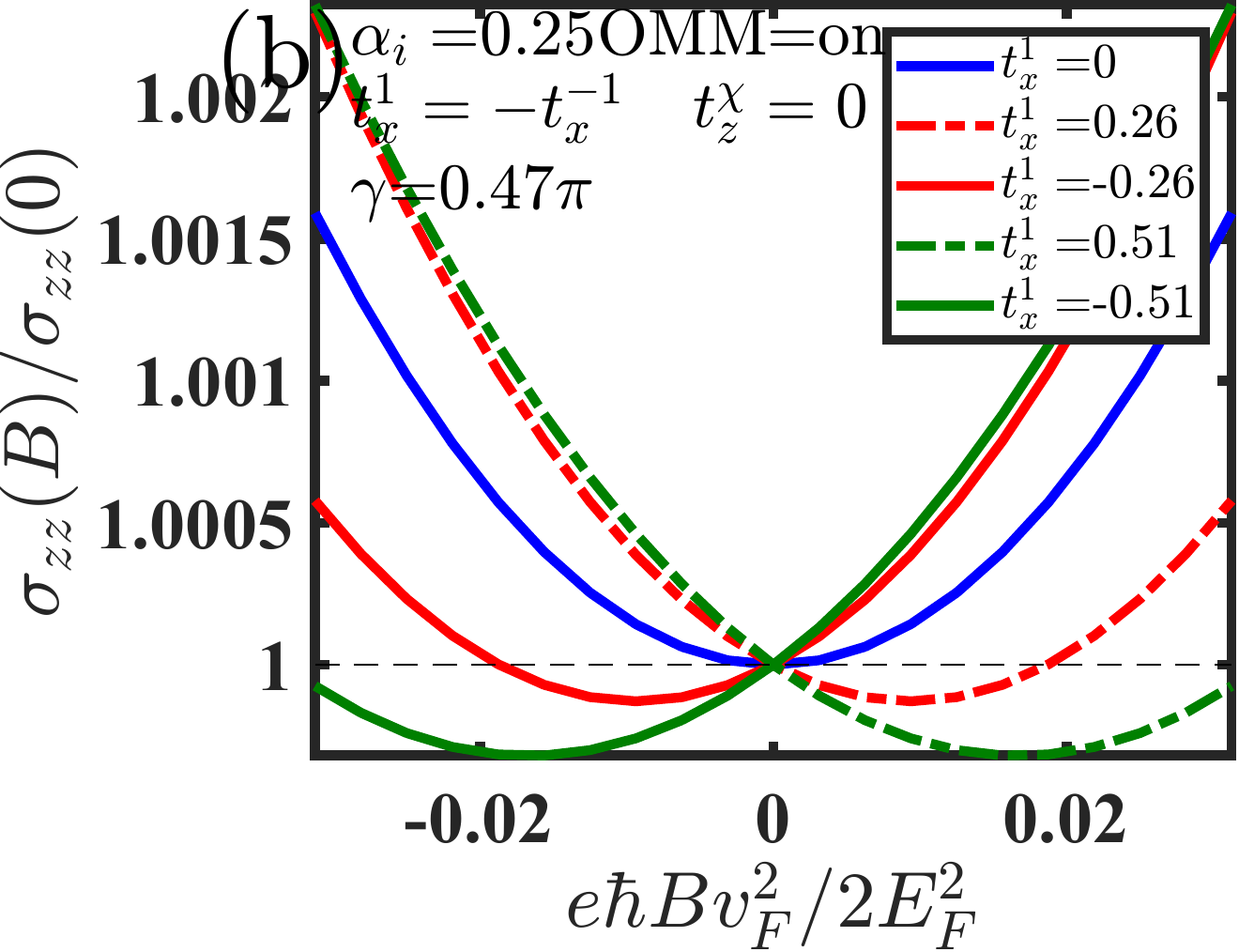}
    \includegraphics[width=0.49\columnwidth]{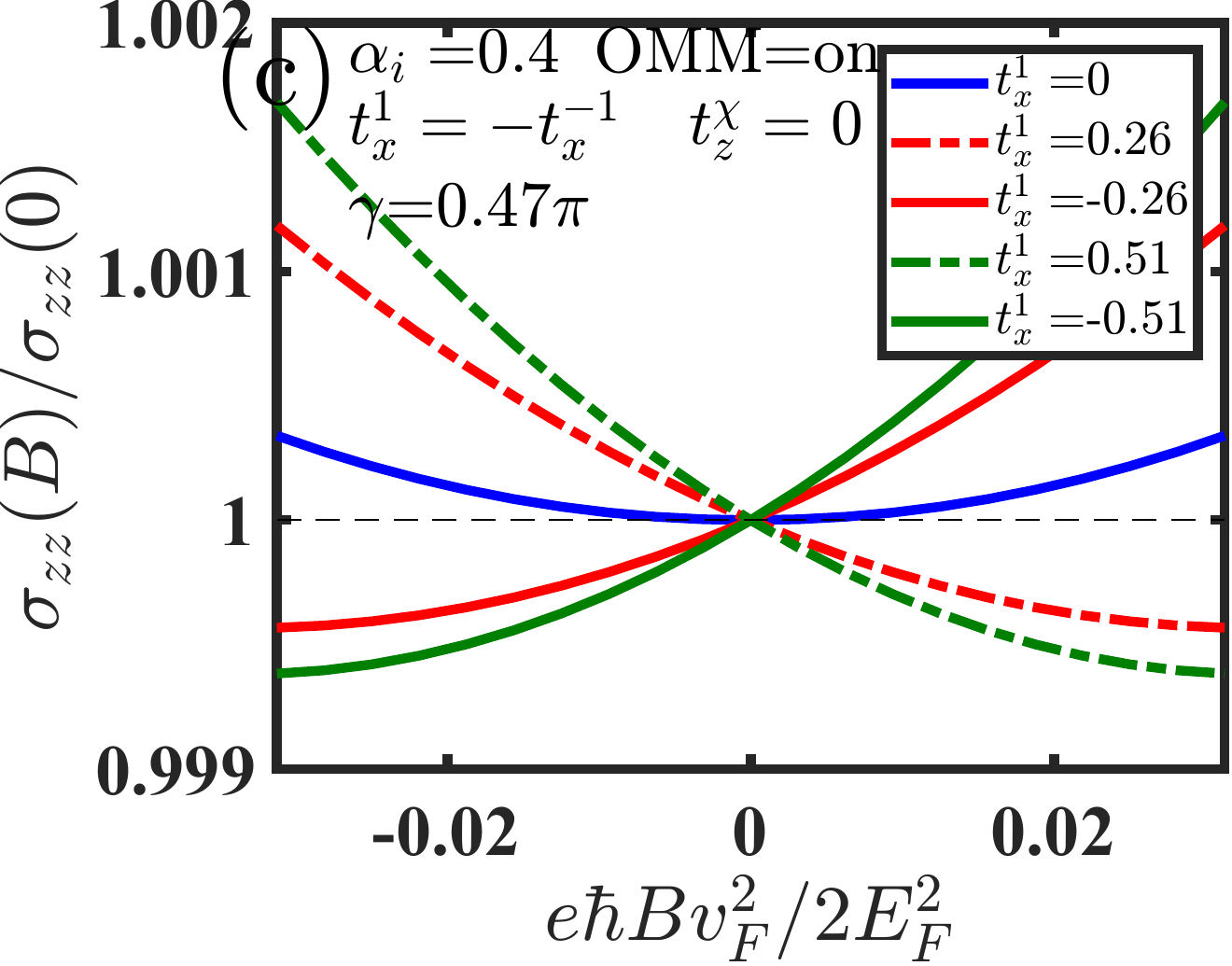}
    \includegraphics[width=0.49\columnwidth]{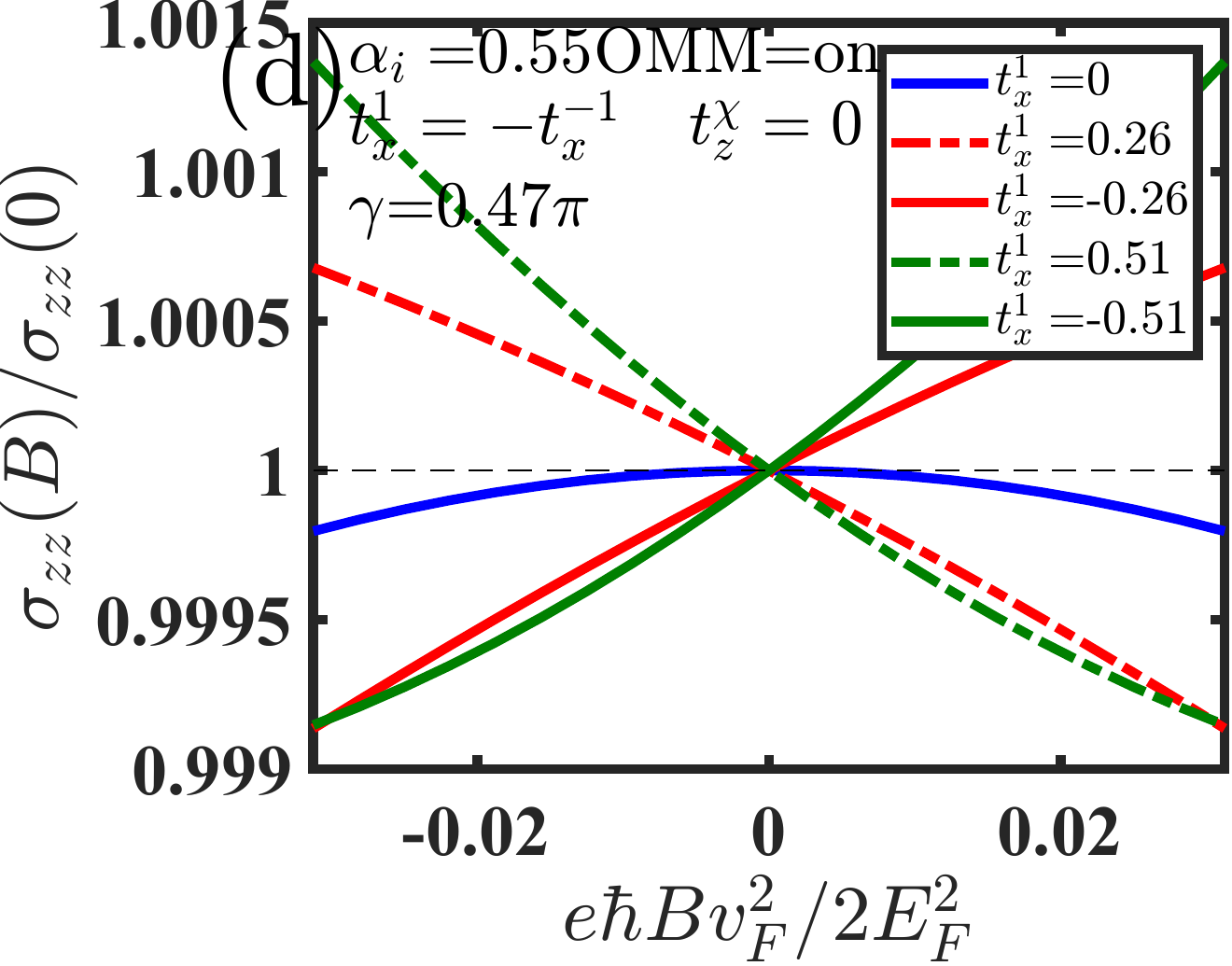}
    \includegraphics[width=0.49\columnwidth]{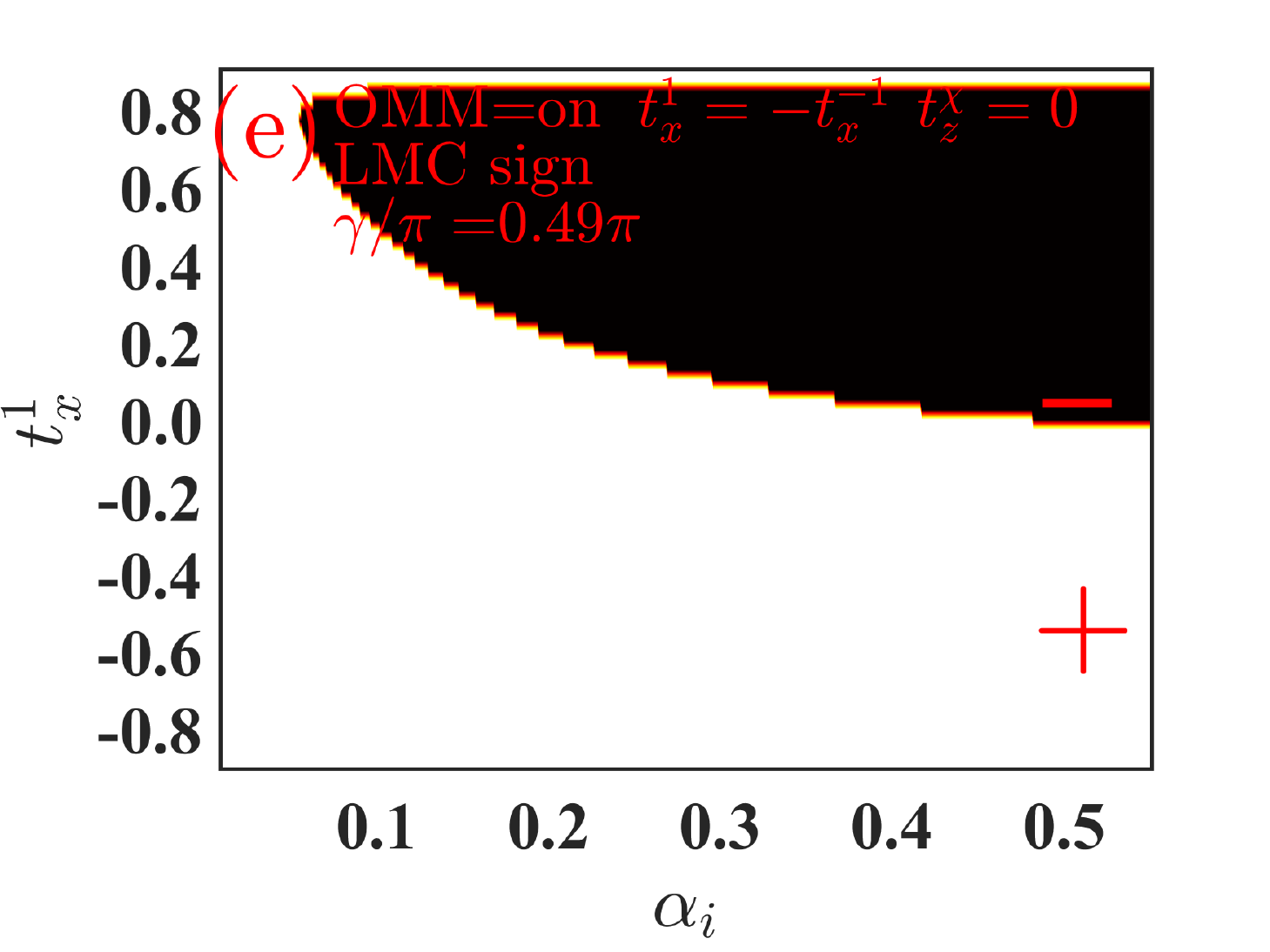}
    \includegraphics[width=0.49\columnwidth]{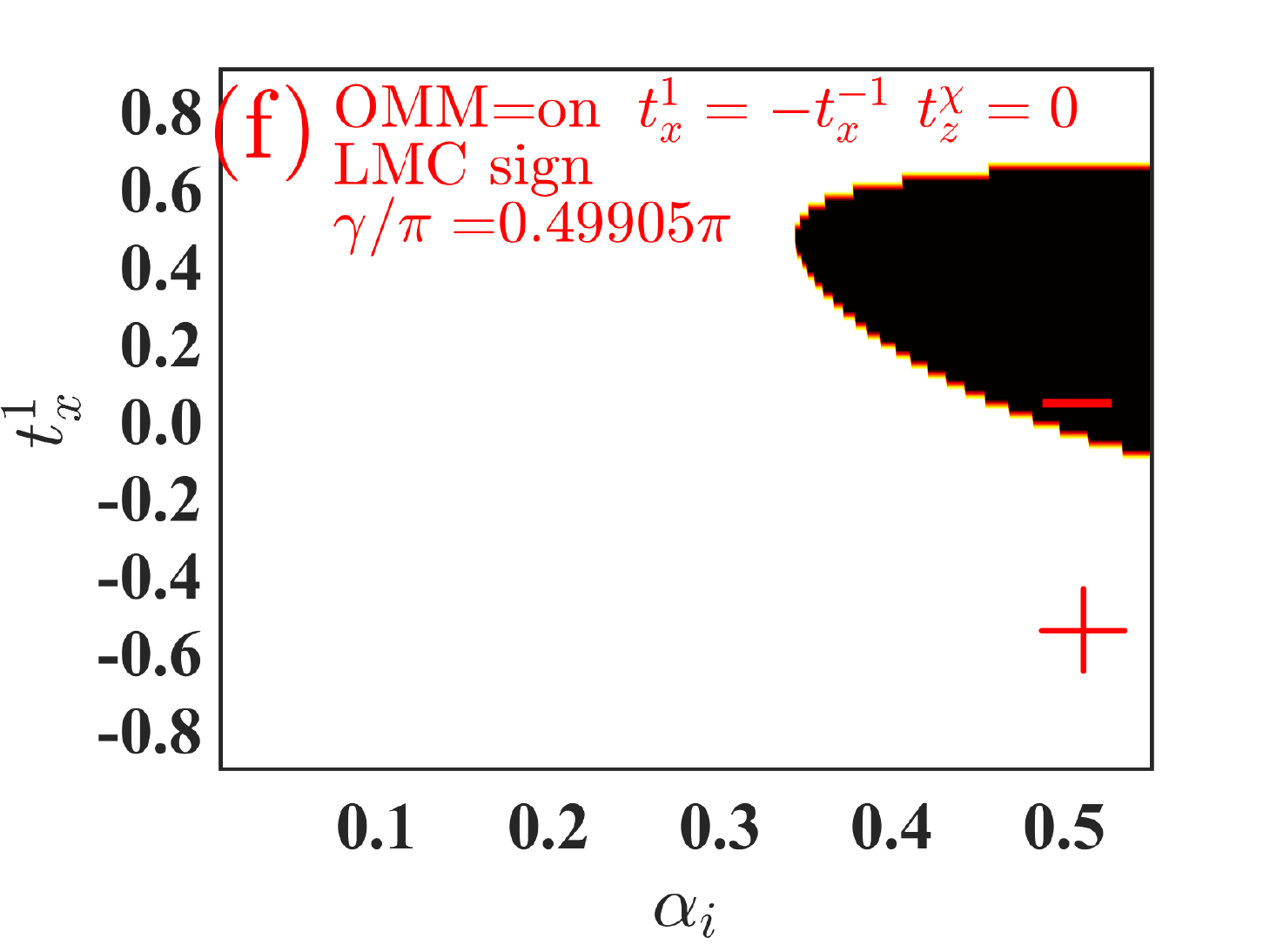}
    \includegraphics[width=0.49\columnwidth]{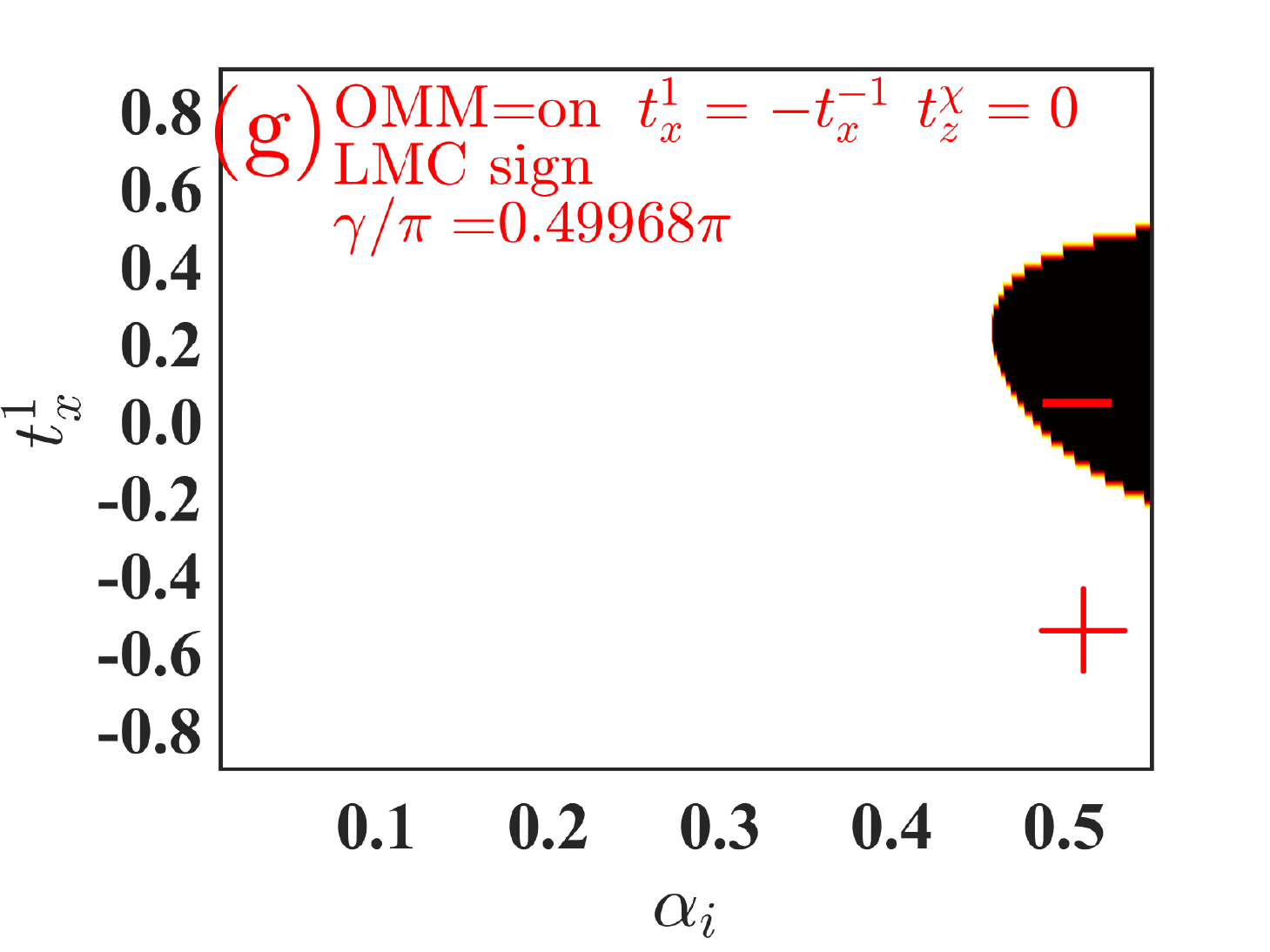}
    \includegraphics[width=0.49\columnwidth]{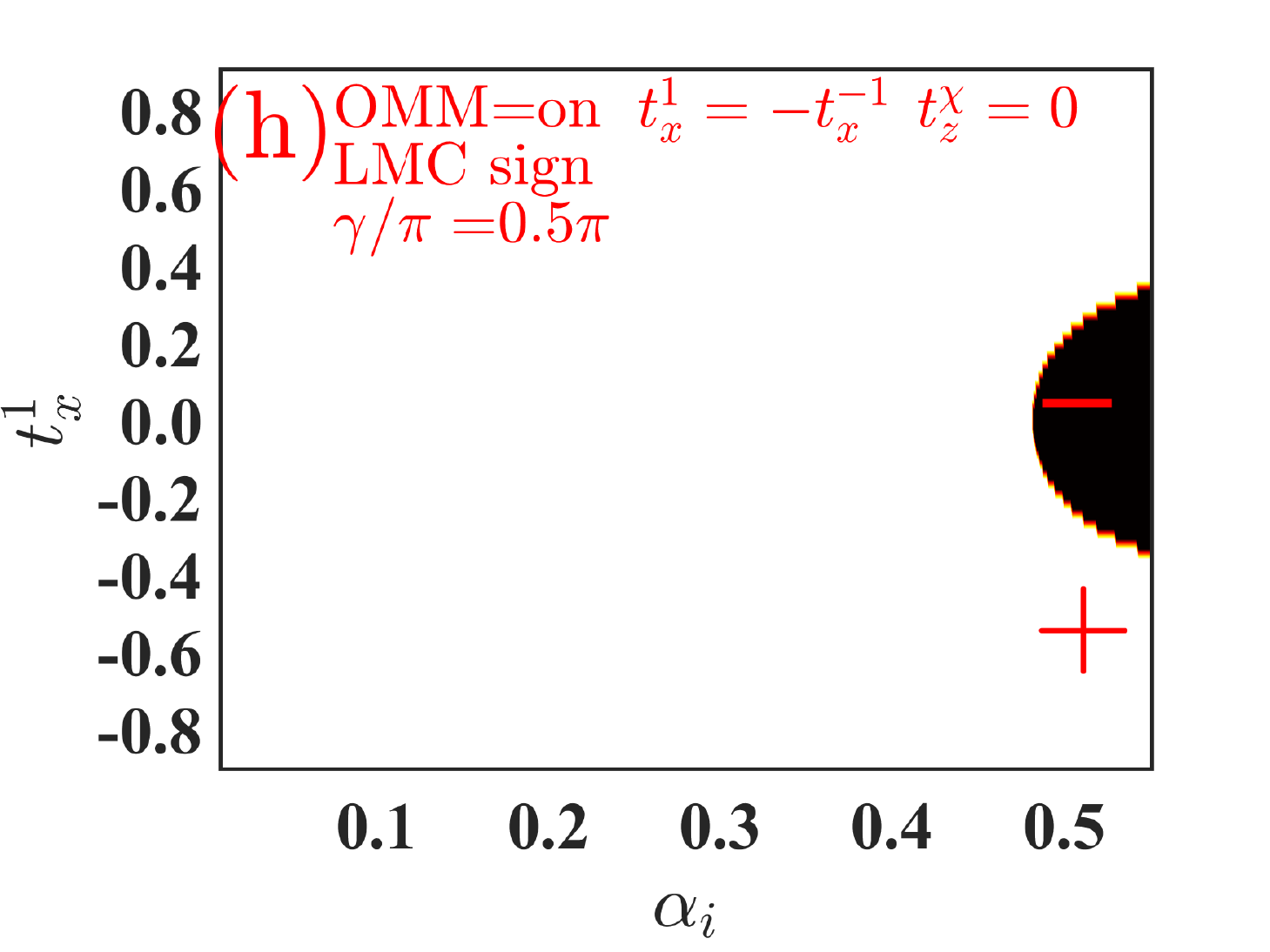}
    \caption{(a) - (d) LMC $\sigma_{zz}$ as a function of the magnetic field when $t_x^1=-t_x^{-1}$ and $t_z^\chi=0$, when the angle of the magnetic field is slightly shifted away from $\pi/2$ ($\gamma = 0.47\pi$). A finite tilt is noted to result in a small linear-in-$B$ contribution that enhances in the presence of intervalley scattering. (e)-(h) The sign of LMC $\sigma_{zz}$ in the $t^1_x - \alpha_i$ parameter space in the limit $B\rightarrow 0^+$ shows drastic variation around $\gamma=\pi/2$.}
    \label{Fig_szz_vs_B_tiltx_opp_gamma}
\end{figure*}

\begin{figure*}
    \centering
    \includegraphics[width=0.49\columnwidth]{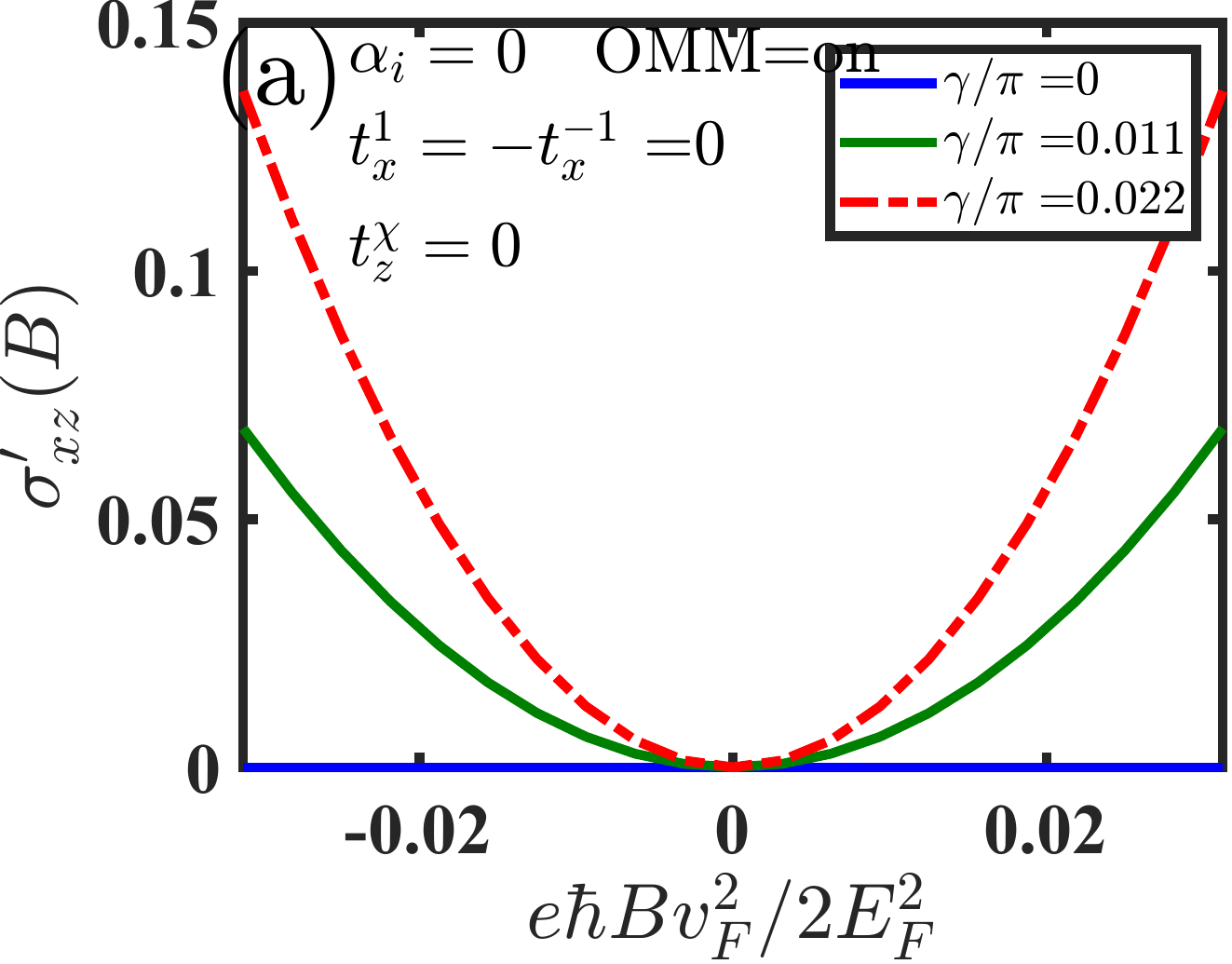}
    \includegraphics[width=0.49\columnwidth]{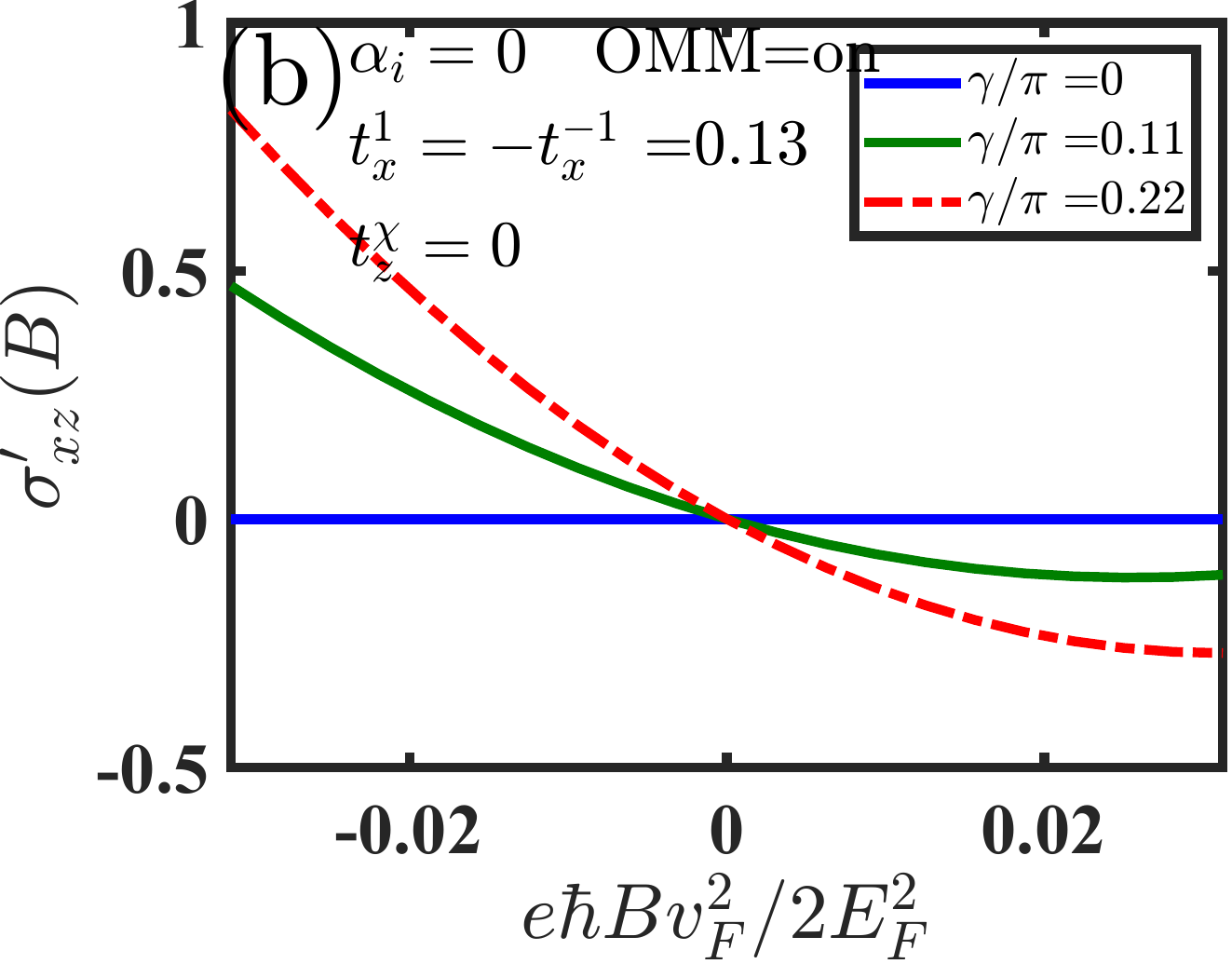}
    \includegraphics[width=0.49\columnwidth]{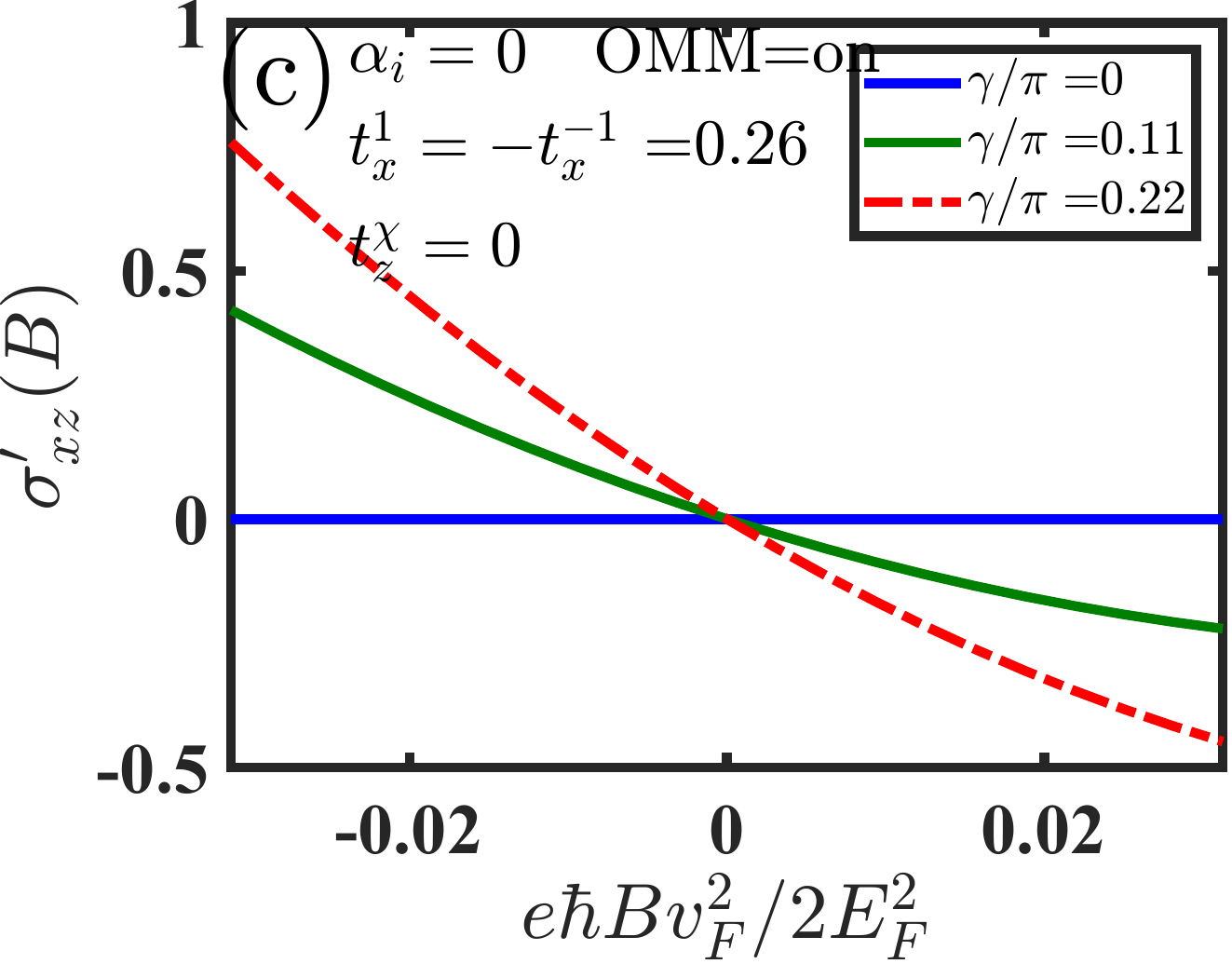}
    \includegraphics[width=0.49\columnwidth]{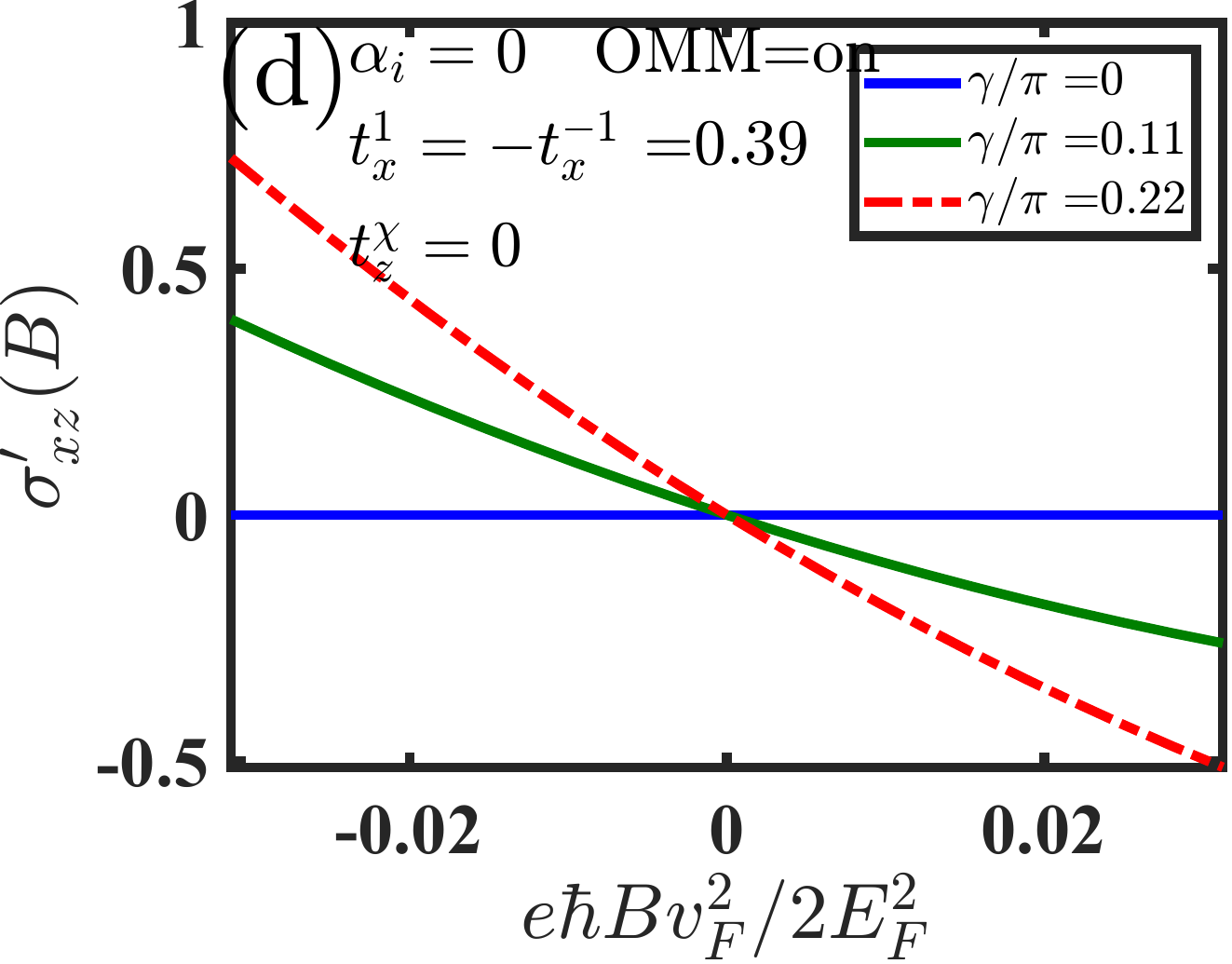}
    \includegraphics[width=0.49\columnwidth]{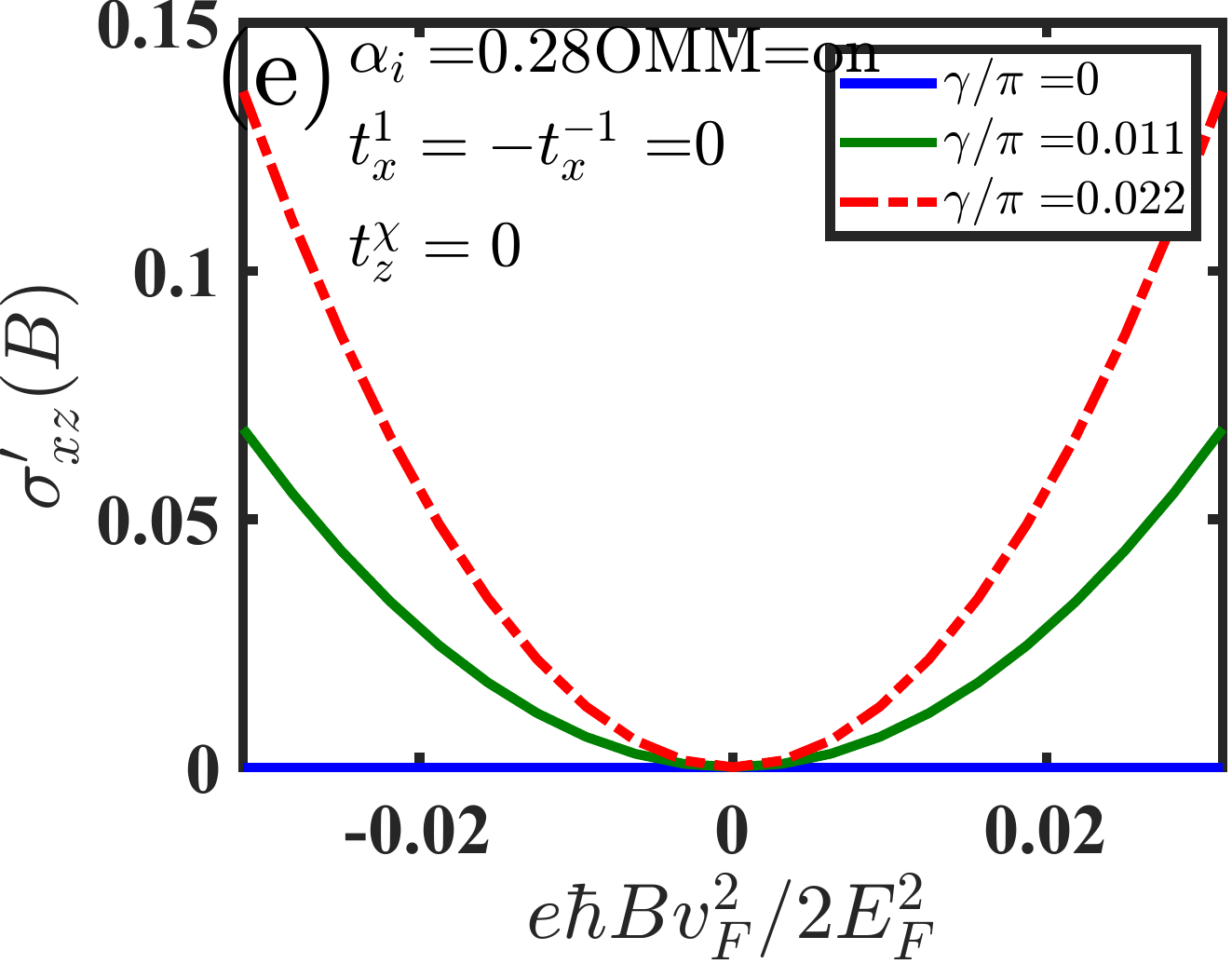}
    \includegraphics[width=0.49\columnwidth]{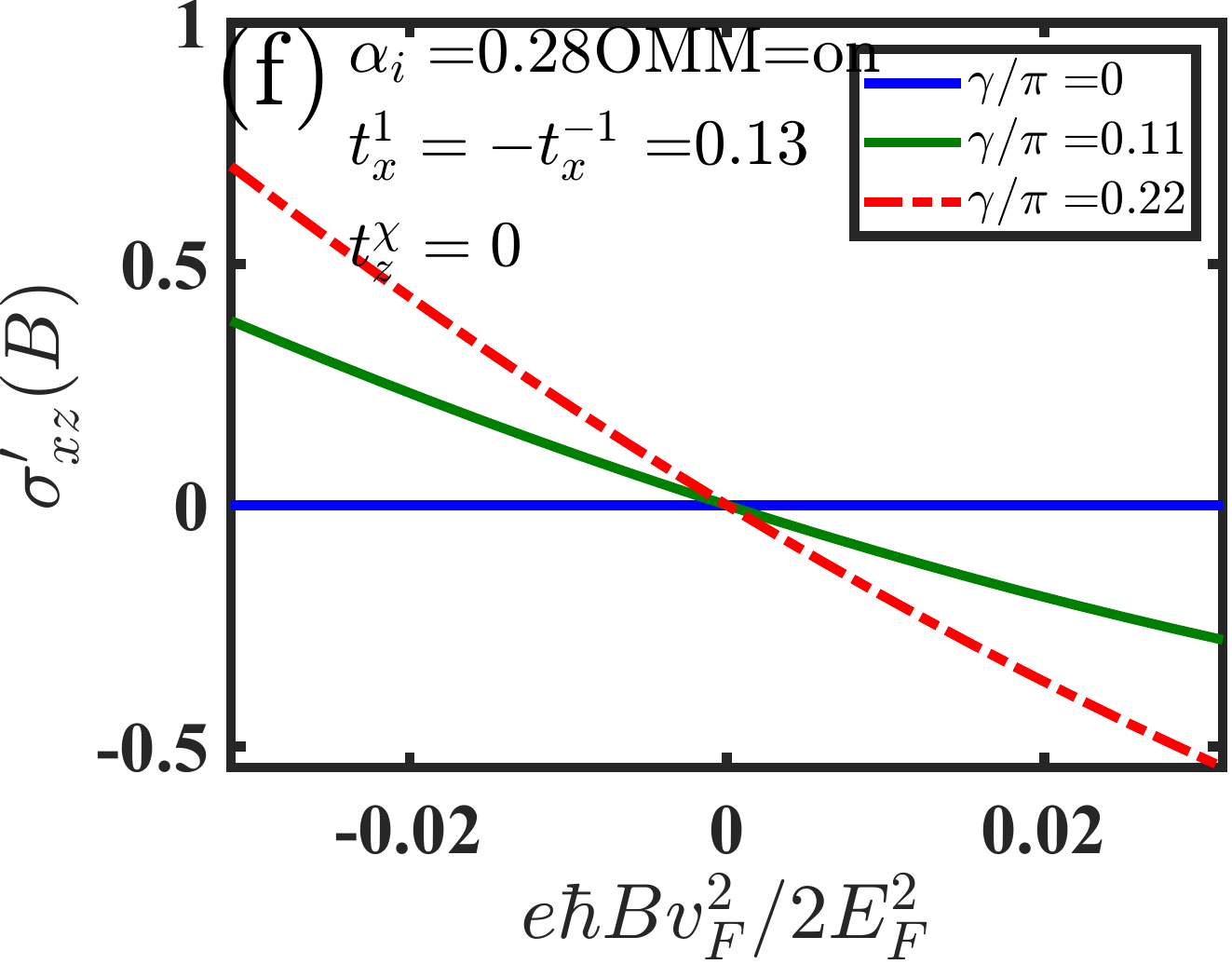}
    \includegraphics[width=0.49\columnwidth]{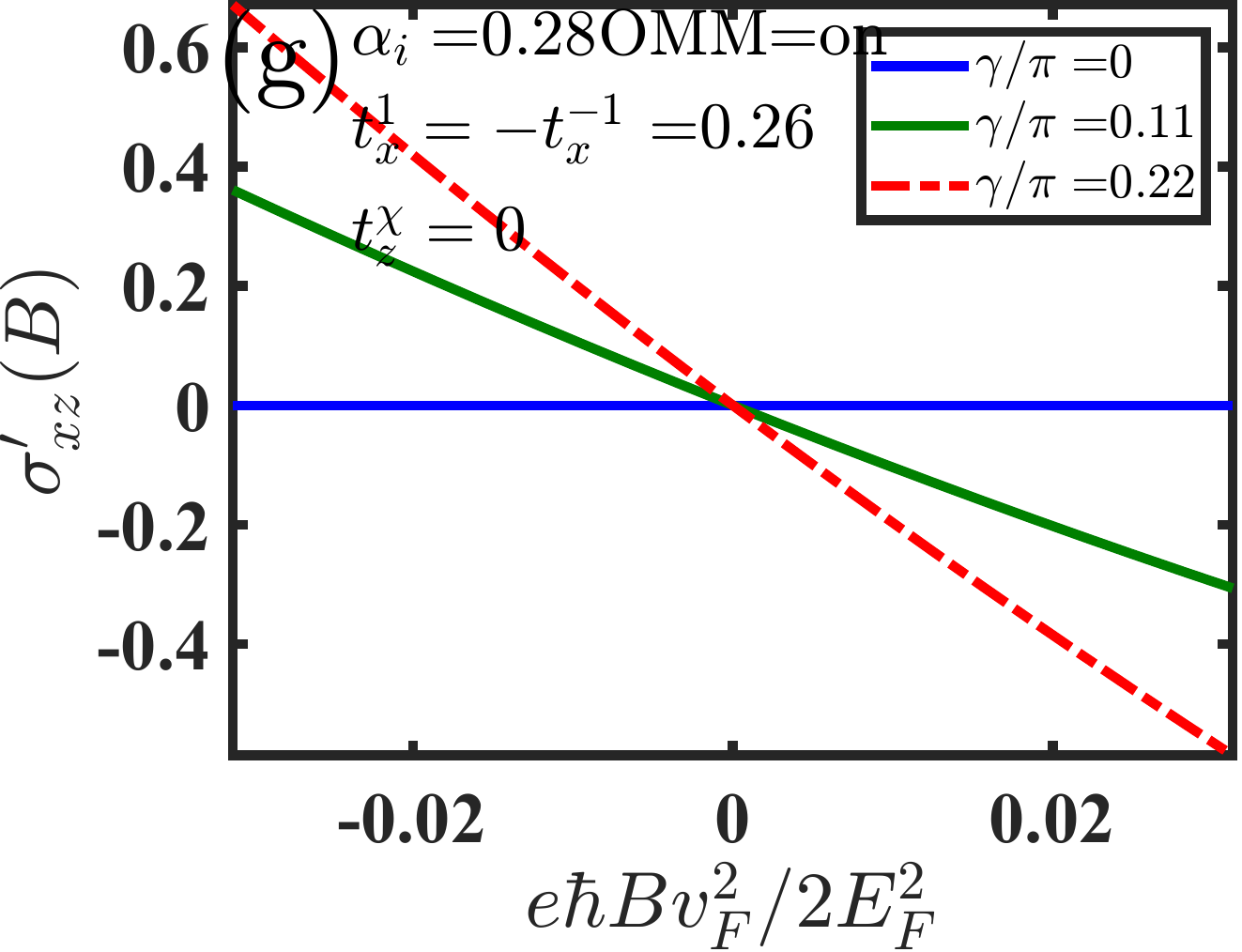}
    \includegraphics[width=0.49\columnwidth]{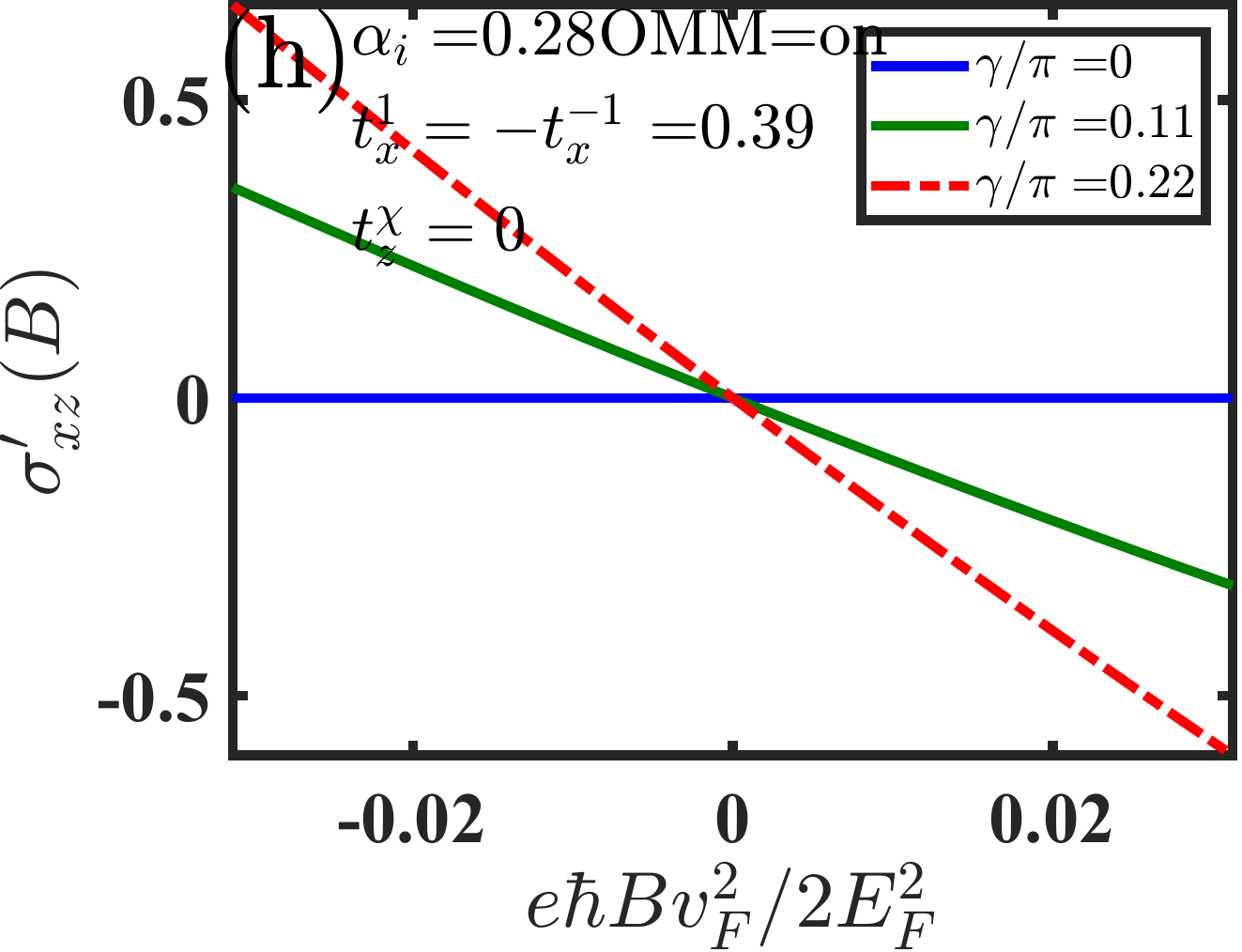}
    \caption{Normalized planar Hall conductivity $\sigma_{xz}'$ for oppositely tilted Weyl fermions along the $k_x$ direction (prime indicates normalization w.r.t. magnetic field at 0.5T). (a)-(d) In the absence of intervalley scattering, a small tilt adds a linear-in-$B$ component. (e)-(h) In the presence of intervalley scattering strength, the linear-in-$B$ component is enhanced, but only in the presence of a finite tilt.}
    \label{Fig_sxz_vs_B_tiltx_opp_gm}
\end{figure*}

\begin{figure*}
    \centering
    \includegraphics[width=0.49\columnwidth]{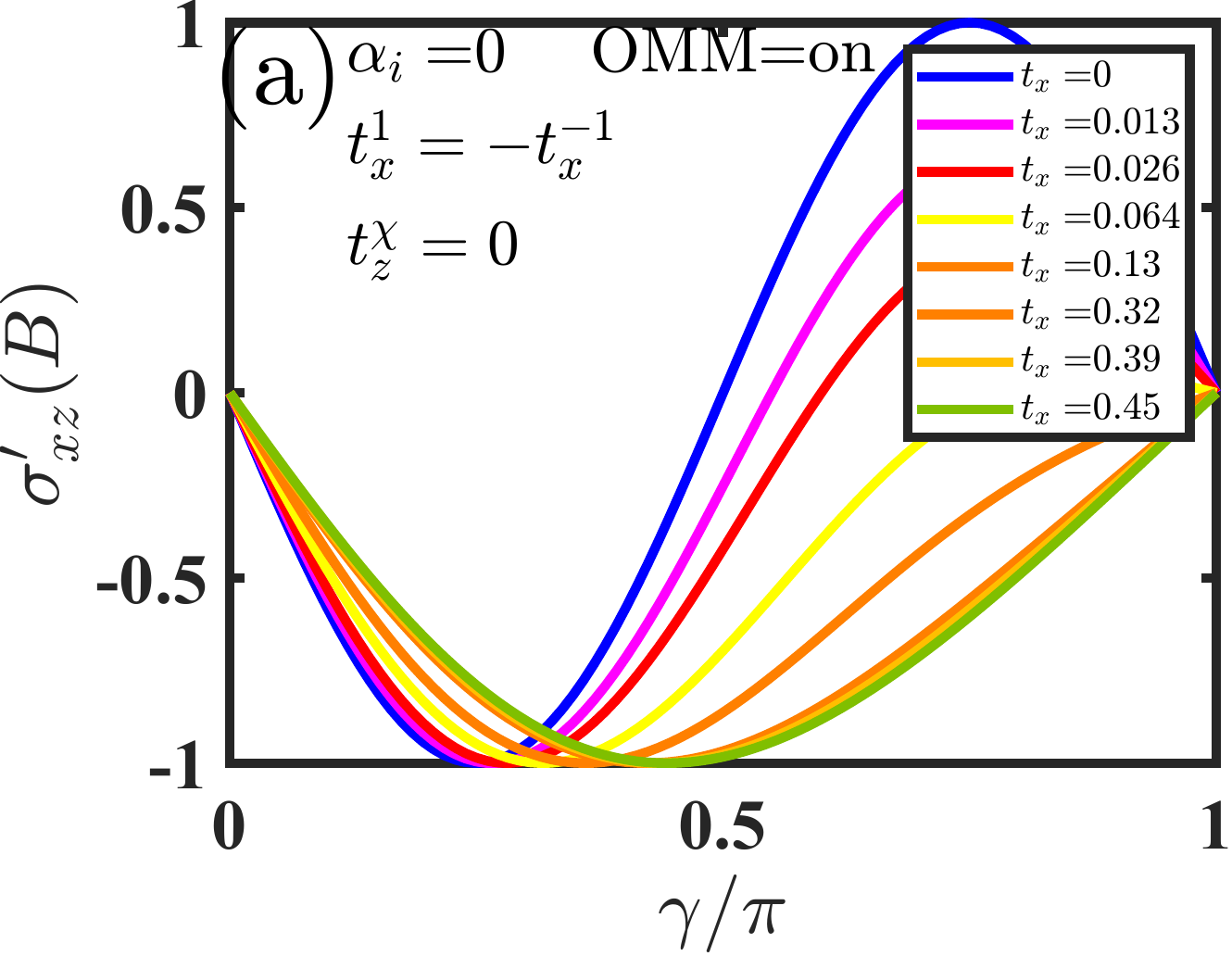}
    \includegraphics[width=0.49\columnwidth]{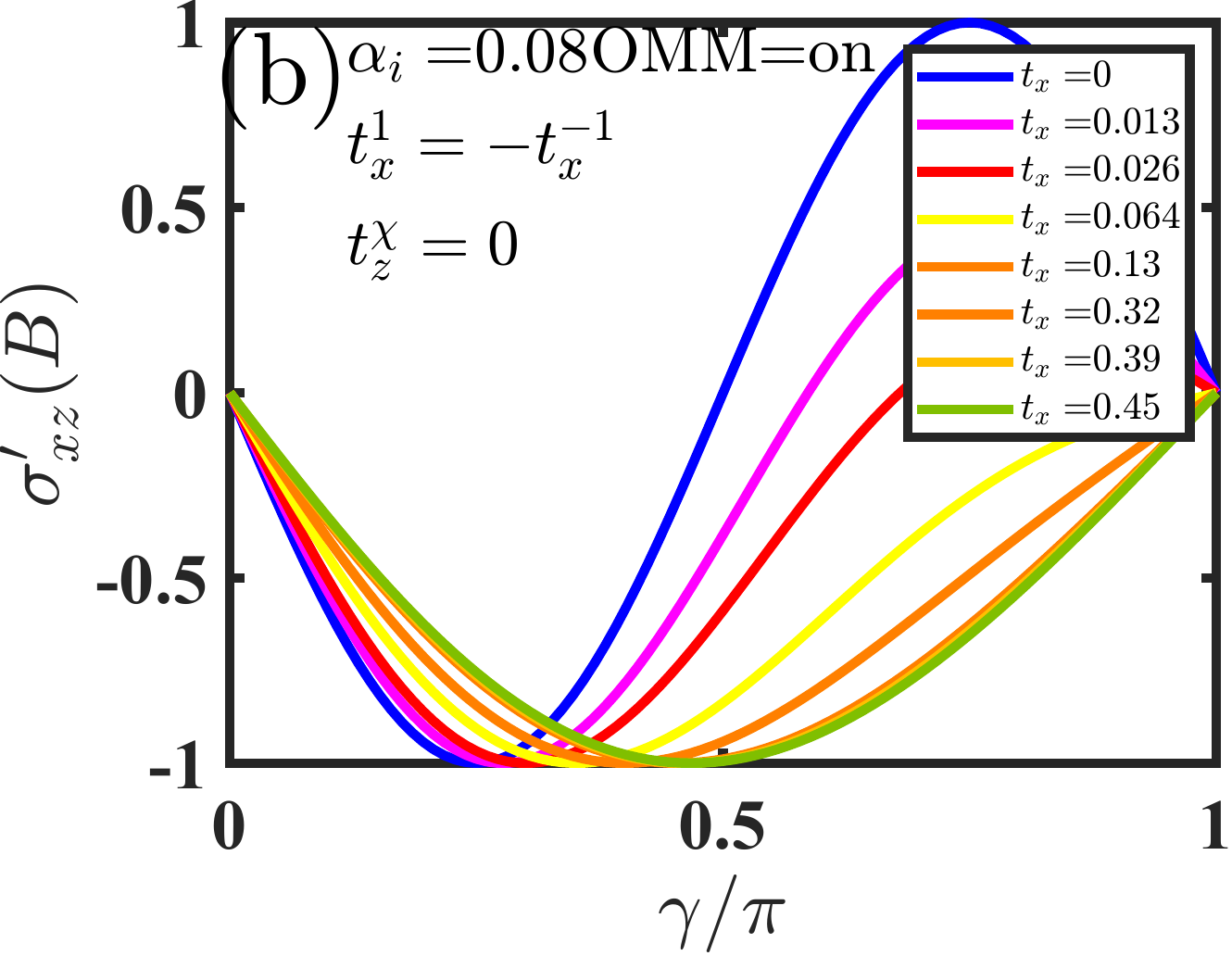}
    \includegraphics[width=0.49\columnwidth]{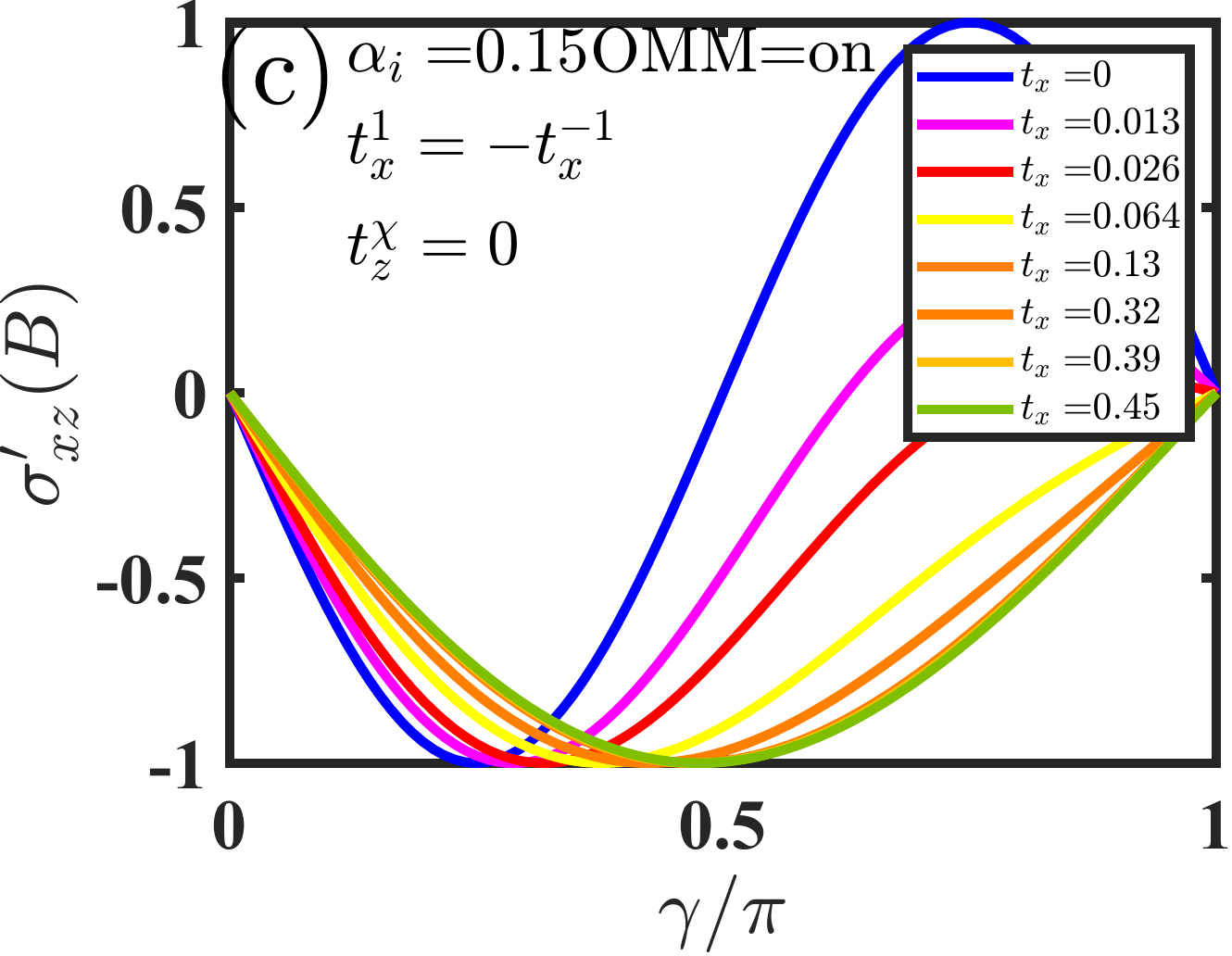}
    \includegraphics[width=0.49\columnwidth]{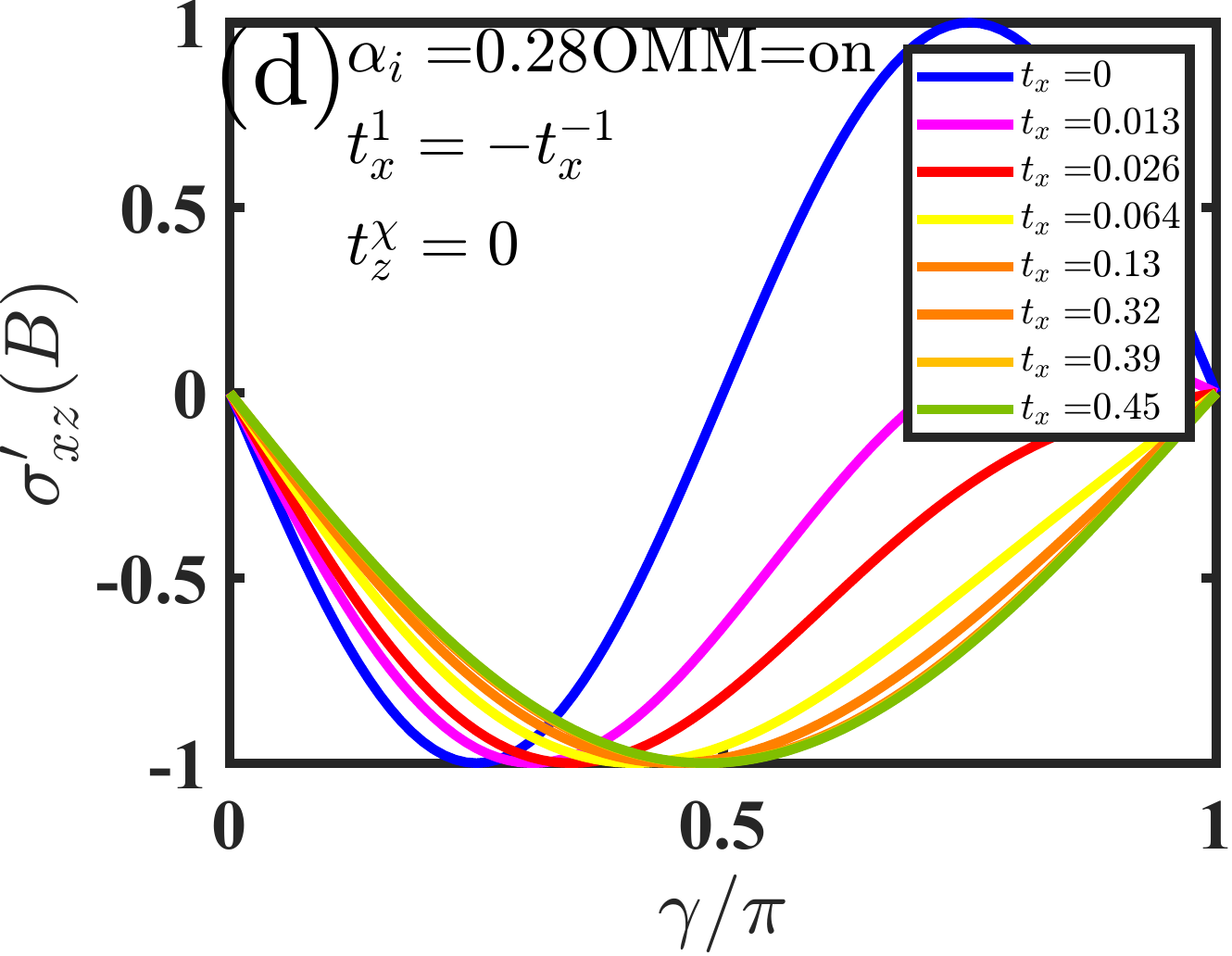}
    \caption{Normalized planar Hall conductivity ($\sigma_{xz}'$) as a function of the angle $\gamma$ for several values of tilt parameter $t_x$ for oppositely tilted Weyl cones. In the absence of tilt the behavior follows the trend $\sin(2\gamma)$, while in the presence of tilt, a $\sin\gamma$ component is added. Beyond a critical $t_x^c$, the $\sin\gamma$ term dominates and $\sigma_{xz}'(\pi/2 + \epsilon)$ changes from positive to negative, where $\epsilon$ is a small positive angle.
    A finite intervalley scattering further enhances the $\sin\gamma$ trend (however only in the presence of a finite tilt). It's effect is to lower the critical tilt $t_x^c$ where the sign change occurs.}
    \label{Fig_sxz_vs_gamma_tiltx_opp_omm_on}
\end{figure*}

\subsubsection{The case when $t_z^\chi = 0$ and $t_x^\chi\neq 0$}
In Fig.~\ref{fig:szz_tiltx_opp_ommon} we plot the results when the Weyl coned are tilted orthogonal to the direction of the magnetic field and oppositely oriented, i.e. $t_x^1 = -t_x^{-1}\neq 0$ and $t_z^\chi=0$. The qualitative trend is very similar to the previously discussed case of $t_x^\chi=0$ and $t_z^1 = t_z^{-1}\neq 0$. When the Weyl cones are tiled in the same direction w.r.t. each other, i.e., $t_x^1 = t_x^{-1}\neq 0$ and $t_z^\chi=0$, the qualitative trend is again observed to be similar and hence is not explicitly plotted. 

\subsection{LMC and PHE in tilted Weyl semimetal, when $\gamma\neq \pi/2$}
When $\gamma \neq \pi/2$, we will have a finite planar Hall contribution along with the longitudinal magnetoconductance. As we shall examine shortly, the LMC in this case can exhibit non-trivial signatures depending on the angle $\gamma$.
\subsubsection{The case when $t_z^1=-t_z^{-1}\neq 0$ and $t_x^\chi=0$}
In this case the longitudinal magnetoconductance isn't qualitatively different from the case of $\gamma=\pi/2$ as shown in Fig.~\ref{fig:szz_tiltz_opp_ommon}. The sign of the LMC is mainly governed by the linear coefficient $\sigma_{zz1}$. The values of the coefficient decreases when $\gamma\neq \pi/2$, but the qualitative trend resembles that of Fig.~\ref{fig:szz_tiltz_opp_ommon} (d). Therefore we do not explicitly plot this behavior. 

In Fig.~\ref{fig:sxz_vs_B_tiltz_opp_omm_on} we plot the normalized planar Hall conductivity $\sigma_{xz}'$ as a function of the magnetic field for different values of the tilt parameter $t_z^\chi$ (for oppositely tilted Weyl cones) and at angles $\gamma$. When the intervalley strength is zero, a finite tilt is observed to add a $B$-linear component that shifts the minima of $\sigma_{xz}'$ away from $B=0$. For a higher values of tilt, the behavior is almost linear for all relevant range of magnetic field. A finite intervalley strength $\alpha_i$ enhances the $B$-linear contribution, however, only in the presence of a finite tilt. In Fig.~\ref{Fig_sxz_vs_gamma_tiltz_opp_omm_on} we plot the normalized planar Hall conductivity ($\sigma_{xz}'$) as a function of the angle $\gamma$ for several values of tilt parameter $t_z$ for oppositely tilted Weyl cones. In the absence of tilt the behavior follows the trend $\sin(2\gamma)$, while in the presence of tilt, a $\cos\gamma$ component is added. Beyond a critical $t_z^c$, the $\cos\gamma$ term dominates and $\sigma_{xz}'(\pi/2 + \epsilon)$ changes from positive to negative, where $\epsilon$ is a small positive angle.
A finite intervalley scattering further enhances the $\cos\gamma$ trend (however only in the presence of a finite tilt). Its effect is to lower the critical tilt $t_z^c$ where the sign change occurs.

\subsubsection{The case when $t_z^1=t_z^{-1}\neq 0$ and $t_x^\chi=0$}
In this case the linear-in-$B$ behavior of the LMC vanishes since the tilts are oriented along the same direction. Thus the sign of the quadratic coefficient corresponds to the sign of LMC. One would therefore expect that the qualitative behavior in this case would again be similar to that observed in Fig.~\ref{fig:szz_tiltz_same_ommon}, however, we find that this is not the case. 
In Fig.~\ref{Fig_lmc_sign_t1z_ai_tiltz_same_omm_on_gm} we plot the sign of LMC as a function of the tilt $t_z^1$ and intervalley scattering $\alpha_i$ for various values of the magnetic field angle $\gamma$. As $\gamma\rightarrow\pi/2$ (parallel $\mathbf{E}$ and $\mathbf{B}$ fields), we recover the result presented in Fig.~\ref{fig:szz_tiltz_same_ommon}, i.e. the shape of contour where LMC is zero is like a ${U}$. Specifically, when $|t_z^1|\lesssim 0.6$ critical value of $\alpha_i$ where the sign change occurs is around 0.5. When $|t_z^1|\gtrsim 0.6$, the sign change does not occur.  
When $\gamma$ is directed away from $\pi/2$ the shape of the zero LMC contour looks like a curved trapezoid instead of $U$. The critical value $\alpha_i^c$ where the sign change first occurs is seen to reduce and elongate its region from $|t_z|\approx 0.5$ when $\gamma=\pi/2$ to $|t_z|\approx 1$ as $\gamma\rightarrow 0$. This feature can be understood as a combination of two factors: when $\gamma=\pi/2$, a finite tilt $t_z^1$ and $\alpha_i$ drives the system to change the sign of LMC from positive to negative (as seen in Fig.~\ref{fig:szz_tiltz_same_ommon}), and secondly when $\gamma \neq \pi/2$ along with a finite $\alpha_i$ (even when $t_z^1=0$) drives the system to change LMC sign from positive to negative much below $\alpha_i=0.5$~\cite{sharma2020sign}. The combination of these two assisting factors shapes the zero LMC contour in the current scenario.

The planar Hall conductance, on the other hand shows expected behavior, i.e., quadratic in the magnetic field and $\sin(2\gamma)$ angular dependence. Therefore we do not explicitly plot this behavior. 

\subsubsection{The case when $t_x^1= t_x^{-1}\neq 0$ and $t_z^\chi= 0$}
If we compare Fig.~\ref{fig:szz_tiltz_same_ommon} and Fig.~\ref{fig:szz_tiltx_opp_ommon}, it is suggested that the qualitative behavior of the three scenarios (a) \{$t_z^1=t_z^{-1}\neq 0, t_x^\chi=0$\}, (b) \{$t_x^1=t_x^{-1}\neq 0, t_z^\chi=0$\}, and (c) \{$t_x^1=-t_x^{-1}\neq 0, t_z^\chi=0$\} is similar to each other when $\gamma=\pi/2$. Therefore, rotating the magnetic field along the $xy$-plane (shifting $\gamma$ away from $\pi/2$) is naively not expected to change any qualitative behavior. However, we find that this is not the case. Consider the two scenarios (a) and (b), which display similar behavior when $\gamma=\pi/2$, i.e. (i) LMC is quadratic in $B$, (ii) LMC switches sign when $\alpha_i>\alpha_i^c(t_{x/z}^1)$, (iii) LMC always remains positive if $t_{x/z}^1$ is too large ($\gtrsim 0.6$) also suggested by the shape of zero-LMC contour ($U$-shaped). If $\gamma\neq \pi/2$, then in scenario (a), the zero-LMC contour assumes the form of an curved trapezoid (Fig.~\ref{Fig_lmc_sign_t1z_ai_tiltz_same_omm_on_gm}, while the zero-LMC contour is much different in scenario (b), as seen in Fig.~\ref{Fig+lmc_sign_t1x_ai_tiltx_same_omm_on_gm}, where the region of negative LMC expands in the parameter space along with the reduction of the critical intervalley strength $\alpha_i^c$ where the sign change first occurs. The reduction of the critical intervalley strength can again be understood as a combination of the two factors like the previous case (i) a finite tilt $t_z^1$ and $\alpha_i$ (when $\gamma=\pi/2$) drives the system to change the LMC sign from positive to negative (as seen in Fig.~\ref{fig:szz_tiltx_opp_ommon}), and secondly $\gamma \neq \pi/2$ along with a finite $\alpha_i$ (when $t_z^1=0$) drives the system to change LMC sign from positive to negative much below $\alpha_i=0.5$~\cite{sharma2020sign}. The different shape of the contour (negative LMC filling out the parameter space instead of a curved trapezoid) is essentially because the cones are now tilted along the $x$-direction and the magnetic field has an $x$-component to it, which is qualitatively different from the tilt occurring in the $z$-direction.

\subsubsection{The case when $t_x^1=- t_x^{-1}\neq 0$ and $t_z^\chi= 0$}
From Fig.~\ref{fig:szz_tiltx_opp_ommon} and our earlier discussion we noted that scenarios (b) and (c) are qualitatively similar, at least when $\gamma=\pi/2$, i.e., the field is directed along the $z$-direction. Directing the magnetic field even slightly away from the $z$-axis changes the qualitative behavior when $t_x^1=-t_x^{-1}$, as a $B$-linear component is added in the LMC response. This is because the magnetic field now has a $x$-component and the tilts are oppositely oriented to each other (though tilted along the $x$-axis). Fig.~\ref{Fig_szz_vs_B_tiltx_opp_gamma} presents the plot of LMC $\sigma_{zz}$ as a function of the magnetic field when the angle of the magnetic field is slightly shifted away from $\pi/2$ ($\gamma = 0.47\pi$). A finite tilt results in a small linear-in-$B$ contribution that is enhanced in the presence of intervalley scattering. 

The presence of a finite $B-$linear component also modifies the planar Hall conductivity $\sigma_{xz}$ in the current case. Fig.~\ref{Fig_sxz_vs_B_tiltx_opp_gm} plots the normalized planar Hall conductance $\sigma_{xz}'$ as a function of the magnetic field. Even in the absence of intervalley scattering, a finite tilt of the Weyl cones along the $x$-direction causes the planar Hall conductivity to be $B$-linear showing asymmetry around $B=0$. The presence of intervalley scattering further enhances the $B$-linear contribution. A difference between this and Fig.~\ref{fig:sxz_vs_B_tiltz_opp_omm_on} (tilt along the $z$-direction) is that the planar Hall conductivity remains zero when $\gamma=0$, i.e., when the magnetic field points along the $x$-direction because the direction is parallel with the direction of tilts in the Weyl cone. On the other hand, for $\gamma=0$, the planar Hall conductivity becomes finite and linear when the Weyl cones are tilted along the $z$-direction. We also plot $\sigma_{xz}'$ as a function of $\gamma$ in Fig.~\ref{Fig_sxz_vs_gamma_tiltx_opp_omm_on}.  In the absence of tilt the behavior follows the expected trend of $\sin(2\gamma)$, while in the presence of tilt, a $\sin\gamma$ component is added.  Beyond a critical value of the tilt ($t_x^c$), the $\sin\gamma$ term dominates the behavior $\sigma_{xz}'$ never changes sign as a function of the parameter $\gamma$. A finite intervalley scattering further enhances the $\sin\gamma$ trend (however only in the presence of a finite tilt).  Its effect is to lower the critical value of the tilt $t_x^c$. 

\section{Discussions and Conclusions}
The linearity or nonlinearity of the bands is alone not sufficient to produce a finite longitudinal magnetoconductance (positive or negative) or a planar Hall effect in materials. It is in fact the topological nature of the bands that gives rise to finite LMC or PHE in Weyl semimetals. The topological nature of the bands is manifest in the Berry curvature and the orbital magnetic moment of the Bloch electrons. Even though the bands no longer disperse linearly away from the Weyl node, their topology is nevertheless preserved, as also demonstrated by exact expressions for Berry curvature and OMM in our prototype lattice model. 
We solved the Boltzmann equation semi-analytically for a lattice model of Weyl fermions and noted that the inclusion of orbital magnetic moment is crucial in obtaining negative LMC in the limit of vanishing intervalley scattering, just like it is crucial in obtaining negative LMC for strictly linearly dispersing Weyl fermions in the presence of intervalley scattering~\cite{knoll2020negative}. This points out to an important fact that nonlinear lattice effects can produce negative LMC for weak magnetic fields irrespective of the presence or absence of intervalley scattering. Therefore it is inconclusive to state that negative LMC for weak magnetic fields in a Weyl semimetals necessarily points out to the presence of intervalley scattering. 

Since nonlinear lattice effects are intrinsically present in real Weyl materials, likewise, the presence of a finite tilt is also inevitable. Finite lattice effects and effects due to tilting of the cones are largely independent of one another, and thus one can solve the Boltzmann equation for tilted Weyl fermions in the linearized approximation. The overall behaviour is a given by a combination of both factors. 
We constructed several phase diagrams in relevant parameter space that are important for diagnosing chiral anomaly in Weyl materials. Specifically, we examine the longitudinal magnetoconductivity $\sigma_{zz}$ as well as the planar Hall conductivity $\sigma_{xz}$ for tilted Weyl fermions for the four relevant cases when the cones are tilted in the same or opposite direction along or perpendicular to the $z-$direction, i.e., (i) $t_x^1=t_x^{-1}$, and $t_z^\chi=0$, (ii) $t_x^1=-t_x^{-1}$, and $t_z^\chi=0$, (iii) $t_z^1=t_z^{-1}$, and $t_x^\chi=0$, (iv) $t_z^1=-t_z^{-1}$, and $t_x^\chi=0$. Crucially, the LMC is found to depend on the angle $\gamma$ that determines the orientation of the magnetic field w.r.t the electric field. When $\gamma=\pi/2$, the electric and magnetic fields are parallel, and the LMC has a linear-in-$B$ component only for case (iv) that results in its asymmetry around $B=0$. We found that LMC when evaluated in the limit $B\rightarrow 0^+$ switches sign as a function of intervalley scattering $\alpha_i$ and the tilt parameter. For cases (i), (ii), and (iii), LMC is symmetric around $B=0$ and quadratic in magnetic field, however, it changes sign from positive to negative depending on the magnitude of $\alpha_i$ and the tilt parameter. When $\gamma\neq \pi/2$, the phase plots for cases (i), (ii), and (iii) shows non-trivial behavior. In particular, the distinction between cases (i) and (iii) becomes evident due to qualitatively different phase plots in the $\alpha_i-t_x$ space separating negative and positive LMC regions, which however is quadratic in magnetic field. Specifically, the shape of the zero-LMC contour is distinct in the two cases.
Interestingly, for case (ii), a linear-in-$B$ component in LMC is added that vanishes in the limit of parallel electric and magnetic fields. This again results in qualitative different phase plots in the $\alpha_i-t_z$ space as a function of $\gamma$. To summarize, the shape of the zero-LMC contour in $\alpha_i-t_k$ space as a function of the angle $\gamma$ is qualitatively distinct in each of the four cases.

Lastly, we also discuss the planar Hall conductivity $\sigma_{xz}$ for each of the above cases. A linear-in-$B$ component to $\sigma_{xz}$ is added in case (ii) and (iv), which is further enhanced by a finite $\alpha_i$. The distinction between cases (ii) and (iv) comes from the fact that in addition to $\sin (2\gamma)$, a $\cos \gamma$, and a $\sin\gamma$ trend to the planar Hall conductivity is as a function of the angle $\gamma$ for cases (iv) and (ii) respectively. The $\cos\gamma$ and $\sin\gamma$ trends are enhanced due to intervalley scattering. 

\appendix
\section{Lattice Weyl fermion}
The Hamiltonian of a Weyl node with smooth lattice cutoff can be expressed as 
\begin{align}
 H^{\chi}=\chi E_{0}\sin({a\mathbf{k}\cdot\boldsymbol{\sigma}}),
\end{align}
where $k$ is measured from the nodal point, $\chi$ is the chirality index, $E_0$ is an energy parameter, and $a$ is constant with dimensions of length. Using the relations, $\sin{\theta}=({\exp^{i\theta}-\exp^{-i \theta}})/{2i}$, and $    \exp{\{i a (\mathbf{\sigma \cdot k})\}}={{I}} \cos{\theta}+i(\mathbf{\sigma \cdot k})\sin{\theta}$, one can  rewrite down the Hamiltonian in the following form, (with $\theta$ and $\phi$ as polar and azimuthal angles respectively )  
\begin{align}
  H^{\chi}=\chi E_{0}\sin(ak)\begin{pmatrix}
  \cos{\theta} & \sin{\theta} e^{-i\phi} \\
  \sin{\theta} e^{i\phi} & -\cos{\theta} \\
  \end{pmatrix}   
\end{align}
Here we are going to use the property of matrices that for a matrix $M=[...]_{N \times N}$ having eigenvalues,$\lambda_1,\lambda_2,\lambda_3,......,\lambda_N,$ and eigenfunctions $\mathbf{\alpha_1,\alpha_2,\alpha_3,.......,\alpha_N,}$ respectively, then for matrix $C M$, the same will be $C \lambda_1,C \lambda_2,C.\lambda_3,......,C \lambda_N,$ \&
 $\alpha_1,\alpha_2,\alpha_3,.......,\alpha_N$, (where $C$ is constant ). Thus the eigenvalues of the Hamiltonian are 
\begin{align}
     \epsilon(\mathbf{k})=\pm{E_{0}\sin{(ak)}},
\end{align}
and eigenfunctions for positive band with different chirality are
\begin{align}
    \ket{u^+{(k)}}=\begin{pmatrix}
     e^{-i \phi} \cos{\frac{\theta}{2}}\\
      \sin{\frac{\theta}{2}}
    \end{pmatrix}
    \end{align}
\begin{align}
        \ket{u^-{(k)}}=\begin{pmatrix} -e^{-i \phi}
        \sin{\frac{\theta}{2}}\\
        \cos{\frac{\theta}{2}}
  \end{pmatrix}
\end{align}
The expressions for Berry curvature and orbital magnetic moments(OMM) are given by
\begin{align}
    \Omega^{\chi}_{\mathbf{k}}&=i\mathbf{\nabla}_{\mathbf{k}}\times(\bra{u^{\chi}({\mathbf{k}})}\mathbf{\nabla}_{\mathbf{k}}\ket{u^{\chi}{\mathbf{(k)}}})\nonumber\\
    m^{\chi}_{\mathbf{k}}&=\frac{-i e}{2\hbar}\bra{u^{\chi}({\mathbf{k}})}\mathbf{\times}(H^{\chi}(\mathbf{(k)}-\epsilon(\mathbf{k}))\ket{u^{\chi}{\mathbf{(k)}}},
\end{align}
from which one can easily find the expressions for Berry curvature and OMM
\begin{align}
   \mathbf{\Omega}^{\chi}_k&=\frac{-\chi\mathbf{k}}{2 k^3}\nonumber\\
   \mathbf{m}^{\chi}_\mathbf{k}&=\frac{-e \chi E_{0}\sin{(ka)} \mathbf{k}}{2\hbar k^3}
\end{align}
\section{Boltzmann transport equation}
The Boltzmann equation is reduced to the following form 
\begin{align}
    \mathbb{Z} = \mathbb{A}\mathbb{Z} - \mathbb{Y}, 
\end{align}
where 
\begin{align}
\mathbb{Z} = \begin{pmatrix}
\lambda^+\\  a^+\\ b^+\\ c^+\\
\lambda^-\\ a^-\\ b^-\\ c^-\\
\end{pmatrix}
\end{align}
\begin{widetext}
\begin{align}
\mathbb{A}=\begin{pmatrix}
\alpha^{++}F^+&\alpha^{++}G^+&\alpha^{++}I^+&\alpha^{++}J^+&\alpha^{+-}F^-&\alpha^{+-}G^-&\alpha^{+-}I^-&\alpha^{+-}J^-\\
\alpha^{++}G^+&\alpha^{++}O^+&\alpha^{++}P^+&\alpha^{++}Q^+&\alpha^{+-}G^-&\alpha^{+-}O^-&\alpha^{+-}P^-&\alpha^{+-}Q^-\\
\alpha^{++}I^+&\alpha^{++}P^+&\alpha^{++}S^+&\alpha^{++}U^+&\alpha^{+-}I^-&\alpha^{+-}P^-&\alpha^{+-}S^-&\alpha^{+-}U^-\\
\alpha^{++}J^+&\alpha^{++}Q^+&\alpha^{++}U^+&\alpha^{++}V^+&\alpha^{+-}J^-&\alpha^{+-}Q^-&\alpha^{+-}U^-&\alpha^{+-}V^-\\
\alpha^{-+}F^+&\alpha^{-+}G^+&\alpha^{-+}I^+&\alpha^{-+}J^+&\alpha^{--}F^-&\alpha^{--}G^-&\alpha^{--}I^-&\alpha^{--}J^-\\
\alpha^{-+}G^+&\alpha^{-+}O^+&\alpha^{-+}P^+&\alpha^{-+}Q^+&\alpha^{--}G^-&\alpha^{--}O^-&\alpha^{--}P^-&\alpha^{--}Q^-\\
\alpha^{-+}I^+&\alpha^{-+}P^+&\alpha^{-+}S^+&\alpha^{-+}U^+&\alpha^{--}I^-&\alpha^{--}P^-&\alpha^{--}S^-&\alpha^{--}U^-\\
\alpha^{-+}J^+&\alpha^{-+}Q^+&\alpha^{-+}U^+&\alpha^{-+}V^+&\alpha^{--}J^-&\alpha^{--}Q^-&\alpha^{--}U^-&\alpha^{--}V^-\\
\end{pmatrix}
\end{align}
\end{widetext}

\begin{align}
\mathbb{Y}=
\begin{pmatrix}
\alpha^{++}H^++\alpha^{+-}H^-\\
\alpha^{++}N^+-\alpha^{+-}N^-\\
\alpha^{++}L^+-\alpha^{+-}L^-\\
\alpha^{-+}M^+-\alpha^{--}M^-\\
\alpha^{--}H^++\alpha^{-+}H^-\\
\alpha^{--}N^--\alpha^{-+}N^+\\
\alpha^{--}L^--\alpha^{-+}L^+\\
\alpha^{--}M^--\alpha^{-+}M^+\\
\end{pmatrix}
\end{align}

The relevant integrals involved in the above matrices are:
\begin{align}
    \iint d\theta' d\phi'f^{\chi'}(\theta',\phi')&=F^{\chi'}\nonumber\\
    \iint d\theta' d\phi'f^{\chi'}(\theta',\phi') h^{\chi'}&=H^{\chi'}
\end{align}
\begin{align}
    \iint d\theta' d\phi'f^{\chi'}(\theta',\phi')\cos{\theta'}&=G^{\chi'}\nonumber\\
    \iint d\theta' d\phi'f^{\chi'}(\theta',\phi')\sin{\theta'}\cos{\phi'}&=I^{\chi'}
\end{align}
\begin{align}
    \iint d\theta' d\phi'f^{\chi'}(\theta',\phi')\sin{\theta'}\sin{\phi'}&=J^{\chi'}\nonumber\\
    \iint d\theta' d\phi'f^{\chi'}(\theta',\phi')\sin^2{\theta'}\cos^2{\phi'}&=S^{\chi'}
\end{align}
\begin{align}
    \iint d\theta' d\phi'f^{\chi'}(\theta',\phi')h^{\chi'}(\theta',\phi')\cos{\theta'}&=N^{\chi'}\nonumber\\
    \iint d\theta' d\phi'f^{\chi'}(\theta',\phi')h^{\chi'}(\theta',\phi')\sin{\theta'}\cos{\phi'}&=L^{\chi'}
\end{align}
\begin{align}
    \iint d\theta' d\phi'f^{\chi'}(\theta',\phi')h^{\chi'}(\theta',\phi')\sin{\theta'}\sin{\phi'}&=M^{\chi'}\nonumber\\
    \iint d\theta' d\phi'f^{\chi'}(\theta',\phi')\cos^2{\theta'}&=O^{\chi'}
\end{align}
\begin{align}
    \iint d\theta' d\phi'f^{\chi'}(\theta',\phi')\sin{\theta'}\cos{\theta'}\cos{\phi'}&=P^{\chi'}\nonumber\\
    \iint d\theta' d\phi'f^{\chi'}(\theta',\phi')\sin{\theta'}\cos{\theta'}\sin{\phi'}&=Q^{\chi'}
\end{align}
\begin{align}
    \iint d\theta' d\phi'f^{\chi'}(\theta',\phi')\sin^2{\theta'}\cos{\phi'}\sin{\phi'}&=U^{\chi'}\nonumber\\
    \iint d\theta' d\phi'f^{\chi'}(\theta',\phi')\sin^2{\theta'}\sin^2{\phi'}&=V^{\chi'}
\end{align}

\bibliographystyle{apsrev4-1}
\bibliography{biblio.bib}
\end{document}